\documentclass[a4paper,fleqn,usenatbib]{mnras}
\usepackage{graphicx}
\usepackage{amsmath}
\usepackage{amssymb}
\usepackage{txfonts}
\usepackage{subcaption}
\usepackage{tabularx}
\usepackage[usenames,dvipsnames,svgnames,table]{xcolor}
\usepackage{txfonts}
\DeclareSymbolFont{cmletters}{OML}{cmm}{m}{it}
\DeclareMathSymbol{v}{\mathalpha}{cmletters}{"76}
\newcommand{\GG}[1]{}

\newcommand{\RedeclareMathOperator}[2]{\renewcommand{#1}{}\let#1\relax\DeclareMathOperator{#1}{#2}}

\newcommand{\hammer}{{\tt{H-AMR}}}
\newcommand{\harm}{{\tt{HARM2D}}}
\newcommand{\harmpi}{{\tt{harmpi}}}

\newcommand\simless\lesssim
\newcommand\simgreat\gtrsim

\title[AGN jet acceleration]{Accelerating AGN jets to parsec scales using general relativistic MHD simulations}
\author[Chatterjee, Liska, Tchekhovskoy, Markoff]{K. Chatterjee$^{1}$\thanks{E-mail: k.chatterjee@uva.nl}, M. Liska$^{1}$, A. Tchekhovskoy$^{2,3,4,5}$, S.B. Markoff$^{1,6}$
\\
$^{1}$Anton Pannekoek Institute for Astronomy, University of Amsterdam, Science Park 904, 1098 XH Amsterdam, The Netherlands\\
$^{2}$Center for Interdisciplinary Exploration \& Research in Astrophysics (CIERA), Physics \& Astronomy, Northwestern University, Evanston, IL 60202, USA\\
$^{3}$Departments of Astronomy and Physics, Theoretical Astrophysics Center, University of California Berkeley, Berkeley, CA 94720-3411, USA\\
$^{4}$Lawrence Berkeley National Laboratory, 1 Cyclotron Rd, Berkeley, CA 94720, USA\\
$^{5}$Kavli Institute for Theoretical Physics, Kohn Hall, University of California at Santa Barbara, Santa Barbara, CA 93106, USA\\
$^{6}$Gravitation Astroparticle Physics Amsterdam (GRAPPA) Institute, University of Amsterdam, Science Park 904, 1098 XH Amsterdam, The Netherlands
}

\date{Accepted XXX. Received YYY; in original form ZZZ}
\pubyear{2019}
\begin{document}
\label{firstpage}
\pagerange{\pageref{firstpage}--\pageref{lastpage}} 
\maketitle
\begin{abstract}
Accreting black holes produce collimated outflows, or jets, that traverse many orders of magnitude in distance, accelerate to relativistic velocities, and collimate into tight opening angles. Of these, perhaps the least understood is jet collimation due to the interaction with the ambient medium. In order to investigate this interaction, we carried out axisymmetric general relativistic magnetohydrodynamic simulations of jets produced by a large accretion disc, spanning over 5 orders of magnitude in time and distance, at an unprecedented resolution. Supported by such a disc, the jet attains a parabolic shape, similar to the M87 galaxy jet, and the product of the Lorentz factor and the jet half-opening angle, $\gamma\theta\ll 1$, similar to values found from very long baseline interferometry (VLBI) observations of active galactic nuclei (AGN) jets; this suggests extended discs in AGN. We find that the interaction between the jet and the ambient medium leads to the development of pinch instabilities, which produce significant radial and lateral variability across the jet by converting magnetic and kinetic energy into heat. Thus pinched regions in the jet can be detectable as radiating hotspots and may provide an ideal site for particle acceleration. Pinching also causes gas from the ambient medium to become squeezed between magnetic field lines in the jet, leading to enhanced mass-loading of the jet and potentially contributing to the spine-sheath structure observed in AGN outflows.
\end{abstract}


\section{Introduction}
\label{sec:introduction}

Powered by magnetic fields brought inwards by infalling gas, relativistic jets form in a variety of astrophysical black hole systems such as active galactic nuclei (AGN), black hole X-ray binaries (XRBs), tidal disruption events (TDEs), and neutron star mergers. Through the exchange of energy with the ambient medium, jets heat up gas in the interstellar medium (ISM) creating a feedback loop between the AGN and its environment \citep[e.g., ][]{bower2006,Fabian2012}. AGN feedback plays a key role in regulating the growth of galaxies and star formation (e.g., \citealt{SilkRees98,Magorrian98}; for a recent review, see \citealt{har18feed}) and therefore, its implementation in cosmological simulations \citep[e.g., ][]{springel2005,anglesalcazar_2017_FIRE,Weinberger2019_illustris} warrants an accurate understanding of jet energetics \citep[e.g., ][]{Sijacki2007,bourne_arepo2017}. Addressing how jets build up the magnetic and kinetic energy required to explain the radiative emission seen from jet observations is still an open problem. Therefore, by studying how jets accelerate and interact with the ambient gas, we can better understand jet emission and its relation to jet-ISM interactions.

Very long baseline interferometry (VLBI) imaging of AGN jets \citep[e.g., ][]{lister2016} has made it possible to track radio emission features of the jet over long time periods and estimate the bulk kinematic properties. As an example, \citet{asadanak2012} show that the parsec scale jet of the galaxy M87 is roughly parabolic in structure.  At $\simeq 10^5$ gravitational radii away from the central black hole, the jet becomes conical, with the transition appearing to coincide with the bright HST-1 feature \citep{Biret1999}. The HST-1 knot may be a result of self-collimation \citep[e.g., ][]{pol10} or a changing density profile of the ISM \citep{asadanak2012,nak2013}. The change in the confining pressure may cause the jets to over-collimate and activate magnetic instabilities or result in the formation of internal shocks, leading to particle acceleration and the appearance of knots, stationary or moving features \citep{tch16,Duran17}. The same phenomena also seem to occur in the jets of stellar mass black holes in XRBs \citep[e.g., ][]{mar01,mar05,russellD2013,Romero2017}. Accelerated particles produce high energy emission, an important observational probe of these outflows. Due to the intricate relationship between jet structure, kinematics and particle acceleration, a variety of theoretical outflow models have been constructed to meaningfully interpret jet observations \citep[for a review, see][]{BOOK2012bhae.book}.

In spite of the large body of theoretical work on outflows, understanding their dynamics remains a challenge, particularly because the jet structure depends on the interaction of the outflow with the ambient medium in a complex and non-linear way. Understanding this coupling at any scale requires the knowledge of the magnetic field configuration as well as the physical conditions at the jet base and the ambient medium, since both constrain the jet's final energy content and Lorentz factor. However, constructing analytic models of jet acceleration is difficult due to the highly non-linear nature of the governing equations. Idealised semi-analytic models, or SAMs \citep[e.g., ][]{vk03a,bes06,broderick_2006,lyub09,Pu2015_SS}, provide reasonable estimates of jet properties and have been used to constrain the behaviour of the jets both near and far away from the event horizon. 

Both the black hole and the surrounding accretion disc can launch outflows via two popular mechanisms. The Blandford-Znajek \citep[BZ77; ][]{bz77} mechanism taps into the rotational energy of the black hole, via the magnetic field lines connected to the black hole event horizon, and leads to relativistic Poynting flux dominated jets \citep{bk00, kom01, gam03, kom05, mck05, tch10a}. Field lines anchored in the accretion disc can launch sub-relativistic mass dominated winds by means of the Blandford-Payne mechanism (BP82; \citealt{bp82}, see also \citealt{meier97,mizuno04}). While SAMs can explain the basic physics of energy conversion from magnetic to kinetic form in jets, they are time-independent solutions that often neglect accretion disc physics as well as general relativistic effects of the space-time geometry around a (spinning) black hole. Due to these simplifications, SAMs are not able to completely capture the complexities of real jets, which is why general relativistic magneto-hydrodynamic (GRMHD) models are required \citep[e.g., ][]{dev03,mck06jf}. GRMHD simulations typically model accretion starting from an initial gas torus, which, when threaded with large scale vertical or toroidal magnetic flux, launches both powerful BZ77- and BP82-type outflows \citep[e.g., ][]{mck05,hk06,tch11,liska2018b}. 

There is currently a disagreement in the literature regarding jet acceleration. While semi-analytic models \citep[e.g., ][]{bes98,bes06} and idealised jet simulations \citep[i.e. the simulations that model the ambient medium by placing a conducting wall at the jet's outer boundary; e.g., ][]{kom07,kom09,tch10a} found efficient acceleration of jets to nearly the maximum Lorentz factor by utilising the jet's entire energy budget, similar high efficiencies have never been seen in GRMHD simulations of a disc-jet system \citep[e.g., ][]{mck06jf,brom2016,Duran17}, where internal pinch/kink instabilities are seen to convert a considerable amount of the jet's energy content to heat by dissipating magnetic energy \citep[e.g., ][]{eichler_93,spruit97,begelman_pinch_instability_1998,giannios2006}. Does this mean that realistic systems are incapable of producing efficiently accelerating jets? In order to answer this question, we need to evolve jets over large time and distance scales since jet acceleration is an inherently time-dependent process in the presence of an ambient medium. 

In this work, we revisit the problem of jet acceleration and collimation using a new state-of-the-art, GPU-accelerated GRMHD code \hammer{} \citep{liska2018a} to carry out high-resolution axisymmetric disc-jet simulations. This was not possible for a long time, since as jets collimate, their magnetic field lines bunch up towards the jet axis, necessitating an extremely high resolution in the polar region. However, due to several algorithmic improvements, we can produce simulations spanning more than 5 orders of magnitude in both time and distance with unparalleled resolutions, presenting us with a unique opportunity to understand the jet physics. In Sec.~\ref{sec:code-and-setup}, we give an overview of our problem setup. In Sec.~\ref{sec:model}, we describe our initial conditions. In Sec.~\ref{sec:results}, we present our results for disc-jet models. In Sec.~\ref{sec:ideal}, we compare our disc-jet models with an idealised one. In Sec.~\ref{sec:discussion}, we discuss our results and we conclude in Sec.~\ref{sec:conclusions}.

\section{Numerical setup}
\label{sec:code-and-setup}

We use the \hammer{} code \citep{liska2018b,Liska2018C,liska2018a} that builds upon \harmpi{}\footnote{Freely available at \href{https://github.com/atchekho/harmpi}{https://github.com/atchekho/harmpi}} and \harm{} \citep{gam03,nob06} and evolves the GRMHD equations on a fixed spacetime. It uses a Harten-Lax-van Leer (HLL) Riemann solver \citep{har83} to calculate fluxes at cell faces and a staggered grid akin to \citet{gar05} to evolve the magnetic fields. \hammer{} performs third order accurate spatial reconstruction at cell faces from cell centres using a piece-wise parabolic method \citep[PPM; ][]{col84} and is second order accurate in time.  The novelty of \hammer{} lies in the use of advanced features such as adaptive mesh refinement (AMR, not utilised in this work) and a local adaptive time-step (LAT). The LAT reduces the number of conserved to primitive variable inversions and thereby increases the accuracy of the simulation  (Appendix~\ref{sec:LAT}); additionally, it speeds up the code by a factor of $\sim 3{-}5$. In its current version, in addition to CPUs, \hammer{} also runs on graphical processing units (GPUs), achieving $10^8$ zone cycles per second on an NVIDIA Tesla V100 GPU and shows excellent parallel scaling to thousands of GPUs. 

\begin{table*}
	\caption{Simulations carried out for this work. Model names shorthand disc inner radius  $r_{\rm in}$ (R), black hole spin (A; $0.9375$, common for all models) and the $b^2/\rho$ floor value (B; in other words, the maximum magnetisation $\sigma_0$) with additional parameters including S for single loop magnetic field and LR and HR for the low and high resolution versions respectively. We also mention the pressure-maximum radius $r_{\rm max}$, outer grid radius $r_{\rm out}$, grid resolution, simulation time $t_{\rm sim}$, MRI quality factor $Q^x$ at $t=2\times10^4t_g$ (see text in Sec.~\ref{sec:lower-jet:-analytic}) and magnetic field configuration (indicated by the number of field line loops) as well as the use of local adaptive time stepping (LAT).}
    \label{tab:models}
\begin{center}
\begin{tabularx}{1.9\columnwidth}{l c c c c c c c c c c}
\hline
Full Model & Short & $r_{\rm in}$& $r_{\rm max}$ & $r_{\rm out}$ &Floor& Resolution&$t_{\rm sim}$ & $Q$ factor & Initial field loop & LAT\\
name & name & [$r_g$] & [$r_g$]&[$r_g$]& $b^2/\rho$ & $N_{r}\times N_{\theta}$ & [$10^5t_g$] & $Q^{r}, Q^{\theta}$& configuration & \\
\hline
R36A93B10 & B10 & 36 & 73.97&$10^6$&10&$6000\times800$& 5.0 & 62,14 & multiple&on\\
R36A93B10-S & B10-S & 36 & 73.97&$10^5$&10&$4000\times800$& 1.2& 96, 24 &single &on\\
R25A93B10 & B10-R & 25 & 50&$10^5$&10&$3000\times800$& 1.2&36, 10 & multiple&on\\
R36A93B10-S-LR & B10-SLR & 36 & 73.97&$10^4$&10&$640\times256$& 0.8& 20, 4 &single &on\\
R36A93B10-S-LR-L & B10-SLRL & 36 & 73.97&$10^4$&10&$640\times256$& 0.8& 20, 4 &single &off\\
R36A93B10-HR & B10-HR & 36 & 73.97&$10^5$&10&$18000\times1200$& 0.5&  -,- & multiple&on\\
R36A93B3 & B3 & 36 & 73.97&$10^5$&3&$4000\times800$& 1.5& 32,10& multiple&on\\
R36A93B50 & B50 & 36 & 73.97&$10^5$&50&$4000\times800$& 1.9& 32,10& multiple&on\\
R36A93B100 & B100 & 36 & 73.97&$10^5$&100&$4000\times800$& 2.0& 40, 12 &multiple &on\\
\hline
\end{tabularx}

\end{center}

\end{table*}

\subsection{Numerical grid}
\label{sec:Numerical_grid}
We use units such that $G=M=c=1$. This sets both the characteristic timescale, $t_g=GM/c^3$, and spatial scale, $r_g=GM/c^2$, to unity. In fact, our simulations are scale-free, i.e., if we provide the black hole mass $M$ and mass accretion rate $\dot{M}$, we can rescale our simulation to the corresponding black hole system. Our grid is axisymmetric, extending from $0.85r_{\rm H}$ to $10^5 r_g$, where $r_{\rm H}$ is the event horizon radius, $r_{\rm H}=r_g(1+\sqrt{1-a^2})$, where we set the dimensionless black hole spin parameter to $a=0.9375$. We carry out the simulations on a uniform grid in internal coordinates ($x^\mu$, see Appendix~\ref{sec:grid}) that are transformations of the spherical polar coordinates ($t,r,\theta,\phi$) in the Kerr-Schild foliation specifically optimised to follow the collimating jets. To resolve the jets, we typically use a numerical resolution of $3,000-18,000$ cells in the radial direction and $800$ cells in the polar direction (Table~\ref{tab:models}). We use the following boundary conditions (BCs): in $r-$, we use outflow BCs at the inner and outer grid radii; at the poles, we reflect the $\theta-$ component of the velocity and magnetic field.

\subsection{Density floors}
\label{sec:floors}
In the jet funnel matter either falls towards the black hole due to gravity or gets flung out due to magnetic forces depending on its location with respect to the stagnation surface, at which the inward pull of gravity balances the outward centrifugal force. The vacuum region thus created at this surface is a common numerical issue for grid-based MHD code, as gas density drops too low to be handled accurately. To avoid this, we replenish gas density and internal energy if they fall too low. We may physically motivate our floors as approximating the poorly understood processes leading to particle creation around the stagnation surface. Namely, mass flow divergence around the stagnation surface may lead to charge separation followed by particle creation \citep[e.g.,][]{1998ApJ...497..563H,2015ApJ...809...97B,2016ApJ...818...50H,2016A&A...593A...8P,2017PhRvD..96l3006L,2018ApJ...863L..31C,Parfrey2019}. In our simulations, we adopt the approach of \citet{res17} and mass-load the jets in the drift frame of the magnetic field. In this method, the component of the fluid momentum along the magnetic field is conserved. Our floor model consists of setting a minimum rest mass density $\rho_{\rm fl}=\max[b^2/\sigma_0,2\times10^{-4}(r/r_g)^{-2.5}]$ and a minimum internal energy of $u_{g,\rm fl}=\max[b^2/750,2\times10^{-5}(r/r_g)^{-2.5\Gamma}]$, where $b=\sqrt[]{b^{\mu}b_{\mu}}$, $b^{\mu}$, $\sigma_0$ and $\Gamma$ are the co-moving magnetic field strength, magnetic four-vector, maximum magnetisation and the ideal gas law adiabatic index, respectively.

\section{Simulation models}
\label{sec:model}
We have carried out 9 simulations (including 2 models from the Appendix) to elucidate the physics of jet acceleration and interaction with the ambient medium (see Table~\ref{tab:models}). Most of our models ran for $t>10^5 t_g$. In all models we start with a \citet{fis76} (FM76) torus in hydrostatic equilibrium around a rapidly spinning $a=0.9375$ Kerr black hole. We place the torus inner edge at $r_{\rm in}$ and density maximum at $r_{\rm max}$. For all but one of our simulations, we set $r_{\rm in}=36r_g$ and $r_{\rm max}=73.97r_g$. We choose the density scale by setting $\max\rho=1$. We adopt the ideal gas law equation of state and set the adiabatic index to that of a non-relativistic gas $\Gamma=5/3$. We set the magnetic field vector potential as described in Secs.~\ref{sec:MAD} and \ref{sec:fiducial} and normalise it such that $\max p_{\rm g}/\max p_{\rm B} =100$, where $p_{\rm g}$ and $p_{\rm B}$ are the gas and magnetic pressure, respectively. 

\subsection{Single field loop setup}
\label{sec:MAD}

\begin{figure}
	\includegraphics[width=\columnwidth]{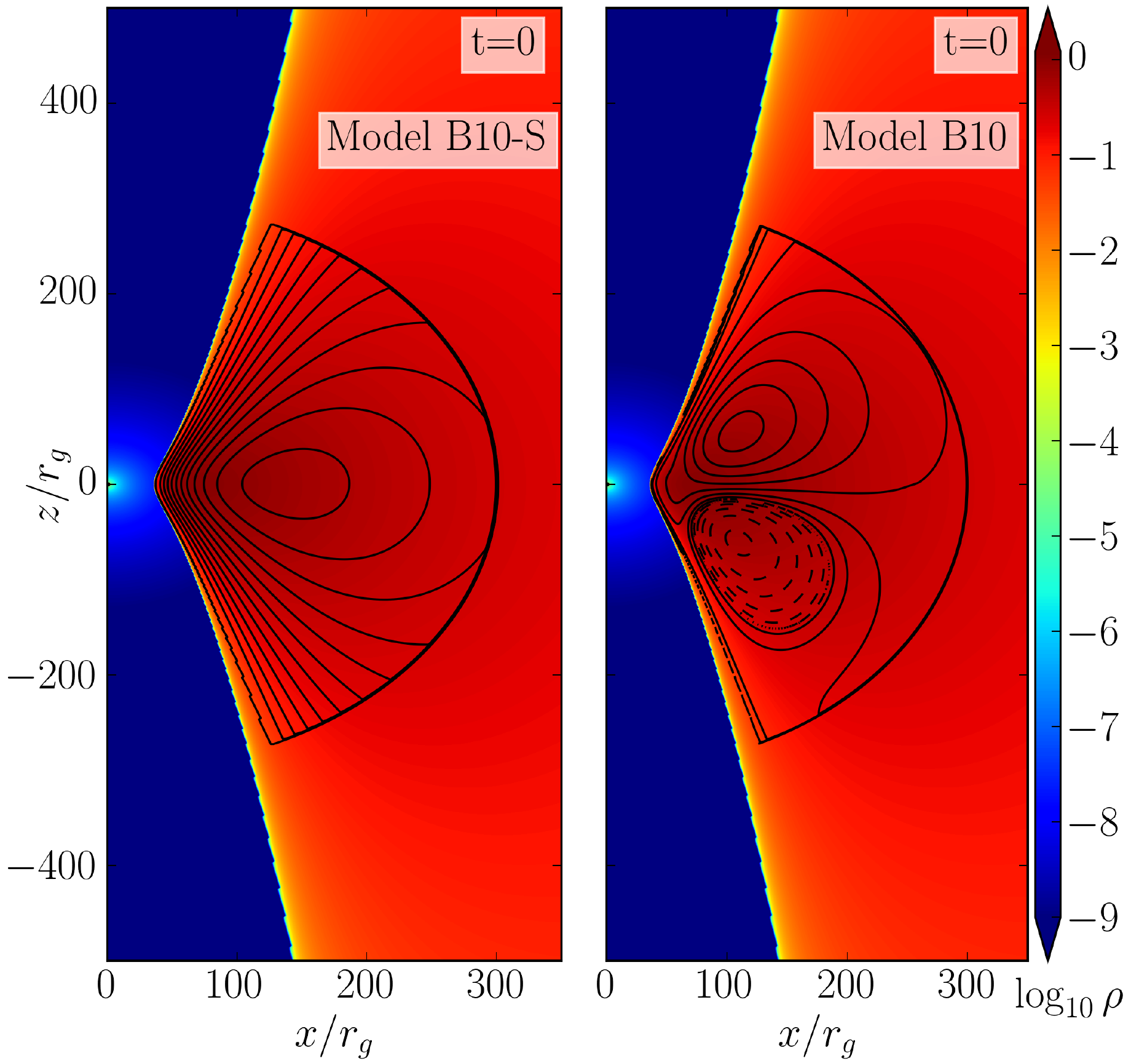}
    \caption{We start out with a magnetised disc around a spinning black hole (at origin). We show two different initial conditions for our accretion disc with the fluid-frame density $\rho$ in colour (red shows high and blue low density; see the colour bar), representing an equilibrium hydrodynamic torus around the spinning black hole, with black lines showing the initial magnetic field configuration of a large poloidal field loop (left panel, model B10-S) and two poloidal loops of opposite polarity (right panel, model B10). Dashed black lines show field lines containing negative magnetic flux. The simulation grid extends out to $10^5 r_g$ or $10^6r_g$, depending on the model (see Table~\ref{tab:models}).}
    \label{fig:test_initial}
\end{figure}

\begin{figure}
	\includegraphics[width=\columnwidth]{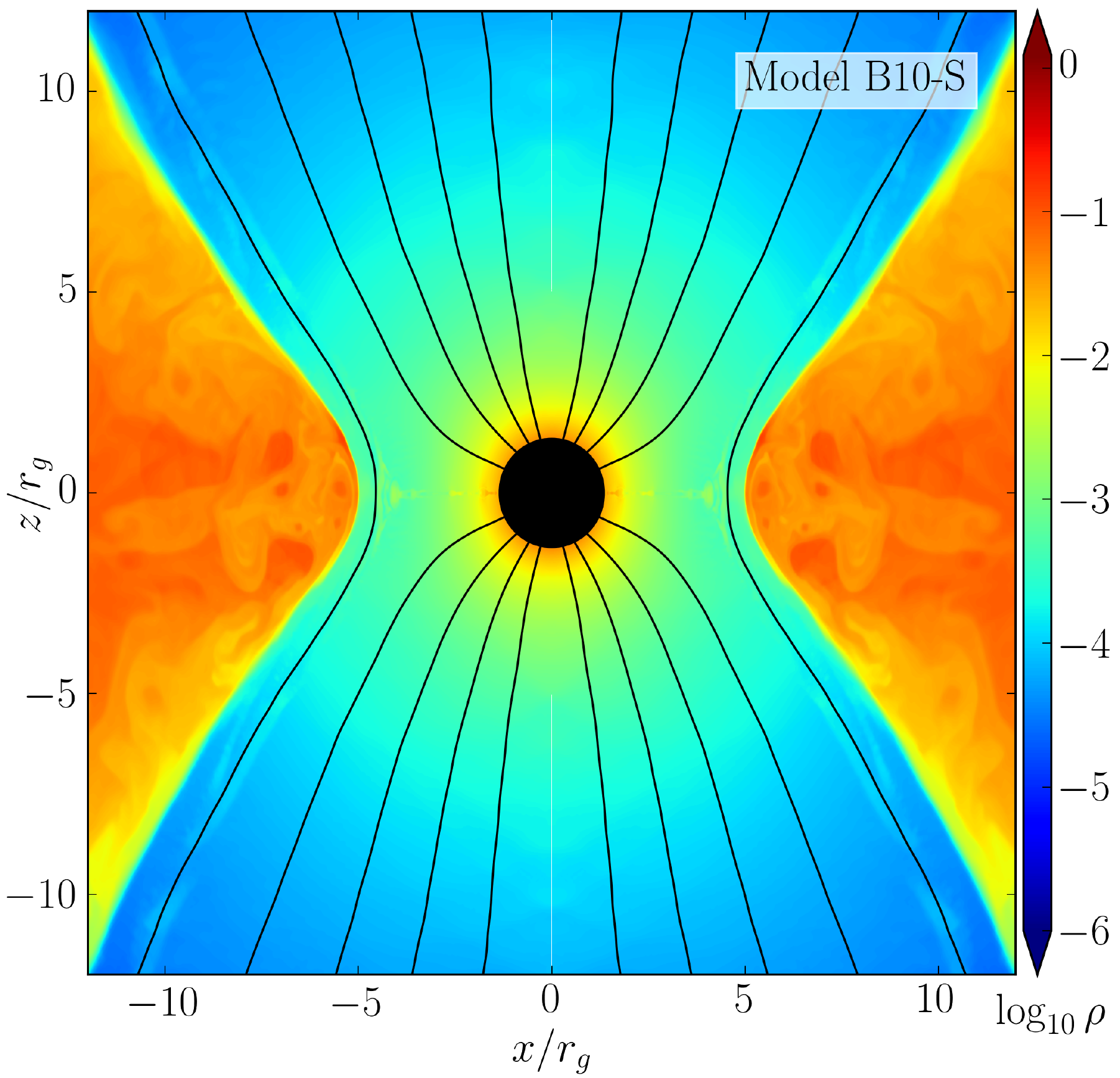}
	\caption{Large magnetic flux near the black hole can stop accretion from the disc. We show a time snapshot of the model B10-S within the innermost ten gravitational radii. The central black hole is shown in black along with the rest mass matter density $\rho$ (in colour) and the magnetic field lines (in black). Black hole gravity attempts to pull in matter from the accretion disc (orange-red region), while the accumulated strong magnetic flux pushes gas away, resulting in the magnetically arrested disc (MAD) state. In axisymmetry, the accretion rate is highly variable for MADs, which in turn affects the jet (Fig.~\ref{fig:mdot}).}
    \label{fig:MAD}
\end{figure}

We first start with a single poloidal magnetic field loop, B10-S model, where `S' stands for `single loop', as seen in the left panel of Fig~\ref{fig:test_initial}. We choose the magnetic vector potential of the form: 
\begin{equation}
\begin{centering}
\label{eqn:Aphi_MAD}
A_{\phi}= \begin{cases}
    (\rho-0.05)^2r^{2}, & \text{if $r<300r_g$ and $\rho>0.05$}. \\
    0, & \text{otherwise}.
    \end{cases}
\end{centering}
\end{equation}

\noindent As the black hole starts accreting the gas, it drags along the magnetic flux, which starts threading the event horizon. Over time, the accumulated magnetic flux becomes large enough to push the gas away and prevent it from accreting, as seen in Fig.~\ref{fig:MAD}. This is an axisymmetric variant \citep{proga06} of the magnetically arrested disc (MAD) state \citep{igu03,nia03,tch11}, which leads to highly efficient outflows: their efficiency, defined as total outflow power normalised by the time-average accretion rate, can exceed unity: $P_{\rm outflow}/\langle\dot{M}c^2\rangle>1$. Here, $P_{\rm outflow}=\dot{M}c^2-\dot{E}$, where $\dot{E}=\iint F_{\rm E}dA_{\theta\phi}$ is the energy accretion rate (defined to be positive when the energy flows into the black hole, see Eq.~\ref{eqn:mu} for $F_{\rm E}$ definition), $dA_{\theta\phi}=\sqrt{-g}d\theta d\phi$ is the surface area element, and $g=|g_{\mu\nu}|$ is the determinant of the metric. Similarly, we can define the magnetic flux on the black hole as $\Phi_{\rm BH}=0.5\iint|B^r|dA_{\theta\phi}$, where the integral is over the entire event horizon. We also define its dimensionless counterpart normalised by the mass accretion rate, $\phi_{\rm BH}= \Phi_{\rm BH}/\langle \dot Mr_g^2c\rangle$.

\begin{figure}
	\includegraphics[width=\columnwidth]{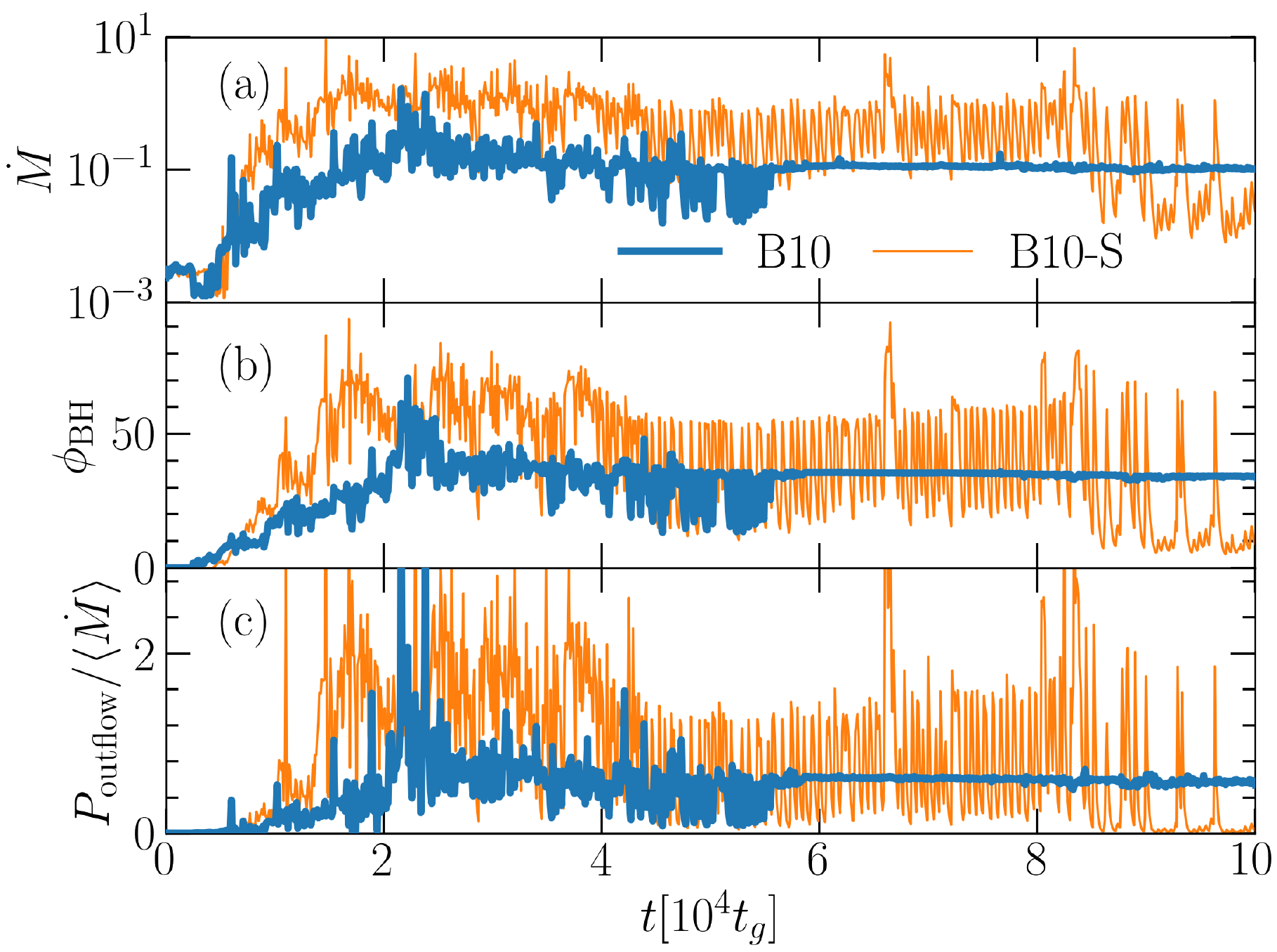}
    \caption{The MAD model B10-S (orange) shows an oscillating mass accretion rate (panel a), magnetic flux (panel b) and total outflow power (panel c). In contrast, model B10 (blue) shows steady behaviour in all. The normalised $\dot{M}$ is computed over $(4-8)\times10^4t_g$. Free of violent variability, model B10 is suitable for studying large scale jet dynamics.}
    \label{fig:mdot}
\end{figure}

In axisymmetry, the magnetic fields on the black hole and the surrounding gas have no way of exchanging places: the magnetic interchange instability requires a third dimension. Thus constrained by 2D symmetry, the fight between gravitational and magnetic forces degenerates into the gas bouncing in and out on top of the magnetic barrier, as seen in Fig.~\ref{fig:MAD}. Figure~\ref{fig:mdot} shows that this results in large (up to an order of magnitude) fluctuations in the mass accretion rate $\dot{M}$ (Fig.~\ref{fig:mdot}a), dimensionless magnetic flux $\phi_{\rm BH}$ (Fig.~\ref{fig:mdot}b) and the dimensionless total outflow power (Fig. \ref{fig:mdot}c). Such oscillations make it difficult to extract the physics from the simulations and thereby make this configuration undesirable.

\subsection{Fiducial setup}
\label{sec:fiducial}

We initialise our fiducial model B10 with a disc threaded with a large enough magnetic flux such that a powerful jet forms while taking care to avoid over-saturating the black hole and getting a MAD (Fig.~\ref{fig:test_initial}, right). The magnetic field configuration in the torus consists of two poloidal field loops described by the following vector potential,
\begin{equation}
\label{eqn:Aphi}
\begin{centering}
A_{\phi}= \begin{cases}
    f, &\text{if $r<300r_g$ and $\rho>0.05$},\\
    0,  &\text{otherwise},
    \end{cases}
\end{centering}
\end{equation}
where $f = 0.1x_1(\rho-0.05)^{1/2} + 0.9x_1^2(\rho-0.05)^{2}\sin^2[\pi(x_1-2)/2]\sin(\pi x_2/2)$.
The first term in $f$ describes a large-scale field loop and the second term a pair of oppositely polarised smaller loops embedded within the large loop (in terms of the internal coordinates: $x_1$ and $x_2$;  Appendix~\ref{sec:grid}). The magnetic fluxes within the pair cancel exactly such that the total magnetic flux is set by the large-scale loop. This cancellation is convenient, because it gives us fine-grained control over the amount of net magnetic flux in the initial conditions and allows us to choose the positive polarity to dominate only slightly over the negative one. Unlike the model B10-S, which showed violent variability in the black hole mass accretion rate $\dot{M}$, dimensionless magnetic flux $\phi_{\rm BH}$ and the total outflow power, model B10 shows a steady behaviour of all three quantities (Fig.~\ref{fig:mdot}), primarily because it has a smaller magnetic flux in the disc. This makes model B10 ideal for studying long duration steady outflows. 

\begin{figure}
	\includegraphics[width=\columnwidth]{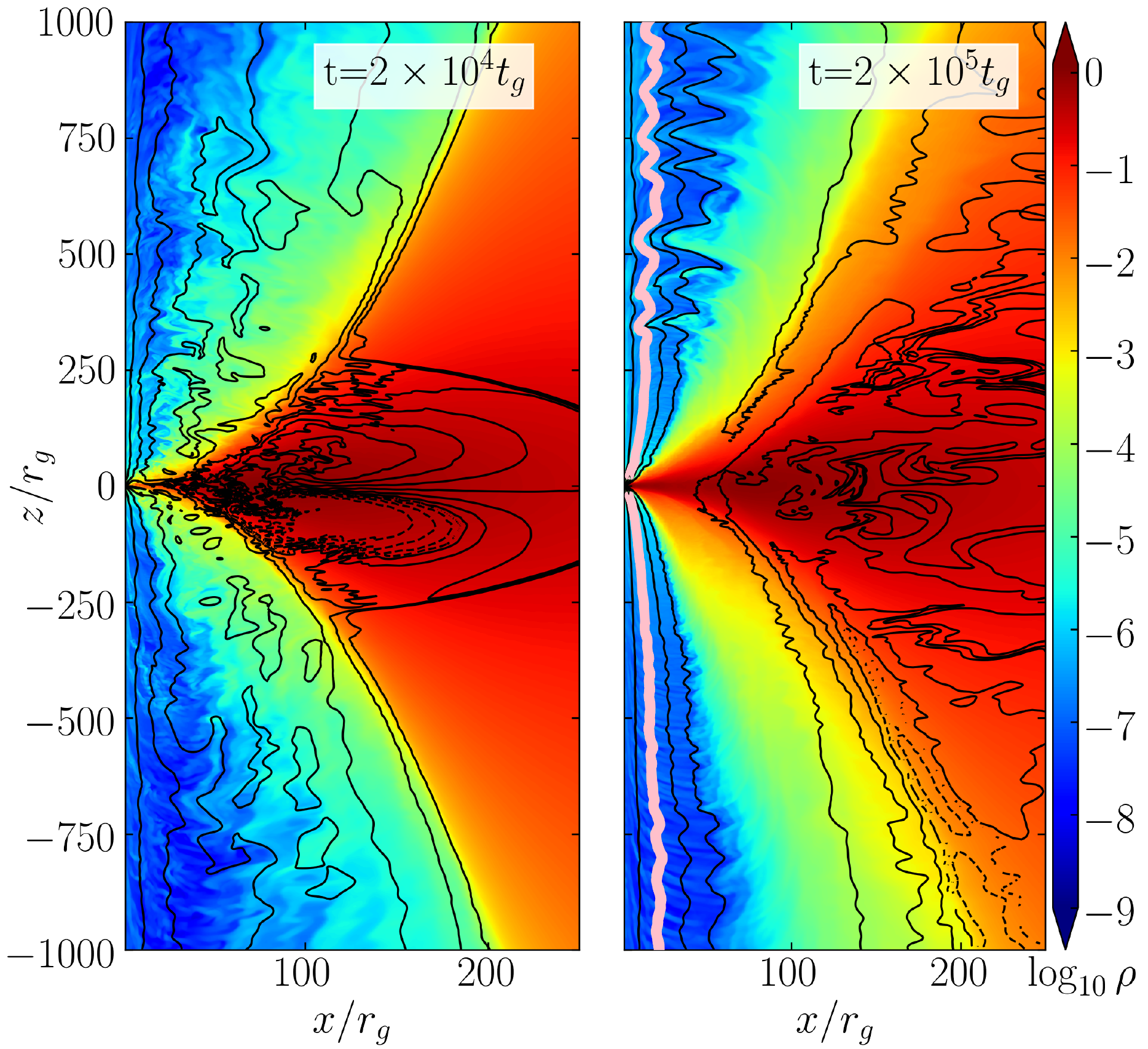}
    \caption{The disc in model B10 launches opposing jets with field lines anchored in the black hole event horizon as well as the disc. Mixing between the jet and the disc-wind mass-loads the jet over time via eddies generated from the wind-jet interaction. We show vertical slices though the density (see the colour bar) at an early time (left panel) and at late time (right panel). Pinch instabilities, in the form of finger-like projections, significantly contribute to mass-loading and plays a vital role in determining jet dynamics. In order to gauge the influence of pinching on the jet, we extract useful information about energetics along a field line (indicated by the pink line) in both jets as shown in Fig.~\ref{fig:evolution_1}.} 
    \label{fig:B10_density}
\end{figure}
\begin{figure}
	\includegraphics[width=\columnwidth]{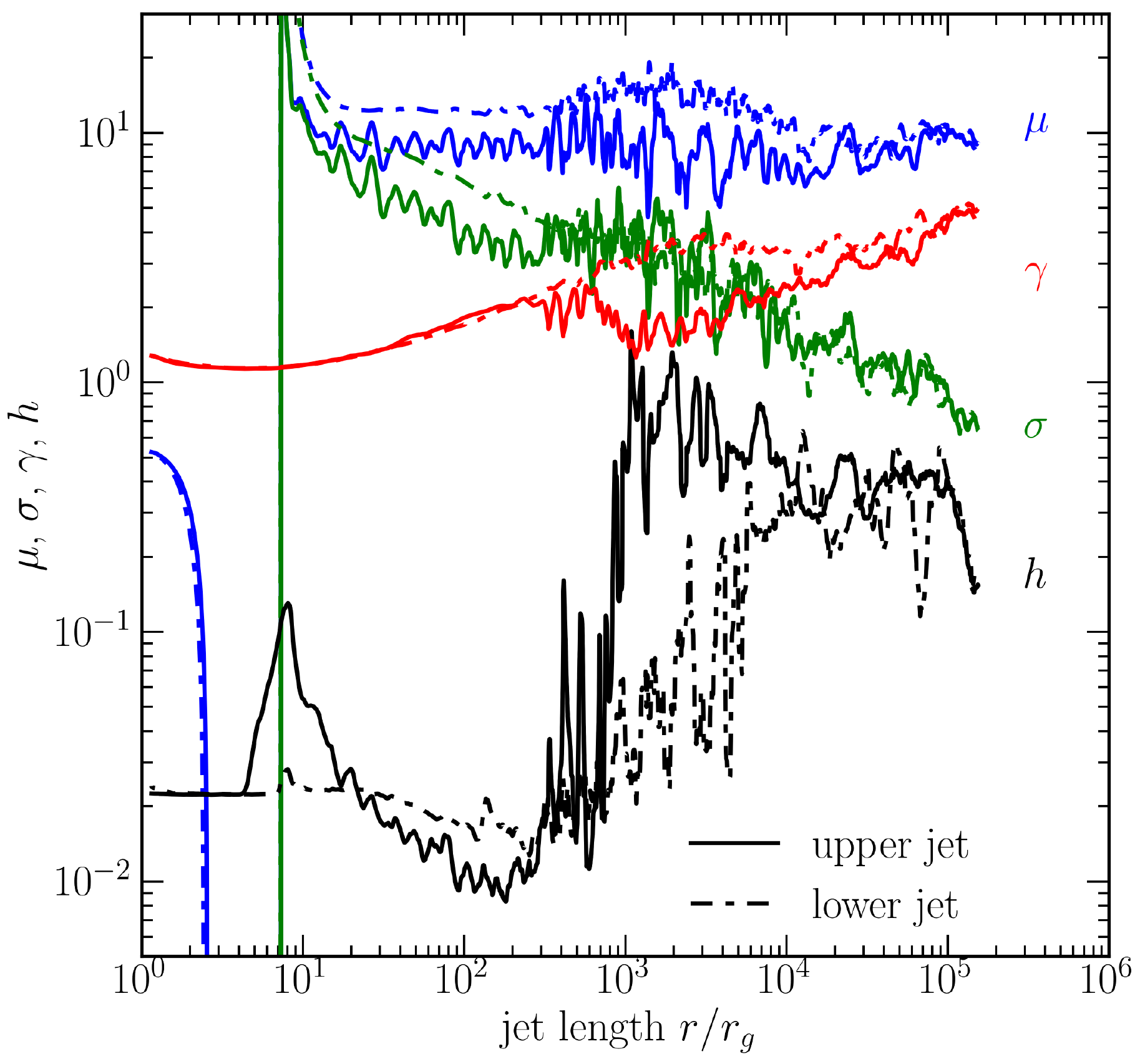}
   \caption{Magnetised jets accelerate by converting Poynting and thermal energy into kinetic energy: Lorentz factor $\gamma$ increases at the expense of decreasing magnetisation $\sigma$ and specific enthalpy $h$, while maintaining a near constant specific total energy flux $\mu$. We show the radial profile of these quantities along a field line with a foot-point half-opening angle $\theta_{j,\rm H}=0.44$~rad (Fig.~\ref{fig:B10_density}, pink) for both the upper (Fig.~\ref{fig:B10_density}, $z>0$) and lower (Fig.~\ref{fig:B10_density}, $z<0$) jets (solid and dash-dotted lines, respectively) of model B10 at $t=2\times10^5 t_g$. The two jets do not have the same acceleration profile, as the upper jet is affected by strong pinches, which lead to gas moving across the field lines in a non-uniform way and contributes to mass-loading the jet. The added inertia in the jet causes a drop in the Lorentz factor and a rise in the specific enthalpy. We radially average the plotted quantities over $\Delta r=0.01r$.}
  \label{fig:evolution_1}
\end{figure}

\section{Fiducial model results}
\label{sec:results}
\subsection{Global evolution}
\label{sec:lower-jet:-analytic}

In this section, we focus on our fiducial model B10. With time, in model B10, the accretion disc develops turbulence through the magneto-rotational instability \citep[MRI; ][]{bal91}, which leads to accretion onto the black hole and launching of the jets on both sides of the disc (Figure~\ref{fig:B10_density}). For accretion to take place, angular momentum needs to be redistributed to the outer parts of the disc via the MRI and therefore, it is important for simulations to properly resolve the MRI turbulence. To quantify this, we calculate the quality factors $Q^{r,\theta}$, where $Q^{i}=\langle2\pi v_A^{i}/(\Delta^{i}\Omega)\rangle_w$ measures the number of cells per MRI wavelength in direction $i=$[$r,\theta$], where $v_A^i$ is the Alfv\'en velocity, $\Delta^{i}$ the cell size, $\Omega$ the angular velocity of the fluid. $Q^{i}$ is averaged over the inner disc ($r<50r_g$) and weighted by $w=\sqrt[]{b^2\rho}$. We achieve $Q^{\theta}\sim 14$ (Table~\ref{tab:models}) at $t=2\times10^4t_g$, fulfilling the numerical convergence criteria \citep[see e.g., ][]{hgk11}. $Q$ values decrease over time as expected for axisymmetric systems \citep[][]{cowling34}, but since we focus on the physics of the jet, it is not a significant concern that we do not resolve the MRI well in the disc at late times. 

Initially, the jet expands as if there were no confinement (i.e., ballistically) until its ram pressure drops below the confining pressure of the disc-wind, which snaps back on the jet. Figure~\ref{fig:B10_density} shows that this unstable interaction between the disc-wind and the jet leads to oscillations of the jet-wind interface: that we refer to as pinches. The pinches also give rise to small scale eddies that mass-load the jet at late times\footnote{Movie showing magnetised jet formation of model B10: \href{https://youtu.be/4MeLZZPYsfc}{https://youtu.be/4MeLZZPYsfc}} (see Sec.~\ref{sec:entrainment} for a more detailed discussion). Interestingly, even at an early time, the jet on one side of the disc shows qualitative differences in behaviour compared to the other jet. Namely, the upper jet ($z>0$ in Fig.~\ref{fig:B10_density}) is strongly affected by the interface instabilities, while the lower jet ($z<0$ in Fig.~\ref{fig:B10_density}) remains much more stable. 

To understand how the oscillating interface affects jet dynamics, we look at how the jet evolves along a field line, which we define as a surface of constant enclosed poloidal magnetic flux, $\Phi(r,\theta)= \int^{\theta}_0 B_p dA_{\theta\phi}$, where $B_p$ is the poloidal field strength. We choose a field line whose foot-point makes an angle of $\theta_{j,\rm H}=0.44$~rad with the black hole spin axis at the event horizon. This makes up about $40$\% of the jet's half-opening angle $\theta_{\rm jet}$. Figure~\ref{fig:evolution_1} shows the evolution of various quantities along our chosen field line at $t=2\times10^5 t_g$, a long enough time for the jet to establish a quasi-steady state solution out to $10^5r_g$. First, we consider the total specific energy $\mu$, which is the maximum Lorentz factor a jet could attain if it converted all forms of energy into kinetic energy, and equals the ratio of the total energy flux, $F_{\rm E}$, and the rest-mass flux, $F_{\rm M}$,
\begin{equation}
\mu = \frac{F_{\rm E}}{F_{\rm M}} = \frac{(\rho+u_{\rm g}+p_{\rm g}+b^2) u^r u_t-b^{r}b_{t}}{-\rho u^r},
\label{eqn:mu}
\end{equation}
where $\rho$ is gas density, $u_{\rm g}$ is internal energy density, $p_{\rm g}=(\Gamma-1)u_{\rm g}$ is gas pressure, $b^{\mu}$ and $u^{\mu}$ are the magnetic and velocity four-vectors respectively. Figure~\ref{fig:evolution_1} shows that the radial profile of $\mu$ stays approximately constant for both jets, with small oscillations due to the lateral movement of gas in response to the jet pushing against the confining pressure of the disc-wind \citep{lyub09,kom2015}. The flip in the sign of $\mu$ around $8r_g$ is caused by the presence of a stagnation surface (see Sec.~\ref{sec:floors}), where $u^r=0$ and thus $\mu\rightarrow{\infty}$. Downstream of the stagnation surface, the floor values (see Sec.~\ref{sec:floors}) set the value of $\mu$. 
To quantify the conversion efficiency of magnetic to kinetic energy, we calculate the ratio of the Poynting flux, $F_{\rm EM}$, to the mass energy flux, $F_{\rm K}$, called magnetisation $\sigma$:
\begin{equation}
\begin{centering}
\sigma=\frac{F_{\rm EM}}{F_{\rm K}}=\frac{b^2u^ru_t-b^{r}b_{t}}{\rho u^{r}u_{t}},
\end{centering}
\label{eqn:sigma}
\end{equation}
where $u_t\sim -\gamma$ for $r\gg r_g$. Figure~\ref{fig:evolution_1} shows that $\sigma$ decreases to below unity as magnetic energy is converted into kinetic energy. This means that even though the jets started out strongly magnetically-dominated near the black hole, the process of acceleration converted their magnetic energy into kinetic form to the point where the jets end up becoming kinetically-dominated. Whether $\sigma$ will keep decreasing even further, leading to unmagnetised jets, or will level off around unity, leading to somewhat magnetised jets, will require a simulation extending to even larger distances. In the following sections, we take a closer look at the jet acceleration profile as well as the dissipation due to interface instabilities.

\begin{figure}
	\includegraphics[width=\columnwidth]{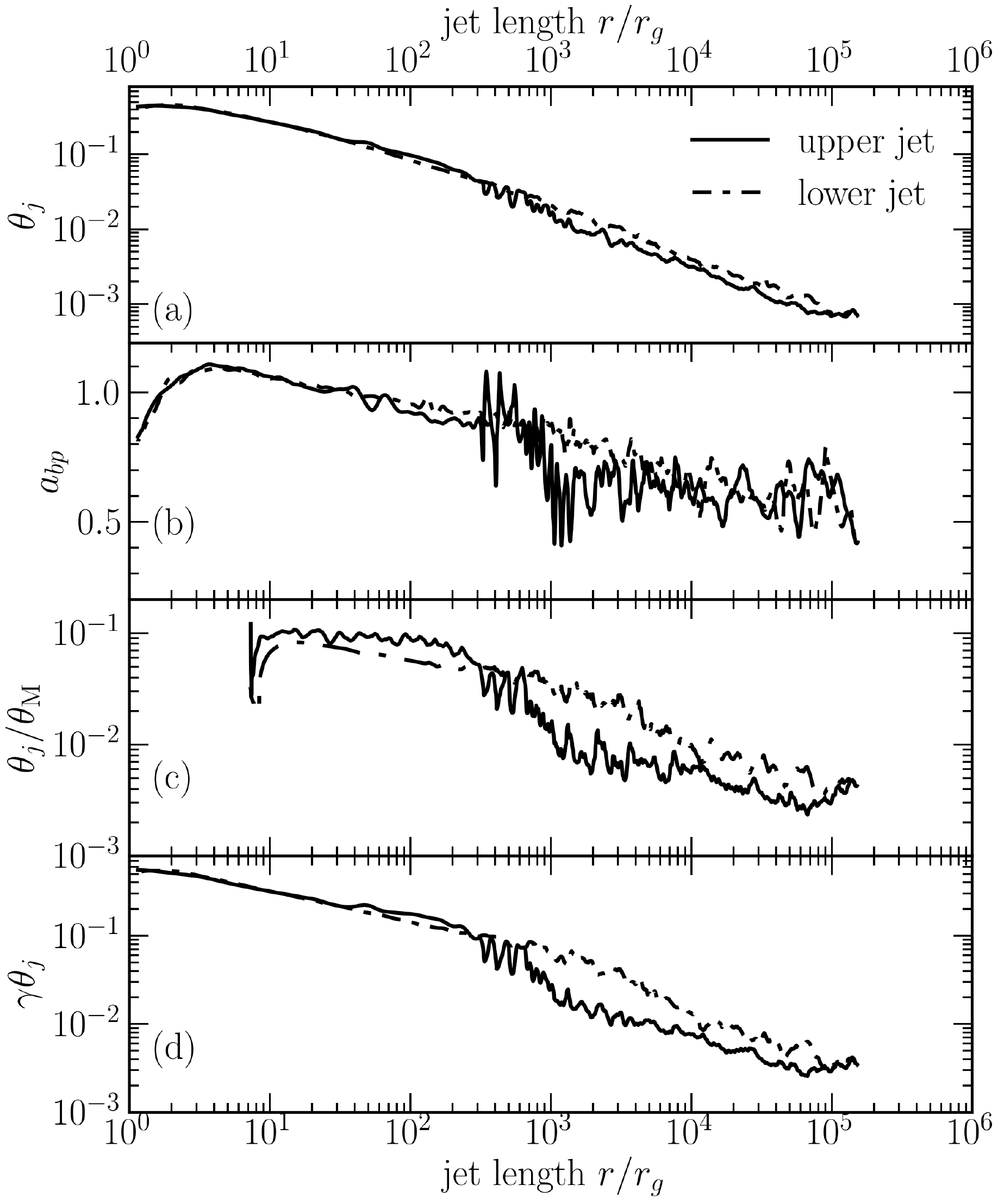}
   \caption{The collimation of the jet has a profound effect on the acceleration profile. Continuing from Fig.~\ref{fig:evolution_1}, here we show more quantities along the field line for model B10. Panel(a) shows the jet half-opening angle $\theta_j$ in radians, (b) the bunching parameter $a_{bp}$ (see text), (c) the transverse causality parameter $\theta_j/\theta_{\rm M}$ and (d) a second causality parameter $\gamma \theta_j$, useful for observed jets. Both jets show a continuous parabolic collimation profile. Strong pinching in the upper jet forces reconnection between two nearby field lines, and result in poloidal flux dissipation. The lower jet is also affected by pinches but they are weaker comparatively, as illustrated by the difference in the Lorentz factor between the two jets at approximately $10^3 r_g$ as shown in Fig.~\ref{fig:evolution_1}. Both jets are casually connected throughout their length, in agreement to VLBI images of AGN jets \citep[e.g., ][]{jorstad_agn_jet_2005}.}
  \label{fig:evolution_2}
\end{figure}
\begin{figure}
	\includegraphics[width=\columnwidth]{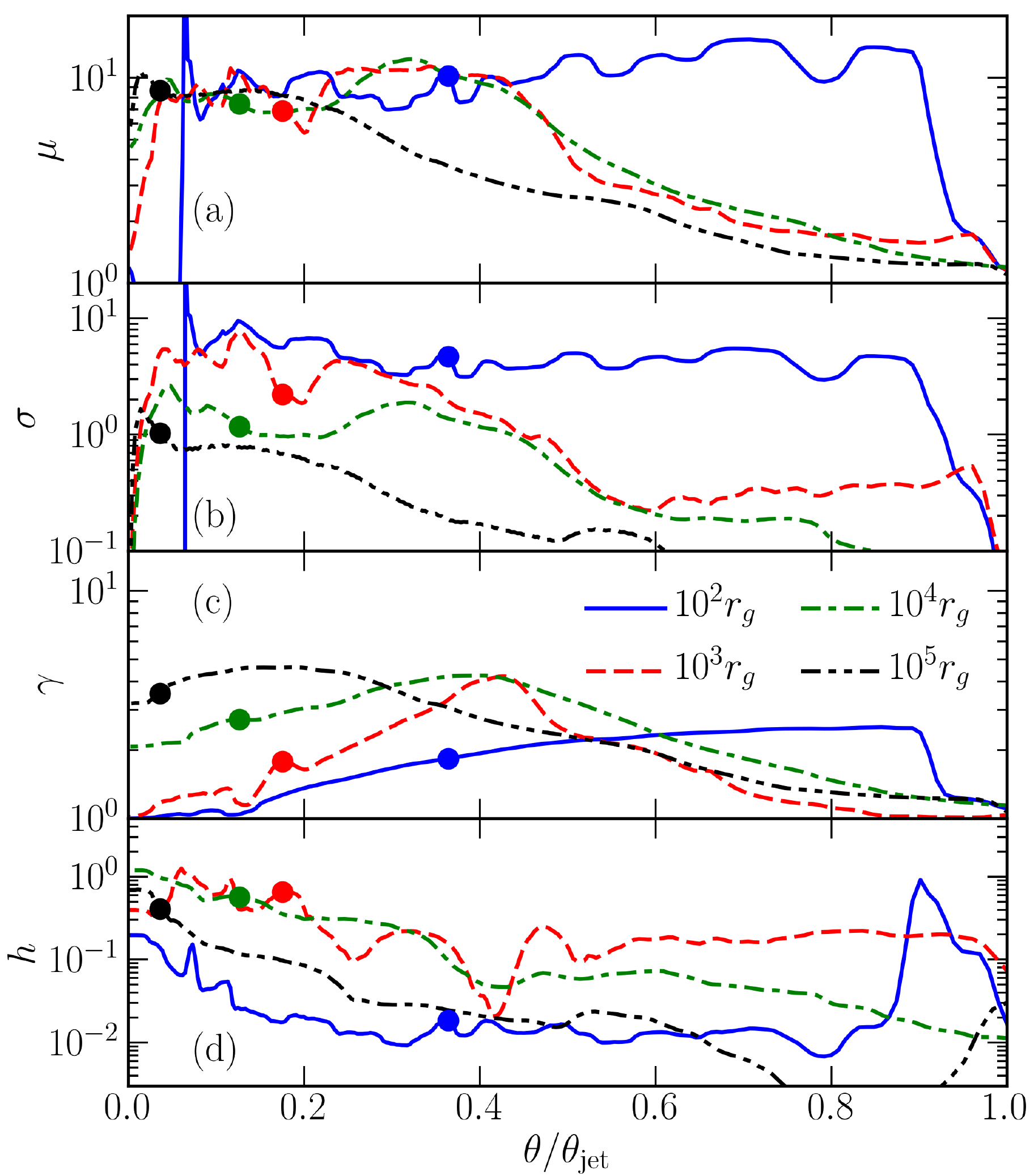}
    \caption{Transverse cross-sections of the jet at different distances (see the legend) show that as the jet accelerates, the poloidal flux surfaces differentially bunch up towards the axis and build up a fast inner jet core. The figure shows (panel a) the specific total energy $\mu$, (panel b) magnetisation $\sigma$, (panel c) Lorentz factor $\gamma$ and (panel d) the specific enthalpy $h$ at $t=2\times10^5 t_g$ at different distances $r=(10^2,10^3,10^4,10^5$)$r_g$. We take the jet-edge to be $\mu=1.2$. The corresponding jet-edge half-opening angles ($\theta_{\rm jet}$) for the different distances are ($0.268$, $0.085$, $0.026$, $0.019$)~rad. We also indicate with circles the opening angle of the field line shown in Fig.~\ref{fig:evolution_1}. The peak in the $\gamma$ profile shifts towards the jet axis with increasing distance as a result of differential field line bunching. The jet-edge experiences mass-loading from the wind and thus with increasing distance, the specific energies decrease at the edge. Pinching causes magnetic dissipation and hence the specific enthalpy tends to increase with distance (see also Fig.~\ref{fig:evolution_1}).}. 
    \label{fig:transverse_1}    
\end{figure}
\begin{figure}
	\includegraphics[width=\columnwidth]{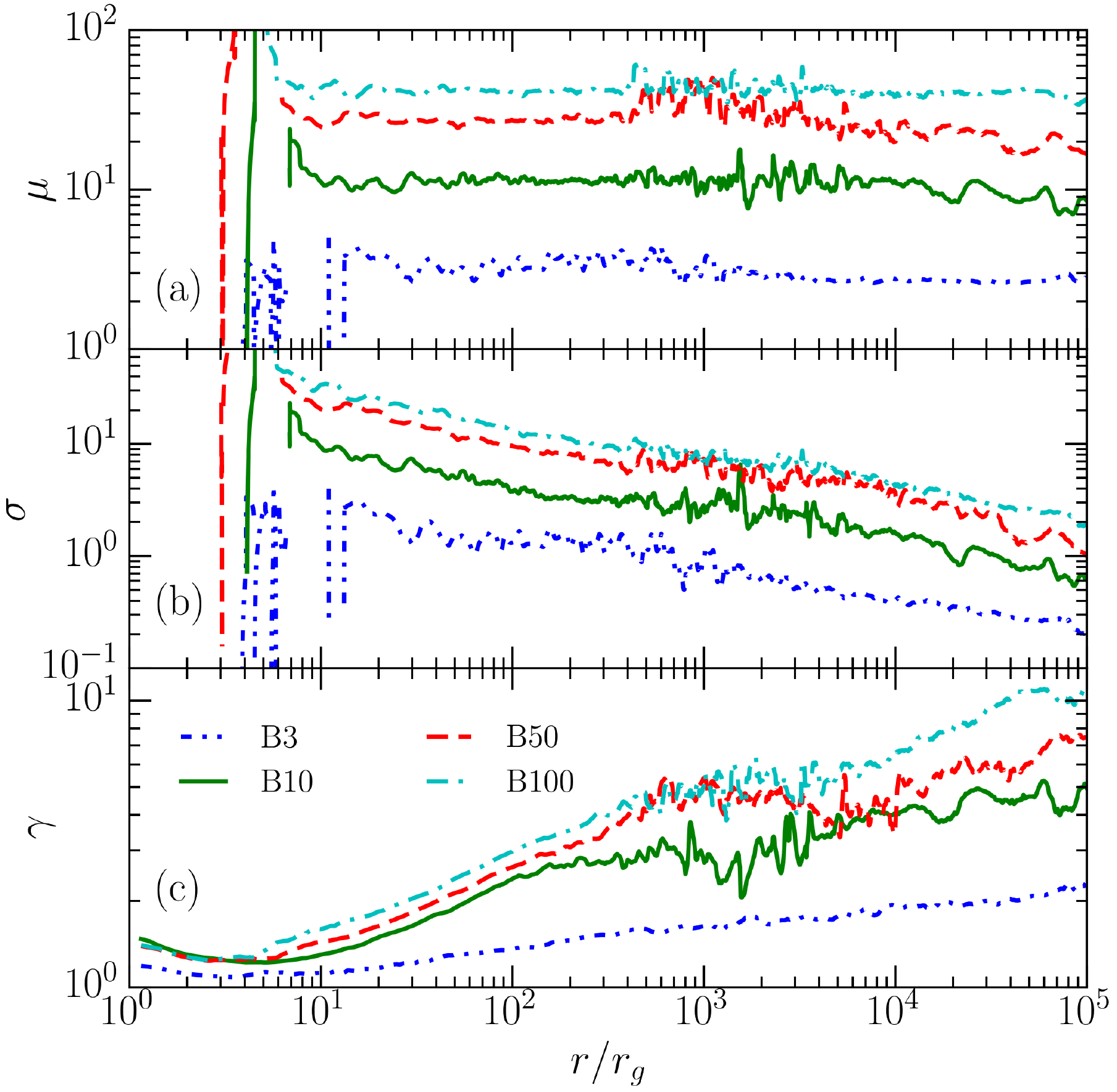}
    \caption{Jets with higher values of $\mu$ accelerate to higher Lorentz factors. We compare the jet acceleration profile of models B3 (dashed double dotted), B10 (solid), B50 (dashed) and B100 (dotted), with maximum magnetisation $\sigma_0$ of 3, 10, 50 and 100 respectively, i.e., only varying the jet base magnetisation (see Sec.~\ref{sec:floors}). We show the evolution of specific total energy $\mu$, magnetisation $\sigma$ and the Lorentz factor $\gamma$ along field lines in the mid-jet ($\theta_{j,\rm H}=0.8$~rad) at $t\approx 2\times 10^5 t_g$. Due to a larger jet base magnetisation, B100 accelerates to $\gamma \sim 10$ while B3 only accelerates to $\gamma \sim 2$. Except for B3, pinching significantly affects all models around $10^3 r_g$. Evidently, the jet base magnetisation plays a role in determining pinch activity.}
    \label{fig:sigma_S36}
\end{figure}

\subsubsection{Jet structure and acceleration}
\label{sec:lower:-acceleration}

Figure~\ref{fig:evolution_2}(a) shows that both the upper and the lower jets collimate similarly from $\theta_j\sim0.3$~rad (or $17.2\deg$) at $8r_g$ to $\theta_j\sim 8\times10^{-4}$~rad (or $0.046\deg$) at $10^5r_g$ while displaying a power-law shape $\theta_j\propto (r/r_g)^{-0.63}$. Remarkably, the outer jet also displays a continuous power-law collimation profile and resembles that of the M87 jet (see Sec.~\ref{sec:M87}).

Beyond the stagnation surface, the Lorentz factor (Fig.~\ref{fig:evolution_1}) increases smoothly until $r\simless 200r_g$ for both the lower and upper jets. Such a $\gamma$ profile is typical for highly magnetised jets in the poloidal field dominated regime \citep[e.g., ][]{bes98,tch08} where the acceleration occurs on the scale of the light cylindrical radius, $R_{\rm L}=c/\Omega$, where $\Omega$ is the conserved field line angular velocity. The Lorentz factor in this regime behaves as $\gamma \approx (1+(R/R_{\rm L})^2)^{1/2}$, where $R$ is the cylindrical radius of the field line. 

Once the jet becomes super-fast magnetosonic (i.e., the jet velocity becomes larger than the fast magnetosonic wave speed), adjacent field lines must shift in the transverse direction in a non-uniform manner for the jet to efficiently accelerate \citep{begelman_asymptotic_1994, chiueh_crab_1998, vla04, tch09, kom09}. This can be thought of as \textit{differential bunching} of field lines towards the jet axis. To quantify this bunching, we can use Eq.~(26) of \citet{tch09},
\begin{equation}
\frac{\gamma}{\mu}\approx 1-\frac{\pi B_p R^2}{\Phi}=1-a_{bp},
\label{eqn:bunch}
\end{equation}
where we define the field line bunching parameter $a_{bp}$ as the ratio of the local poloidal field strength $B_p$ and the mean poloidal field strength, $\Phi/\pi R^2$. For efficient acceleration, the poloidal field must decrease faster with distance than the mean poloidal field, hence creating a pressure gradient that exceeds the hoop stress, which slows down the jet, thereby accelerating the jet. Indeed, Fig.~\ref{fig:evolution_2}(b) shows that the bunching parameter $a_{bp}$ does decrease and therefore, $\gamma$ increases to values approaching, within a factor of few, the maximum Lorentz factor $\mu$. The Lorentz factor in this simulation reaches $\gamma \sim 6$, which is consistent with typical AGN jets \citep[e.g., ][]{Pushkarev2017}. 

\subsubsection{Transverse jet causality}
\label{sec:lower:-causality}
As we discussed in Sec.~\ref{sec:lower:-acceleration}, for efficient jet acceleration, magnetic field lines need to move across the jet and bunch up towards the axis. This requires the jet to be laterally causally connected \citep{tch09,kom09}. We approximate that a field line is in lateral causal connection with the jet axis as long as its Mach cone (i.e., the range of directions that a point on the field line can communicate with) crosses the axis of the jet. Thus, for efficient communication with the axis, the jet half-opening angle $\theta_j$ must be smaller than the Mach cone half-opening angle ($\theta_{\rm M}=1/M_f$, where $M_f$ is the fast Mach number). Using the definition of the fast magnetosonic wave velocity $v_f$ and Lorentz factor $\gamma_f$ \citep[e.g., ][]{gam03}, 
\begin{equation}
\begin{centering}
\gamma_fv_f=\left(\frac{b^2}{\rho}\right)^{1/2},
\end{centering}
\label{eqn:fast}
\end{equation}
and focusing on the asymptotic regime of the jet, i.e., $R\gg R_{\rm L}$ and $b_t\approx 0$, we can compute the Mach cone half-opening angle as,
\begin{equation}
\begin{centering}
\theta_{\rm M}=\frac{1}{M_f}= \frac{\gamma_f v_f}{\gamma v} \overset{\eqref{eqn:fast}}{=} \frac{\sqrt{b^2/\rho}}{\gamma v} \overset{\eqref{eqn:sigma}}{\approx} \frac{\sqrt{\sigma}}{\gamma v}.
\end{centering}
\label{eqn:mach_angle}
\end{equation}
\noindent Approximating the flow velocity $v\approx 1$ (for a relativistic jet), we have $\theta_{\rm M}\approx\sqrt{\sigma}/\gamma$. Figure~\ref{fig:evolution_2}(c) shows the transverse causality parameter $\theta_j/\theta_{\rm M}$, the ratio of the jet and Mach cone half-opening angles. For $\theta_j/\theta_{\rm M} >1$, we expect acceleration to slow down as causal contact is lost \citep{tom94,bes98,tch09}. However Fig.~\ref{fig:evolution_2}(c) demonstrates that the flow along the field line always remains in causal contact with the polar axis, which is consistent with VLBI observations of AGN jets \citep[e.g., ][]{jorstad_agn_jet_2005,clausenB13} that show $\gamma \theta_j \sim 0.1-0.3$ (Fig.~\ref{fig:evolution_2}d). The acceleration slow down occurs as the jet becomes matter-dominated, i.e., $\sigma<1$, with additional deceleration due to pinching which we discuss next.

\subsubsection{Toroidal pinch instabilities}
\label{sec:upper:-pinch}
The upper jet has a distinctly different acceleration profile compared to the lower jet. As the jet propagates through the ambient medium, it expands and adiabatically cools down. Due to this, the specific enthalpy $h=(u_{\rm g} + p_{\rm g})/ \rho$ (Fig.~\ref{fig:evolution_1}, black line) decreases initially, slightly at $r\lesssim 100r_g$. However, beyond $100r_g$, pinch instabilities cause magnetic dissipation that raises the jet specific enthalpy to order unity, effectively creating a thermal pressure gradient directed against the direction of the flow at $r\lesssim10^3r_g$. This pressure gradient significantly slows down the outflow to $\gamma\simless2$ by $r\approx10^3r_g$. At larger radii, $h$ drops, and the jet re-accelerates under the action of both the magnetic and thermal pressure forces. We can interpret this behaviour also through energy conservation. Using the definition of magnetisation (Eq.~\ref{eqn:sigma}), specific enthalpy and approximating $u_t\sim -\gamma$, from Eq.~\eqref{eqn:mu} we get 
\begin{equation}
\begin{centering}
\mu=\gamma(\sigma+h+1),
\end{centering}
\label{eqn:energy}
\end{equation}
a useful form of the energy equation. It clearly shows that for $\sigma\simeq$constant, an increase in enthalpy to $h\sim 1$ results in $\gamma$ decreasing, which is seen for the upper jet around $10^3r_g$. The upper jet shows stronger pinch activity and thus collimates slightly more than the lower jet: $\theta_j$ is smaller by a factor of $\lesssim 2$; Fig.~\ref{fig:evolution_2}a, solid line), and other quantities like the maximum Lorentz factor $\mu$ and bunching parameter $a_{bp}$ strongly oscillate. 

\subsubsection{Energetics across the jet}
\label{sec:upper:-transverse}

Figure~\ref{fig:transverse_1} shows how the different components of jet energy flux, $\mu$, $\sigma$, $\gamma$ and $h$, vary across the jet at different distances along the upper jet. A fiducial field line with the foot-point at $\theta_{j,\rm H}=0.44$~rad (see also Fig.~\ref{fig:evolution_1}), shown with filled circles, collimates faster than the jet-edge, indicated by $\theta_{\rm jet}$. We take the jet-edge to be at $\mu=1.2$, which is reasonable since $\mu$ should drop rapidly at the jet-edge \footnote{We also note that $\mu$ is roughly constant throughout the width of the jet as opposed to increasing as $\mu \propto \sin^2\theta$, seen in previous simulations \citep[e.g.,][]{kom07,tch08}, which might be a consequence of the density floor model coupled to the stagnation surface.}. This means that there is internal reconfiguration of the flow within the body of the jet that leads to the formation of a fast magnetised inner core. At the jet-edge, mass-loading via pinching (Sec.~\ref{sec:entrainment}) causes specific total energy $\mu$, the magnetisation $\sigma$ and the Lorentz factor $\gamma$ to drop gradually, forming a slower sheath that surrounds the core, resulting in a structure similar to the spine-sheath seen in AGN jets.  
\begin{figure}
	\includegraphics[width=\columnwidth]{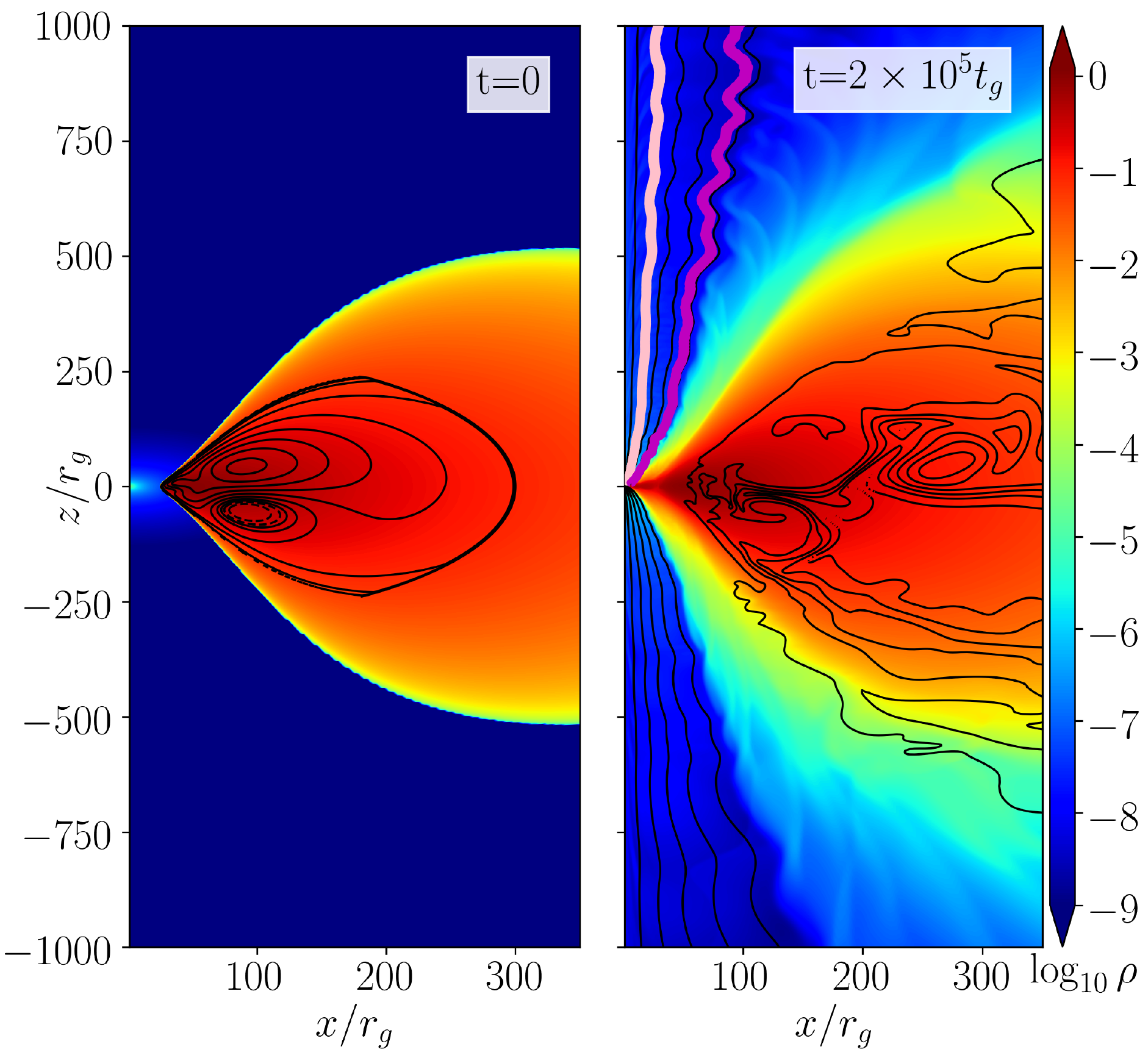}
    \caption{Vertical slices through the density profile model B10-R, which has a small disc, show that smaller discs have wider jets and broader wind regions as compared to larger discs (compare to Fig.~\ref{fig:B10_density}). Density and field lines are labelled as in Fig.\ref{fig:B10_density}. The initial disc (left panel) is much smaller than in Fig.~\ref{fig:B10_density}, enabling the jet to freely expand laterally into the low density ambient medium as seen in the right panel at $t=2\times10^5t_g$. As the jet undergoes rapid lateral expansion, pinches do not have enough time to develop and therefore, the jet mass-loading is much lower (see the main text). We highlight two representative field lines with pink and magenta colours and show their properties in Fig.~\ref{fig:B10-R_evolution}.} 
    \label{fig:S25_density}
\end{figure}

\subsection{Dependence on mass-loading at the stagnation surface}
\label{sec:mu model}
In Sec~\ref{sec:upper:-transverse}, we showed that pinching instabilities lead to dissipation and therefore, reduces jet acceleration efficiency. Would the acceleration efficiency change if the jet had a different specific total energy? In order to answer this question, we compare results from four models different only by the density floors (Sec.~\ref{sec:floors}; also see Table~\ref{tab:models}), namely the maximum magnetisation $\sigma_0$ values of 3 (model B3), 50 (model B50) and 100 (model B100), along with model B10. Figure~\ref{fig:sigma_S36} shows that in all models the Lorentz factor $\gamma$ increases steeply until $\sim 10^3 r_g$, beyond which the acceleration slows down. Models with higher jet base magnetisation (and hence, larger $\mu$ values) accelerate slightly faster and reach higher Lorentz factors, but get affected by pinch instabilities at roughly the same distance as models with lower $\mu$. For model B3 the pinch instability is weaker and leads to smaller oscillations in, for example, the Lorentz factor (see discussion in Sec.~\ref{sec:pinch:-origins}). Similar to \citet{kom09}, we find that models with smaller $\mu$ achieve smaller $\sigma$ at large distances. For instance, model B3 achieves $\sigma\simeq 0.2$ at $r=10^5r_g$. It is interesting to note that efficient heating via relativistic shocks requires $\sigma\lesssim0.1$  \citep[e.g.,][]{kennel_confinement_crab_1984, kom2012}. Such low $\sigma$ values are very difficult to achieve for collimated flows as $\sigma$ drops very gradually in the toroidally dominated regime \citep[known as the $\sigma$ problem; e.g.,][]{tch09,kom09}. This suggests that magnetic reconnection might be a more efficient mechanism for particle acceleration in jets.

\begin{figure}
	\includegraphics[width=\columnwidth]{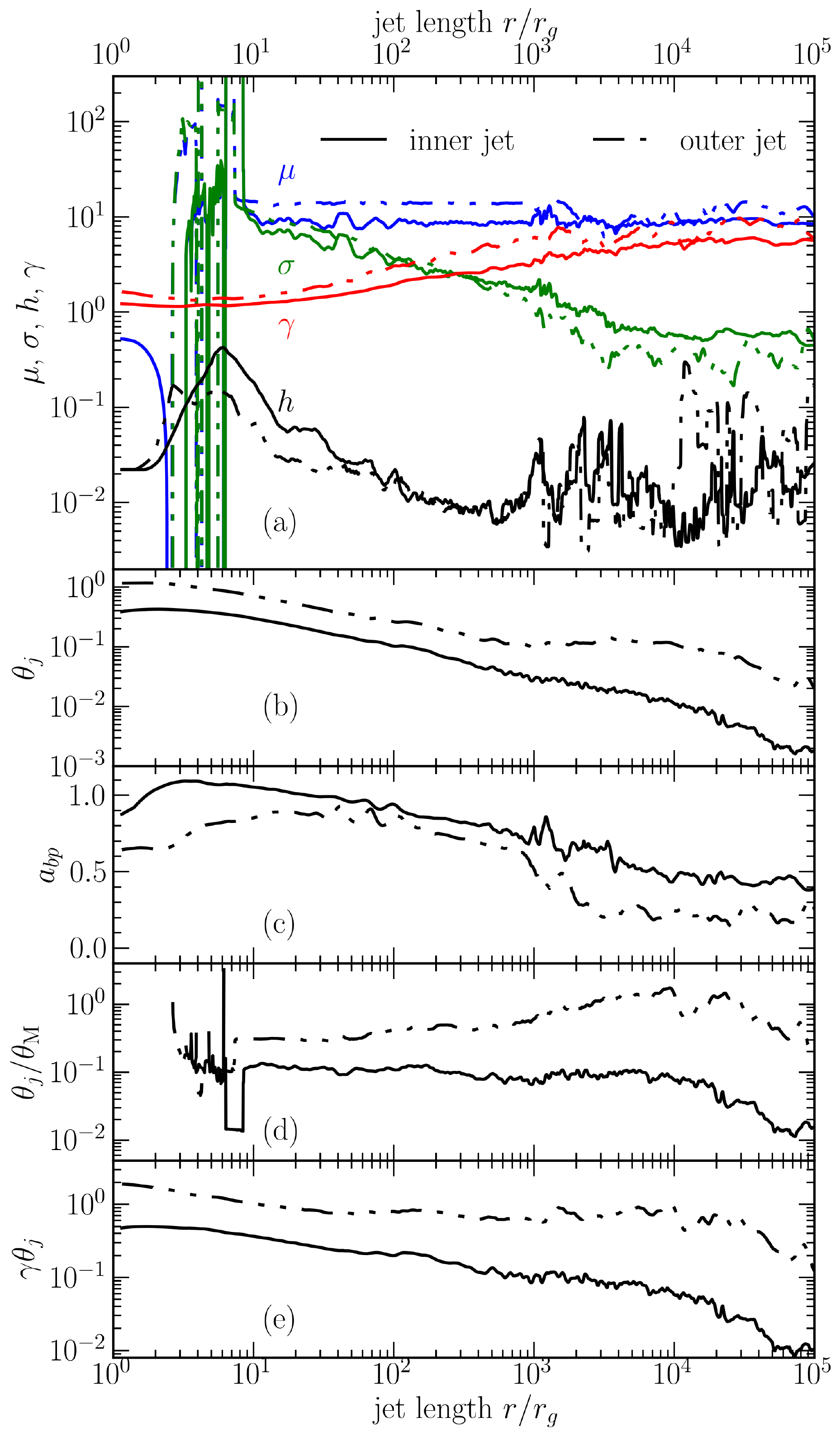}
    \caption{Radial profiles of quantities along two magnetic field lines in the upper jet at $t=2\times10^5 t_g$ for model B10-R, one in the inner jet ($\theta_{j,\rm H}=0.41$~rad, solid lines; the field line is highlighted in pink in Fig.~\ref{fig:S25_density}, right panel) and the other in the outer jet ($\theta_{j,\rm H}=1.17$~rad, dashed-dotted; magenta in Fig.~\ref{fig:S25_density}, right panel). Refer to Fig.~\ref{fig:evolution_1} and \ref{fig:evolution_2} for the notations used. Panel(a): The outer field line accelerates slightly faster than the inner one. The enthalpy $h$ remains $<0.1$ on average beyond $10^3 r_g$. Panel(b): The jet is initially parabolic and becomes conical beyond $10^3 r_g$, with $\theta_j$ becoming constant for the outer field line. This expansion causes the outer jet to exhibit (panel c) a sudden drop in the bunching parameter and (panel d) loss of causal contact in the lateral direction as $\theta_j/\theta_{\rm M}\simeq 1$. Panel(e): The inner jet remains causally connected while the outer jet becomes conical with $\gamma\theta_j\gtrsim1$. Thus, deconfinement has a notable influence on jet dynamics, reinforcing the notion that jet acceleration is coupled to the jet collimation profile.}
    \label{fig:B10-R_evolution}
\end{figure}
\subsection{Acceleration of a jet collimated by a small disc}
\label{sec:smalltorus}
So far, we discussed models with large discs that collimate the jets out to large distances. Here, we consider model B10-R with a small disc. In this case, the disc wind collimates the jets out to smaller distances\footnote{Movie showing the difference in collimation and acceleration between B10 and B10-R: \href{https://youtu.be/2C4re4aiuQM}{https://youtu.be/2C4re4aiuQM}}. This happens because larger, more radially extended discs launch disc-winds over an extended range of radii: the winds launched from small radii collimate off of those launched further out, and off of the disc itself, leading to a radially extended collimation profile of the jets. B10-R, with a smaller disc extending only up to $500r_g$, is embedded with the same magnetic field configuration as model B10 (Eq.~\ref{eqn:Aphi}). The left panel in Fig.~\ref{fig:S25_density} shows the vertical slice through the initial conditions and the right panel shows the system at $t=2\times10^5t_g$. Model B10-R, compared to B10 (Fig.~\ref{fig:B10_density}, right), has a much wider jet as the weaker confining pressure of the disc-wind enables the jet to expand laterally.  

In Fig.~\ref{fig:B10-R_evolution} we show the radial profiles of quantities along two magnetic field lines in the upper jet, one located in the inner jet (foot-point half-opening angle of $\theta_{j,\rm H}=0.41$~rad; highlighted in pink in Fig.~\ref{fig:S25_density}, right panel) and the other in the outer jet ($\theta_{j,\rm H}=1.17$~rad; magenta in Fig.~\ref{fig:S25_density}, right panel), at $t=2\times10^5 t_g$. Initially, similar to our fiducial model B10, the disc and disc-wind collimate the jet into a parabolic shape. However, beyond $10^3 r_g$, the confining pressure of the disc drops and the outer field lines in the jet become conical (Fig.~\ref{fig:B10-R_evolution}b). The deconfinement leads to a drop in the confining pressure, causing field lines to diverge and experience a quicker acceleration due to the outwards pressure gradient. This boost in $\gamma$ due to smooth deconfinement of the jet has been shown to occur by previous idealised simulations \citep[e.g., ][]{tch10b,kom2010}, though the increase in acceleration is not quite as significant as in Fig.~2 of \citet{tch10b}. Beyond $10^3r_g$, the quick lateral expansion of the jet suppresses pinch instabilities to a large extent \citep[consistent with e.g., ][]{Moll2008,Granot2011,porth2015} and adiabatically cools the jet leading to an order of magnitude smaller enthalpy $h$ compared our fiducial model B10 (Fig.~\ref{fig:evolution_1}). 

Even though the outer field line expands ballistically and maintains an approximately constant opening angle, the inner field lines continue to collimate off of the outer ones into a parabolic shape similar to our fiducial model (Fig.~\ref{fig:evolution_1}). The outer jet experiences a relatively larger change in the bunching parameter $a_{bp}$ compared to the inner field line (Fig.~\ref{fig:B10-R_evolution}c), in accordance with Eq.~\eqref{eqn:bunch} and reaches $\sigma\simeq0.2-0.3$ (Fig.~\ref{fig:B10-R_evolution}a). Upon loss of collimation, the outer jet also loses transverse causal connection ($\theta_j/\theta_{\rm M}>1$; Fig.~\ref{fig:B10-R_evolution}d), and the acceleration ceases (see Sec.~\ref{sec:lower:-causality}). From Fig.~\ref{fig:B10-R_evolution}(e), $\gamma \theta_j\sim0.7-1$ for the outer jet, while $\gamma\theta_j$ is between 0.1 and 0.4 within $1000 r_g$ for the inner jet. See Sec.~\ref{sec:M87} for further discussion of $\gamma\theta_j$ values in our models.

\section{Comparisons to idealised jet simulations}
\label{sec:ideal}
Here we aim to study jet dynamics in the absence of pinching instabilities by constructing smooth idealised outflows and maintaining fine control over the jet shape by confining the flow using a conducting collimating wall \citep{kom07,kom09,tch10a}. Such a setup also removes the shear-induced turbulence and dissipation at the jet edge-disc wind layer.

\begin{figure}
	\includegraphics[width=\columnwidth]{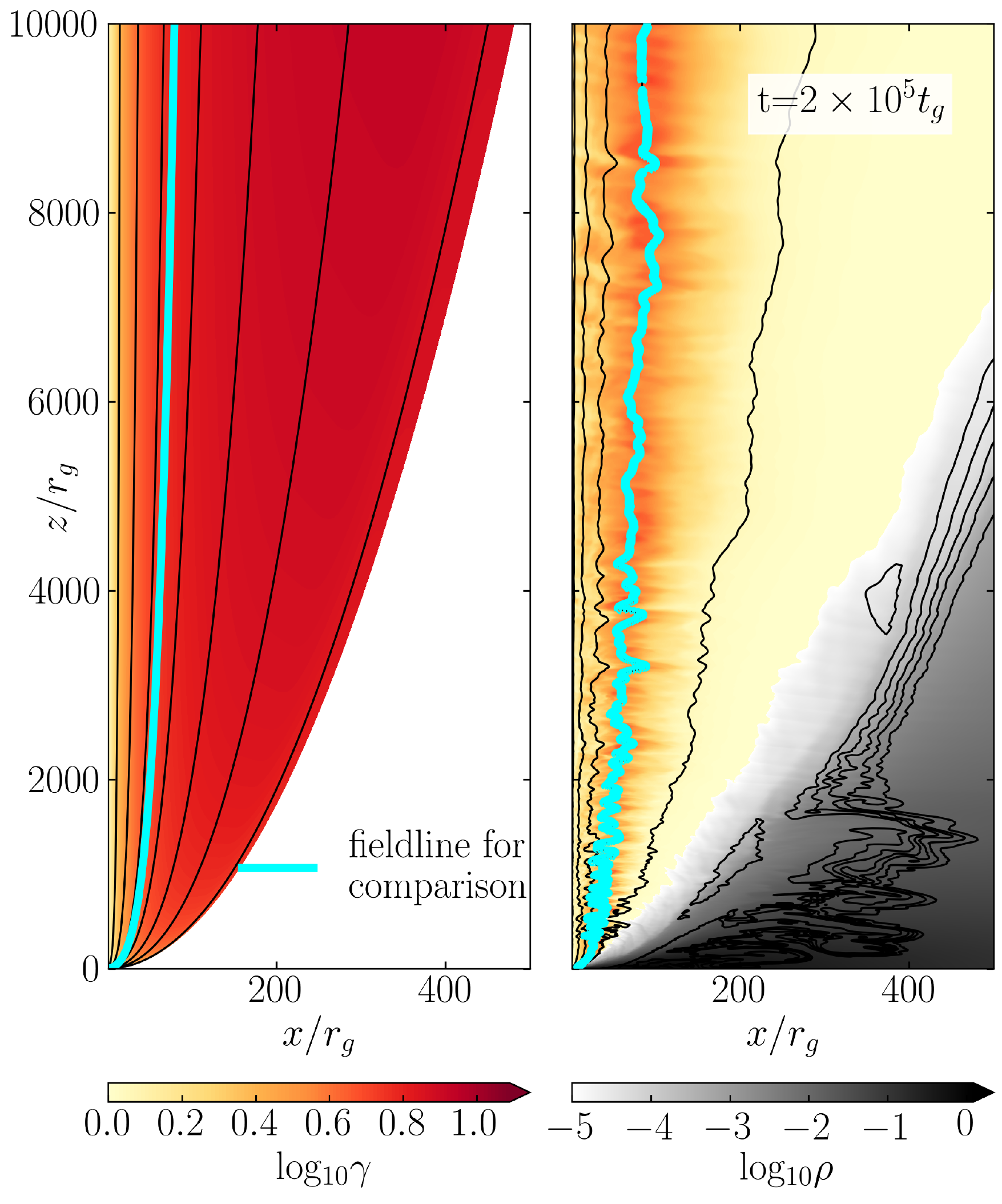}
    \caption{Comparison between idealised wall-jet and disc-jet simulations shows that instabilities at jet-disc interface slow down the outflow. \textit{Left}: Lorentz factor $\gamma$ plot of a jet bound by a rigid parabolic wall. \textit{Right}: combined Lorentz factor and density plot for disc-jet model B10. For the case of the wall-jet, the field line shape is smooth and the outflow quickly accelerates. However, for the disc-jet, the pinch instabilities distort the shape of the field line and slow down acceleration. This is better seen when we compare quantities along the indicated field lines (cyan) in Fig.~\ref{fig:compare_ideal_1}.}
    \label{fig:wall}
\end{figure}
\begin{figure} 
	\includegraphics[width=\columnwidth]{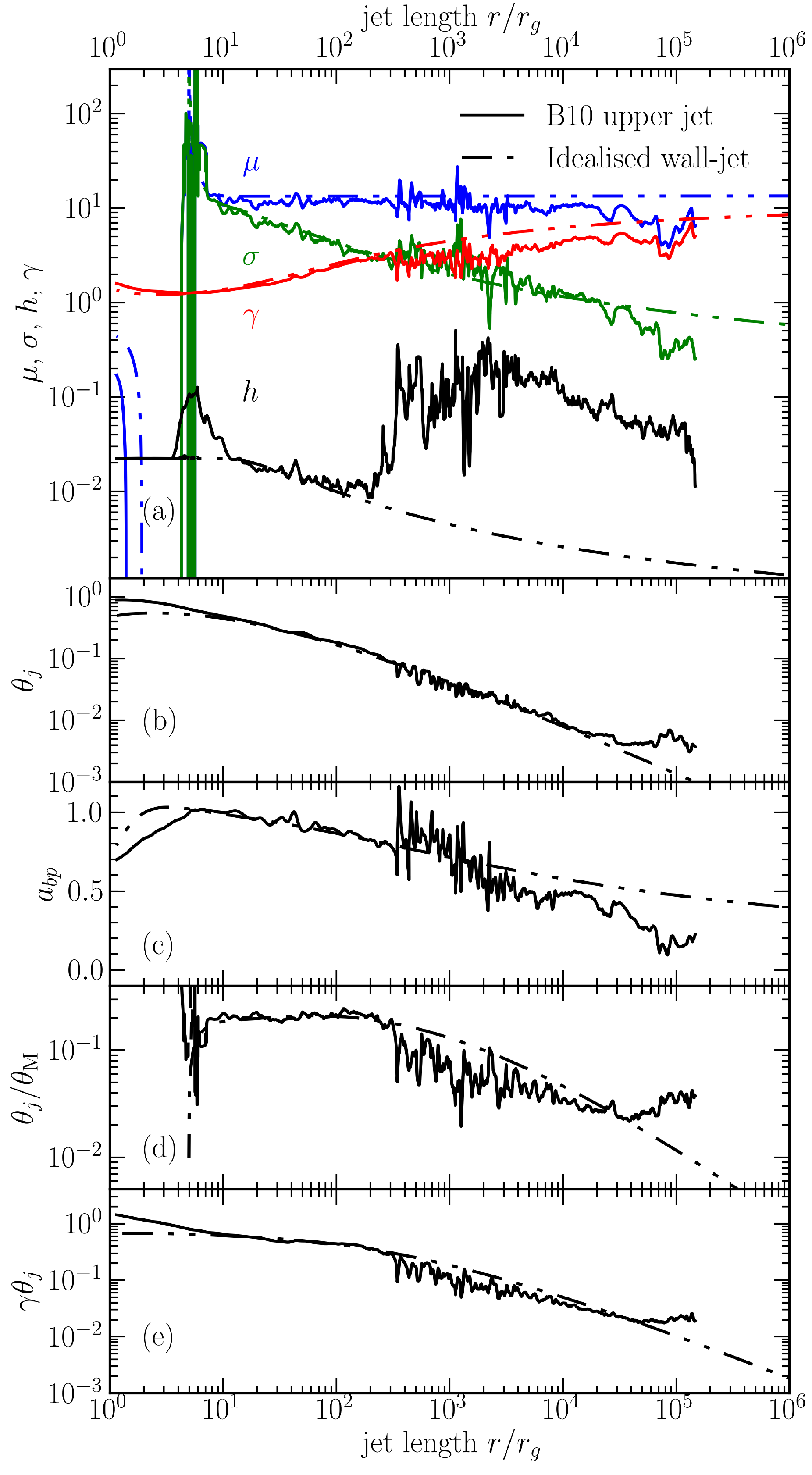}
    \caption{Comparing the radial profile of quantities along field lines for the upper jet in disc-jet model B10 (solid lines, $\theta_{j,\rm H}=0.52$~rad) and the idealised wall-jet model (double dot-dashed lines, $\theta_{j,\rm H}=0.9$~rad) shows that the two models agree apart from the pinch instabilities that slow down the disc-jets and dissipate magnetic fields into heat. The disc-jet simulation is shown at $t=2\times10^5 t_g$, while the wall-jet is in steady state. Panel(a): The specific energy profiles ($\mu$, $\sigma$, $h$ and $\gamma$) for B10 match the idealised model reasonably well, with deviations arising in the pinched region of model B10, especially in the enthalpy $h$. Panel(b): Both field lines have similar collimation profiles. Panel(c): Pinching causes the poloidal field in model B10 to fluctuate and dissipate into heat, thereby decreasing the bunching parameter.The jet mass-loading also plays a part as $\mu$ decreases with respect to the wall-jet $\mu$. Panels(d and e): The values of $\theta_j/ \theta_{\rm M}$ and $\gamma \theta_j$ remain below 1, indicating lateral causal connection for both jets. Overall, the radial profiles between the two models match well except in the disc-jet pinched region.}
    \label{fig:compare_ideal_1}
\end{figure}
\subsection{Model setup}
\label{sec:ideal:-setup}
We set up an outflow bound by a perfectly conducting wall, mimicking a jet collimated by an external medium. We refer to this setup as a wall-jet simulation, in contrast to disc-jet simulations in which the disc-wind collimates the jets. The field lines threading the event horizon initially follow the shape of the wall that collimates in a parabolic fashion:
\begin{equation}
\begin{centering}
1-\cos\theta=\left(\frac{r+r_0}{r_{\rm H}+r_0}\right)^{-\nu}.
\end{centering}
\end{equation}
This gives us the initial poloidal field configuration:
\begin{equation}
\begin{centering}
A_{\phi}=\left(\frac{r+r_0}{r_{\rm H}+r_0}\right)^{\nu}(1-\cos\theta).
\end{centering}
\label{eqn:ideal_psi}
\end{equation}

Here, the outermost field line touching the wall is given by $A_{\phi}=1$: it starts out at the intersection of the event horizon ($r=r_{\rm H}$) and the equatorial plane ($\theta=\pi/2$), is initially radial for $r\lesssim r_0$ and asymptotically collimates as $\theta\approx r^{-\nu/2}$. In this setup, $\nu=0$ results in a monopolar field shape ($\theta_j=$constant), while $\nu=1$ gives us the parabolic field shape ($\theta_j\propto r^{-1/2}$). We set a transitional radius $r_0=10r_g$ and employ $\nu=0.8$ as these values give a good match to the disc-jet shape, as we discuss below. We define the physical coordinates ($r$, $\theta$) as functions of the internal coordinates ($x_1$, $x_2$) as $r= \exp(x_1)$ and $x_2=\mathrm{sign}(\theta)|A_{\phi}|^{1/2}$. We use a resolution of $12800\times 400$ cells. Our computational domain range extends radially from $0.85r_{\rm H}$ to $10^6 r_g$. We employ the same polar reflective boundary conditions at $x_2=0$ as for the disc-jet simulations (see Sec.~\ref{sec:Numerical_grid}). At the wall, $x_2=1$, the boundary conditions are also reflective so that the gas and fields follow the wall \citep{tch10a}. The density floors are the same as in model B10. The left panel in Fig.~\ref{fig:wall} shows the resulting wall-jet solution. For comparisons with the disc-jet model B10, we choose a field line in the inner jet  of B10 model as field lines near the jet-edge are strongly affected by mass-loading (Fig.~\ref{fig:wall}, right). At time $t=0$, we start the wall-jet simulation with a purely poloidal magnetic field given by Eq.~\eqref{eqn:ideal_psi}, which then develops a toroidal component due to the rotation of the black hole. Our simulation time extends to $10^7t_g$, which ensures that the outflow reaches a steady state up to at least a distance of $10^6 r_g$. To speed up the simulation, from $t=1000t_g$ onwards, we freeze out cells that reached steady state, i.e., the cells located at $r<0.1ct$, where $t$ is the simulation time (similar to \citealt{tch08}; see also \citealt{kom07}). 

\subsection{Disc-jets vs. idealised wall-jets}
\label{sec:ideal:-comparison}
Figure~\ref{fig:wall} shows the comparison between the ideal wall setup (left) and disc-jet setup (right). For the disc-jet setup, the presence of a pressure imbalance between the jet and the accretion disc-wind gives rise to oscillations in the jet shape. In contrast, for the idealised wall-jet, the boundary is rigid and hence, the pinches are absent. The energy flux components, $\mu$, $\sigma$, and $\gamma$ agree between the disc-jet and the wall-jet rather well (Fig.~\ref{fig:compare_ideal_1}a), showing that wall-jet models capture most of the time-average steady state dynamical properties of disc-jet models with the same shape (Fig.~\ref{fig:compare_ideal_1}b), especially in the absence of pinches. For the wall-jet, the specific enthalpy $h$ decreases with increasing $r$ as expected due to adiabatic jet expansion. For the disc-jet, $h$ increases substantially at $r\sim 200r_g$ due to the onset of the pinch instabilities that convert the poloidal field energy into enthalpy. Free of pinches, the wall-jet smoothly accelerates as $\gamma \propto R$ until a few times $10^3 r_g$, followed by the slower acceleration as the field lines slowly become cylindrical when they enter the jet core (Fig.~\ref{fig:compare_ideal_1}a). The acceleration is more rapid for field lines closest to the wall as field lines in this region diverge away from each other more (Fig.~\ref{fig:wall}, left). For the disc-jet, the presence of the pinches causes $B_p$ to dissipate (Fig.~\ref{fig:compare_ideal_1}c), along with a slight drop in $\theta_j/\theta_{\rm M}$ (Fig.~\ref{fig:compare_ideal_1}d). The product of the Lorentz factor and jet opening angle $\gamma\theta_j\gtrsim0.1$ for $r<3000r_g$(Fig.~\ref{fig:compare_ideal_1}e), similar to the values found for the inner jet (see Sec.~\ref{sec:lower:-causality}).

\section{Discussion}
\label{sec:discussion}
\subsection{Comparison to the M87 jet}
\label{sec:M87}
\begin{figure}
	\includegraphics[width=\columnwidth]{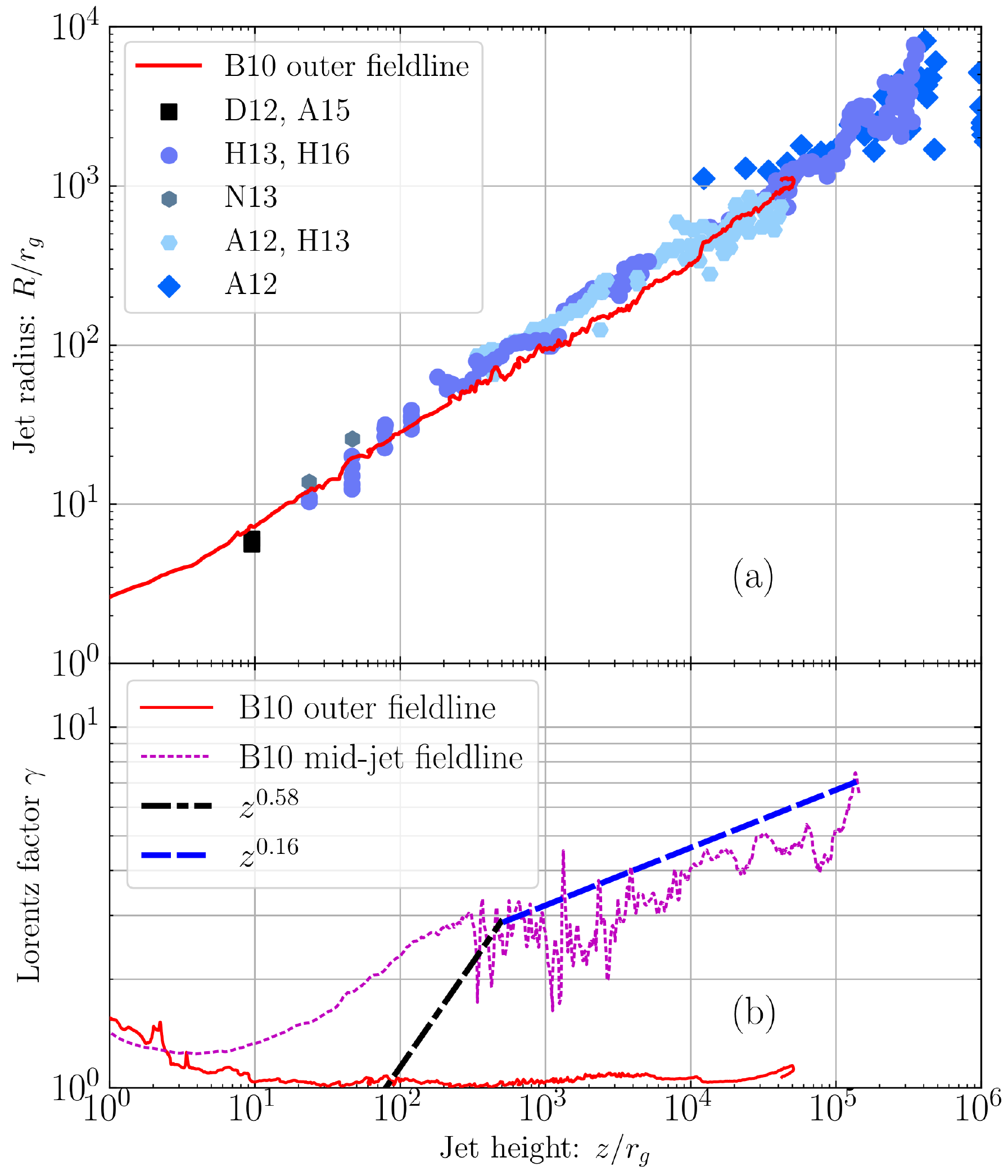}
    \caption{Comparison of the shape and Lorentz factor of a simulated jet with M87 observations shows remarkable resemblance. (Panel a) Jet radius along a field line near the jet-edge ($\theta_{j,\rm H}=1.53$~rad) for model B10 at $t=2\times10^5t_g$. The data points are read off from Fig.~15 of \citet{Nakamura_2018} and consist of data from \citet{doel12} (D12), \citet{asadanak2012} (A12), \citet{Hada2013} (H13), \citet{nak2013} (N13), \citet{Akiyama_2015} (A15) and \citet{Hada2016} (H16). The jet from model B10 fits very well with the M87 parabolic jet shape up to $10^5r_g$. The de-projected distance is calculated with M87 black hole mass $M=6.2\times10^9M_{\odot}$ and observer viewing angle of $14^{\circ}$. (Panel b) Lorentz factor along panel(a) field line as well as a field line in the mid-jet ($\theta_{j,\rm H}=0.77$~rad) compared to the broken power-law profile for the M87 jet Lorentz factor as measured by VLBI \citep{mertens2016}. There is a large distribution in $\gamma$ across the jet, similar to Fig.~15 of \citet{mertens2016}.} 
    \label{fig:M87:-data}
\end{figure}
We find that all of our simulated jets with large discs propagate with a parabolic shape over at least 5 orders of magnitude in distance.  In this section, we consider whether our fiducial model jet behaves the same way as the jets seen in nature, looking at the shape and acceleration profiles inferred from multiple Very Large Array (VLA)/ Very Long Baseline Interferometry (VLBI) observations of the M87 jet. Figure~\ref{fig:M87:-data}(a) compares the jet geometry for our fiducial model B10 with the observed M87 jet. The shape of the field line near the jet-edge fits very well with the observed data, displaying a parabolic collimation profile up to $10^5 r_g$, close to the location of HST-1 \citep{asadanak2012, Hada2013, mertens2016, Kim_M87_2018,Nakamura_2018}. 

\citet{mertens2016} showed using VLBI measurements that the acceleration profile of M87 follows $\gamma \propto R \propto z^{0.58}$ till $10^3 r_g$, changing to $\gamma \propto z^{0.16}$ up to the HST-1 knot. These power-law profiles agree reasonably well with the Lorentz factor profile along a field line in the mid-jet as can be seen in Fig.~\ref{fig:M87:-data}(b). The discrepancy at small radii may result from systematic measurement errors in the VLBI observations, the jet opening angle becoming comparable to the viewing angle or our preference for a particular field line. Additionally, in Fig.~\ref{fig:M87:-data}(a), the D12 \citep{doel12} and A15 \citep{Akiyama_2015} data points representing the Event Horizon Telescope Core at 230 GHz depend considerably on the assumed black hole mass and viewing angle along with additional uncertainties on the position \citep{Nakamura_2018}. Hence, it is possible that instead of the 230GHz core being smaller than the jet interior as the figure suggests, the emission might come from the disc, i.e., outside the jet-edge, where the Lorentz factor $\sim1$. The outer field line Lorentz factor is close to 1.1 (Fig.~\ref{fig:transverse_1}), similar to the velocities found for the M87 jet sheath \citep{mertens2016} and agrees with previous GRMHD simulations modified for M87 \citep{Nakamura_2018}. Indeed, as the jet gets mass-loaded via the jet-wind interaction, we expect a gradual decrease in the Lorentz factor as we go from the inner jet to the jet-edge, which may explain the wide distribution of the Lorentz factors across the M87 jet in Fig.~15 of \citet{mertens2016}. 

The HST-1 knot in M87 is a region where the jet is deemed to over-collimate and transitions from parabolic to conical structure \citep[][]{asadanak2012}. Unfortunately, we do not find such a dissipative feature in any of our simulations, nor do we see the jet turn conical around $10^5 r_g$. One possible reason may be that HST-1 lies very close to the Bondi radius of M87 \citep[$\sim7.6\times10^5r_g$; ][]{nak2013} where the shallow density profile of the ISM prevails. If there is an increase of confining pressure from the ISM beyond $10^5 r_g$, it is possible that the jet becomes over-pressured, perhaps forming a re-collimation feature. Further, as is the case for model B10-R (see Sec.~\ref{sec:smalltorus}), if the jet pressure subsequently becomes larger than the confining pressure, the jet would open up and turn conical. This suggests that it is important to consider a more realistic ISM pressure profile in future work \citep[e.g., ][]{Duran17}.

The product of the Lorentz factor and the jet opening angle $\gamma\theta_j$ is an important quantity we use for comparison to AGN jets. It is clear from Fig.~\ref{fig:evolution_2} that the inner jet exhibits very low values of $\gamma\theta_j$ ($<0.01$) for distances larger than $10^3r_g$, compared to those observed ($\sim0.1-0.3$, e.g., \citealt{jorstad_agn_jet_2005, pushkarev2009, clausenB13, jorstad2017}), whereas for the jet-edge, we find $\gamma\theta_j<0.1$. It is possible that the difference in measurements of $\gamma\theta_j$ between our models and observed jets might be a result of the latter assuming a conical jet as well as the uncertainty of attributing the Lorentz factor of the underlying jet flow to emission features (e.g., there might be standing shocks in the jet). In the case of model B10-R, the outer jet becomes conical and causally disconnected: $\gamma\theta_j\gtrsim1$ (see Fig.~\ref{fig:B10-R_evolution}), which may be more applicable for jets in gamma-ray bursts. Additionally, the peak Lorentz factor for many of the observed jets is over 10, a value only one of our models achieves (model B100), suggesting highly magnetised jets. 

\begin{figure}
	\includegraphics[width=\columnwidth]{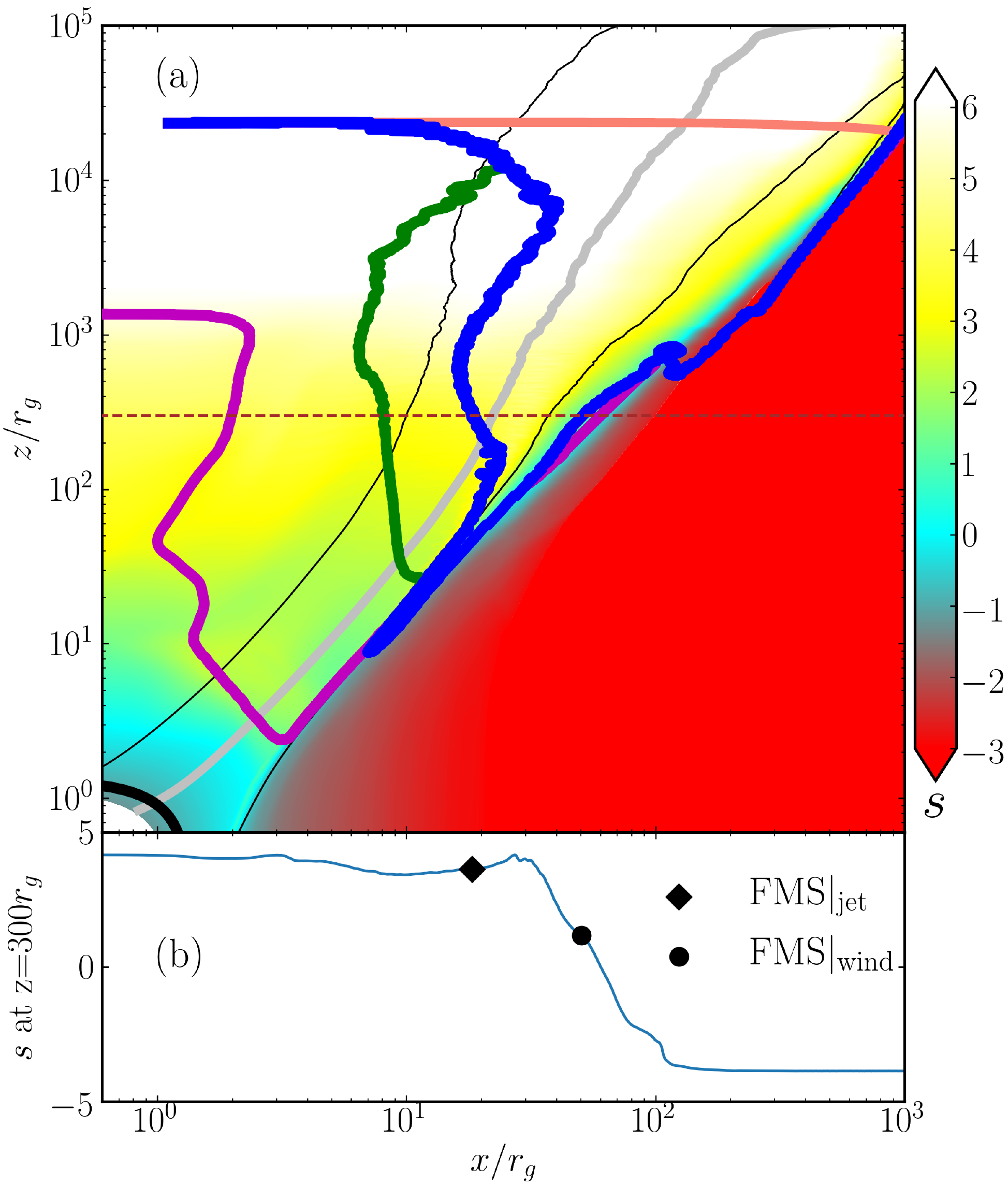}
	
    \caption{(Panel a) Log-log plot of characteristic surfaces along with entropy $s$ in colour (arbitrary units) and magnetic field lines (black) for a highly magnetised jet time-averaged over $(1.1-1.5)\times10^5 t_g$ in model B10. The lines shown are, starting from small radii, the event horizon (thick black), stagnation (magenta), Alfv\'en (green), classical fast magnetosonic surfaces (FMS; blue) and the fast magnetosonic separatrix surface (FMSS; salmon). We highlight a field line in silver and show the variability in its shape due to pinching in Fig.~\ref{fig:opening_angle}. The characteristic surfaces are only shown for the jet and the wind. The FMSS does not appear to coincide with dissipative features. (Panel b) We look at the entropy at $z=300r_g$, indicating the points where the fast surfaces for the jet (diamond) and the wind (circle) crosses the horizontal line, shown in panel(a). The entropy rises smoothly beginning from the sub-fast wind region, right up to the fast surface in the jet, suggesting that while the FMS might be relevant for dissipation in the jet, for the wind, the FMS is not so useful.}
    \label{fig:surfaces}
\end{figure}
  
\begin{figure} 
	\includegraphics[width=\columnwidth]{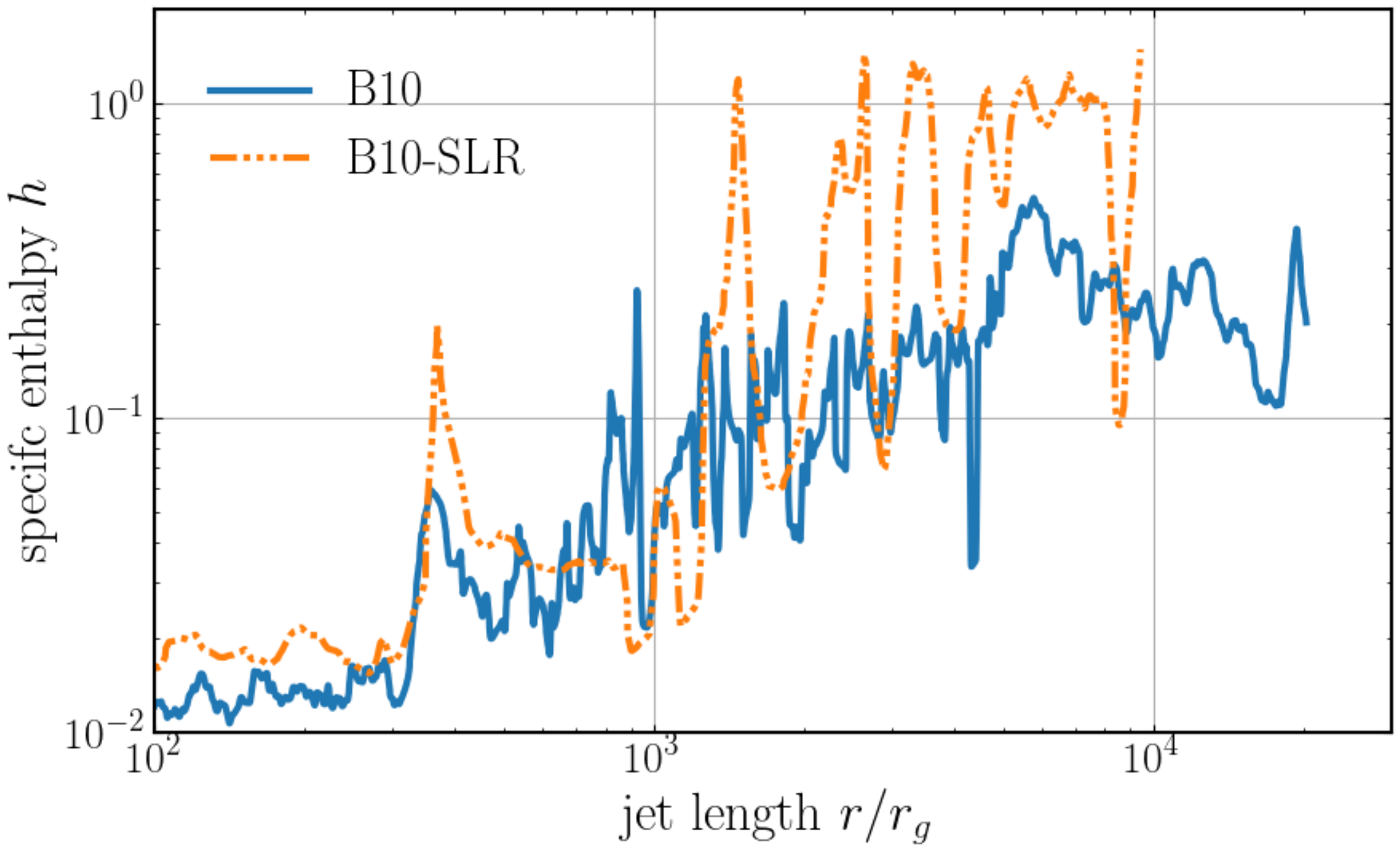}
    \caption{Sufficient resolution is required to properly capture the small scales of the pinches. We show the specific enthalpy $h$ along a field line in the inner jet for our fiducial model B10 and a low resolution model B10-SLR at $t=5\times10^4t_g$. There is larger dissipation, i.e., higher $h$, for B10-SLR as it is unable to resolve the micro-structures in the jet caused by toroidal pinch instabilities.}
    \label{fig:dissipation}
\end{figure}
\begin{figure} 
	\includegraphics[width=\columnwidth]{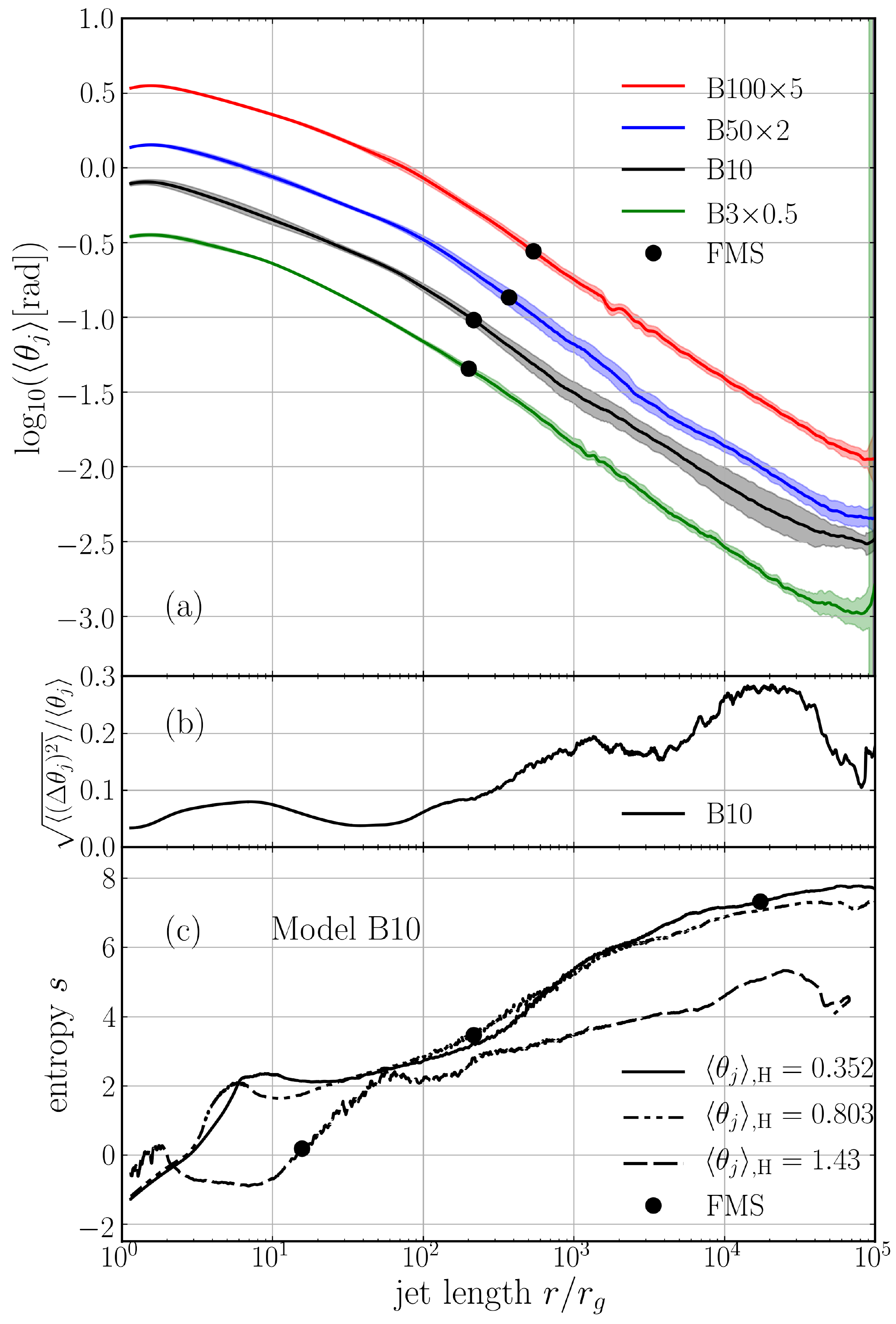}
    \caption{The jet shape changes over time due to the pinching between the jet and the disc-wind. (Panel a) We show the time averaged half-opening angle for field lines from different simulation models, with the same foot-point half-opening angle of $\theta_{j,\rm H}=0.8$~rad, along with their $1\sigma$ standard deviation (shaded area) over $(1.1-1.5)\times10^5 t_g$. To minimise crowding, the curves are shifted vertically. We indicate the position where the field line crosses the fast surface (FMS) with a circle. The field line for model B10 is shown in silver in Fig.~\ref{fig:surfaces}. There is significant time variation in shape of the B10 and B50 field lines beyond the FMS due to the presence of toroidal pinch instabilities, which continue throughout the entire jet. The fast surface moves outward as the jet magnetisation increases \citep[similar to ][]{ceccobello18}. (Panel b) The relative deviation with respect to the time-averaged jet opening angle along the B10 field line shows that there is $>10\%$ deviation in the pinched region. (Panel c) We show the entropy along all 3 field lines from Fig.~\ref{fig:surfaces}, with $\langle \theta_j\rangle_{,\rm H}=(0.352,0.803,1.43)$~rad representing the inner, mid and outer jet respectively. We also indicate where the FMS crosses the field lines with circles. The rise in entropy appears to coincide very well with the FMS, except for the inner jet field line where the entropy rise may be due to round-off errors from machine-precision calculations end up affecting the smallest energy term, i.e., the internal energy, and consequently the gas pressure. }
    \label{fig:opening_angle}
\end{figure}

\subsection{Causal structure of jets}
\label{sec:FMSS}
When a jet becomes super-fast magnetosonic (i.e., downstream of the fast magnetosonic surface), perturbations in the jet cannot be communicated upstream and thus, the jet loses causal connection along its flow. However, beyond the fast surface, globally, causal contact can still be maintained as the jet can communicate upstream via the sub-fast jet axis. Thus, when the flow along the axis turns super-fast, the jet reaches a magnetosonic horizon and causal connection is fully lost. The location where this causal breakdown occurs is the fast magnetosonic separatrix surface \citep[FMSS; for a review, see][]{BOOK2012bhae.book}. Self-similar models \citep[such as e.g.,][]{vla04, pol10, ceccobello18} predict that a jet collapses on its axis once the jet reaches the FMSS and may form a highly radiating hot-spot. Could then bright features in the jets, such as HST-1, be powered by such over-collimation seen in self-similar models?

To test if the FMSS can explain bright jet features, we have developed an algorithm that determines the FMSS location. This algorithm calculates the Mach cone angle \citep[Eq.~D5 in ][]{tch09} for each cell assuming approximate magnetosonic fast wave velocity \citep[][]{gam03}. We track the left and right edges of the Mach cone to check if a fast magnetosonic wave can travel to the sub-fast region near the the jet's axis. Figure~\ref{fig:surfaces}(a), salmon line) shows that the FMSS in model B10  travels inwards across the jet from the outer boundary, before joining with the fast surface at the jet's axis. The FMSS does not coincide with dissipative features, which are shown in Fig.~\ref{fig:surfaces}(a), via the entropy $s$,
\begin{equation}
\begin{centering}
s=\frac{1}{\Gamma-1}\log_{10}\left(\frac{p_{\rm g}} {\rho^{\Gamma}}\right), 
\end{centering}
\label{eqn:entropy}
\end{equation}
a proxy for identifying shocks and magnetic dissipation \citep{Duran17}. Instead, the fast surface (Fig.~\ref{fig:surfaces}(a), blue) coincides with the steady rise in entropy at the jet-edge (Fig.~\ref{fig:surfaces}b). Outside of the jet, in the disc-wind, Fig.~\ref{fig:surfaces}(b) shows that entropy begins increasing in the sub-fast regime (due to shearing between the disc and the wind) and continues to rise smoothly till the jet's fast surface. The above results suggest that the fast surface in the jet plays a role in triggering events which cause dissipation. 

While we can compare the jet structure with radially self-similar models \citep[e.g., ][]{bp82,vla04,pol10,ceccobello18}, the lack of an over-collimation in our simulations suggests that there is a difference in the way the FMSS manifests in the jet within a self-similar approximation. Namely, in radially self-similar models, the FMSS is located where the flow achieves super-magnetosonic speeds towards the polar axis, which our models never reach. That radially self-similar models restrict the radial dependence of quantities to fixed power-laws, which is not the case in our simulations, might be the crucial difference that leads to different nature of FMSS in the self-similar models. Asymptotically in our jets, field lines join with the jet ``core", by which point they become almost cylindrical and stop accelerating efficiently. Perhaps this asymptotic behaviour can be explored in self-similar models by placing the FMSS at infinity \citep[e.g., ][]{li92, vla03a} or via $\theta-$self-similar models \citep[e.g., ][]{Sauty2004}.
\begin{figure*} 
	\includegraphics[width=\textwidth]{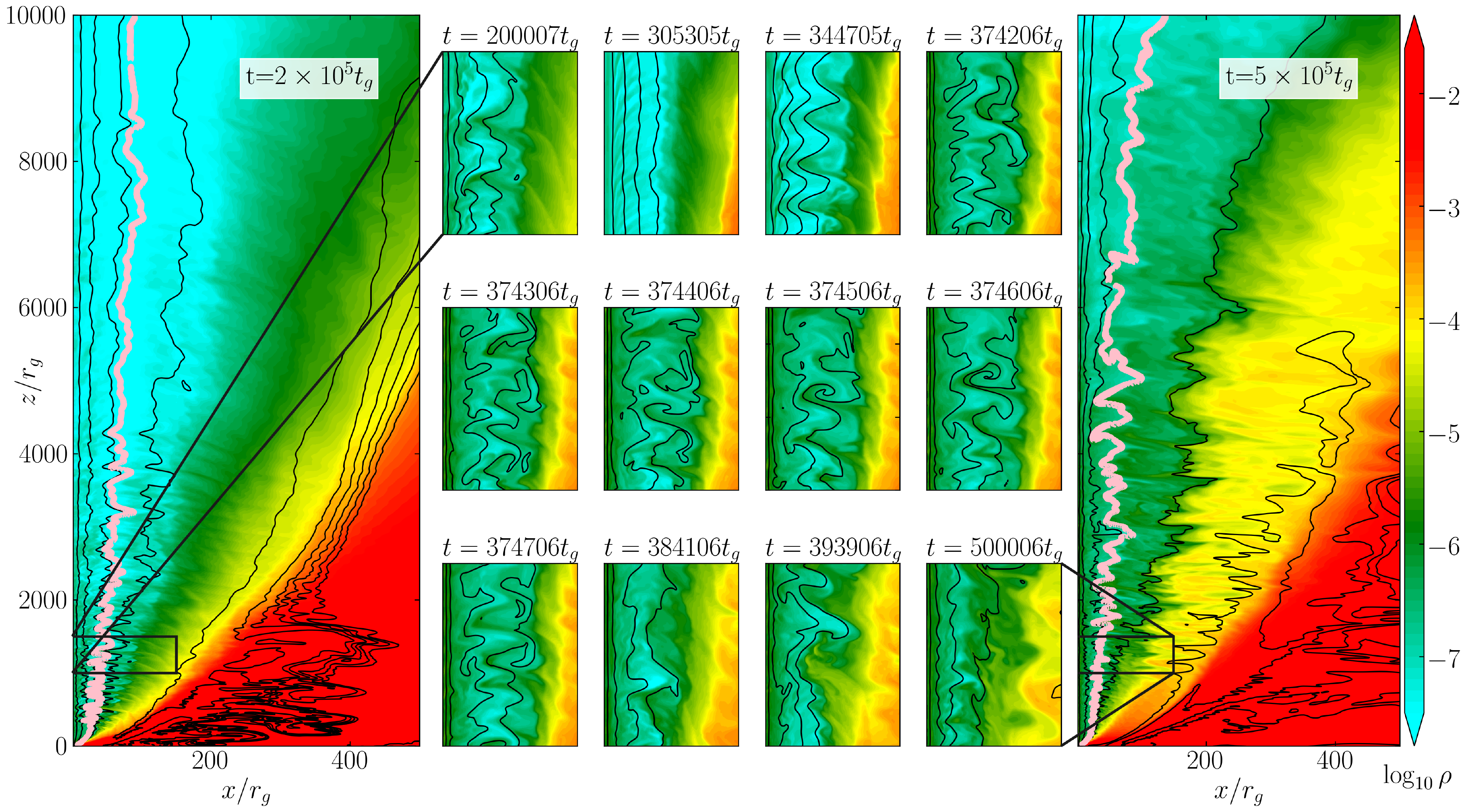}
    \caption{Mixing due to pinch instabilities between the jet and the disc-wind leads to mass-loading of the jet. We show a vertical slice of the B10 upper jet-wind system with density in colour and magnetic field lines in black, at  $t=2\times10^5t_g$ (left) and $t=5\times10^5t_g$ (right). We indicate the field line shown in Fig.~\ref{fig:compare_ideal_1} with pink. Inset panels: Zoom-in snapshots of the jet over $(0-150)r_g\times (1000-1500)r_g$ at different times. Over time, pinch instabilities at the jet-wind interface set off reconnection events and mass-load the jet. The increase in density changes the energy distribution along the jet, reducing its specific energy content and greatly affecting the jet's  acceleration profile (Fig.~\ref{fig:compare_ideal_2}). }
    \label{fig:entrainment}
\end{figure*}
\begin{figure} 
	\includegraphics[width=\columnwidth]{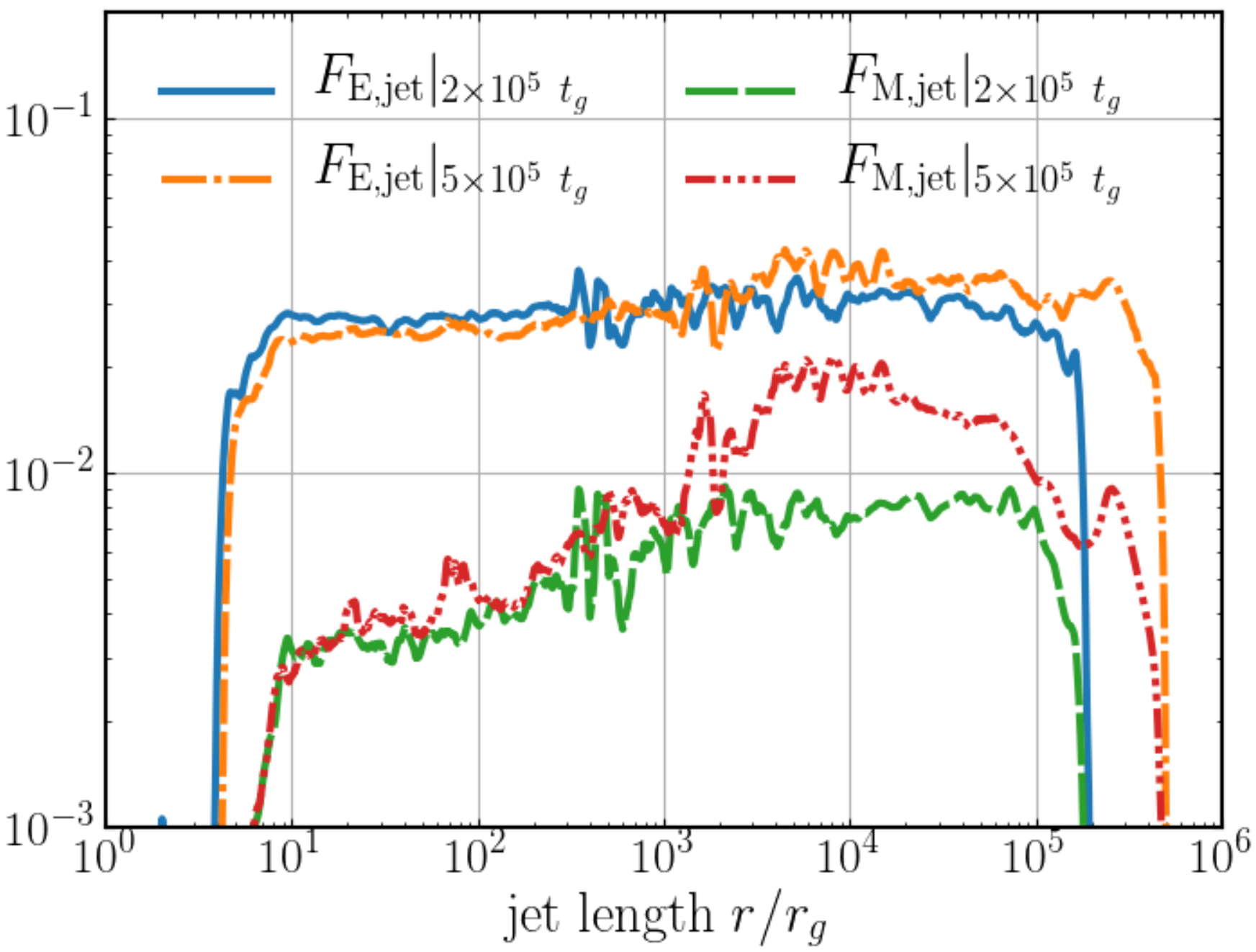}
    \caption{Gas captured from the disc-wind via entrainment increases the mass flux of the jet, while the total energy flux is conserved on average, thus decreasing the specific total energy flux $\mu$ and reducing the jet energy budget. We show the jet total energy flux $F_{\rm E, jet}$ and the rest-mass energy flux $F_{\rm M, jet}$ over the length of the upper jet at $t=2\times10^5t_g$ and $t=5\times10^5t_g$. The jet averaged fluxes are calculated as $F_{x, jet}=\int F_{x}\sqrt{-g}d\theta d\phi$ using the criterion $\mu>1.2$ for the magnetised jet.}
    \label{fig:flux_massload}
\end{figure}

\subsection{Origin of pinch instabilities}
\label{sec:pinch:-origins}
Pinch instabilities forming at the jet-wind interface are easily excited in 2D GRMHD simulations of black hole accretion (see e.g.\ \citealt{mck06jf,bb11,Nakamura_2018}). Whereas previous work found that pinch instabilities do not survive beyond $\sim 10^3 r_g$ \citep[e.g., ][]{mck06jf}, we observe them significantly affecting jet dynamics throughout the length of the jet. We hypothesise that this difference in pinch activity stems from the small disc size in \citet{mck06jf}, as a smaller disc would lead to a conical jet in which pinching instabilities are suppressed (we see the same behaviour for model B10-R, see Sec~\ref{sec:smalltorus}). However, the amount of dissipation seen in \citet{mck06jf} is far larger ($> 2$ orders of magnitude) than in any of our models. To address this discrepancy, we ran an additional low-resolution simulation B10-SLR. Because the B10-SLR model under-resolves the pinches at large radii, this leads to enhanced dissipation at $r \gtrsim 10^3 r_g$ (Fig.~\ref{fig:dissipation}). Additionally, the floor model used in \citet{mck06jf} could also contribute to larger dissipation.

To better quantify the location at which the pinches start to affect the jet, we show in Fig.~\ref{fig:opening_angle}(a) the standard deviation in the jet opening angle along a field line for several models. From Fig.~\ref{fig:opening_angle}(b), we see that the field lines tend to wobble significantly due to pinching, achieving $\sqrt{\langle(\Delta\theta_j)^2\rangle}/\langle\theta_j\rangle>10\%$ close to the fast surface. The fast surface could play a role here since beyond the fast surface, the ram pressure of the wind may become the dominant pressure component and may produce shock-like events at the jet-disc wind interface \citep[e.g., ][]{kom94,Brom2007}. However, in our simulations, we find that the wind remains subsonic in the $\theta-$ direction, suggesting ram pressure in the $\theta-$ direction is not prominent. Indeed, we see a smooth increase of entropy across the jet-disc interface, indicating no prominent shocks (Fig.~\ref{fig:surfaces}). In Fig.~\ref{fig:opening_angle}(c), we see that the inner jet crosses the fast surface at a very large distance, while the entropy begins to rise much earlier. However, the fast surface coincides with the increase of entropy for the mid- and outer-jet. These results suggest that the oscillations in the jet-wind interface (that form very close to the black hole) might give rise to the pinch instabilities, which grow significantly once the jet-edge becomes super-fast. 

However at this point, it is not clear whether the oscillating interface at small radii and pinches at large radii are due to the same underlying physical phenomenon. Both the oscillating interface and the pinches appear to be the response of the jet to the pressure of the surrounding disc-wind \citep{sobacchi_2018a}. Indeed, both the oscillations and pinches disappear in the case of the idealised wall-jet, where the rigid wall prevents a dynamic jet-wind boundary. The jet becomes susceptible to pinch instabilities when the toroidal field dominates over the poloidal field (given by the Tayler criterion: \citealt{Tayler_1957}; also see \citealt{sobacchi_2018b}). We note that the jet-edge field line shown in Fig.~\ref{fig:opening_angle}(c) becomes strongly toroidal at a very small distance. The growth rate of the pinch/kink instability scales with the $\phi-$ component of the Alfv\'en velocity, which is proportional to the toroidal field strength \citep[e.g., ][]{Moll2008}. As \citet{Moll2008} also notes, jet expansion restricts the growth of the pinch, seen in the case for model B10-R, where the small disc allows rapid jet de-collimation, and hence, the jet exhibits weak pinching.

Interestingly, pinches begin to noticeably heat up the jet within $200-800r_g$ from the central black hole \citep[agreeing with e.g., ][]{giannios2006}. This is similar to the distances at which the synchrotron break is estimated to occur for both AGN and X-ray binary jets \citep[e.g., ][]{mar01,mar05,russellD2013,lucchini19_BLLac}. The break arises when the synchrotron emission of a compact jet shifts from its characteristic power-law profile to a flat/inverted spectrum due to self-absorption, transitioning from an optically thin regime at higher frequencies to optically thick (\citealt{blandfordkonigl79}; for a review, see \citealt{Markoff2010,Romero2017}) and is generally attributed to non-thermal emission from particle acceleration caused by e.g., shocks \citep[e.g., ][]{Sironi2009} or magnetic reconnection \citep[e.g., ][]{spruitdaignedrenkhan_2001,Drenkhahn_spruit_2002,Sironi_2014,sironi2015} in the jet. Given that pinching sets off magnetic reconnection in our simulations, we suggest that the start of the pinch region may potentially be an ideal site for particle acceleration to occur for the first time and hence, can manifest itself as a break in the synchrotron spectrum. Additionally, pinching may cause variation in the optical depth for the synchrotron self-absorption, leading to variability in the observed jet depth at a given frequency over time, which might have consequences for the radio core shift in AGN jets (\citealt{blandfordkonigl79,plavin2019}; for M87: \citealt{hada2011}). Using GRMHD simulations that extended out to $100r_g$, \citet{Nakamura_2018} found higher values of Lorentz factor in pinched regions and suggested that if such compressions lead to dissipation, pinches can be associated with superluminal blobs observed in the jets. 

As the jet base magnetisation increases (see Sec.~\ref{sec:mu model}), Fig.~\ref{fig:opening_angle}(a) shows that that the fast surface moves out, away from the launchpoint of the jet (consistent with results from radially self-similar models: Fig.~11, left panel of \citealt{ceccobello18}). Starting from Eq.~\eqref{eqn:fast}, we can derive the approximate distance at which the fast surface resides. We have,   
\begin{equation}
\begin{centering}
\gamma_f^2 v_f^2 \overset{\eqref{eqn:sigma}}{\approx} \sigma \overset{\eqref{eqn:energy}}{\approx} \frac{\mu}{\gamma}-1, 
\end{centering}
\label{eqn:radii_fast_1}
\end{equation}

\noindent assuming a cold jet, i.e., specific enthalpy $h\ll 1$. At the fast surface ($\gamma=\gamma_f$), we then have for the jet, 

\begin{equation}
\begin{centering}
\gamma_f\approx\mu^{1/3}.
\end{centering}
\label{eqn:radii_fast_2}
\end{equation}
Assuming that $\gamma\approx\Omega R$ and $\theta_j=C(r/r_g)^{-\zeta}$, we arrive at the location of the fast surface, 

\begin{equation}
\begin{centering}
r_f \approx \left(\frac{\mu^{1/3}}{C\Omega}\right)^{1/(1-\zeta)}.  
\end{centering}
\label{eqn:radii_fast_3}
\end{equation}

\noindent From Eq.~\eqref{eqn:radii_fast_3}, we can indeed say that $r_f$ increases with increase in $\mu$ for a given field line. For $\mu=10$, Eq.~\eqref{eqn:radii_fast_2} gives $\gamma_f=2.154$ which is about $25\%$ off from the simulation value ($\approx2.7$). However, the assumption of $\gamma\approx\Omega R$ is not valid in the outer jet, due to stronger effects of mass-loading. The time variability of $\theta_j$ decreases for model B10 through to model B100, which suggests that larger magnetisation stabilises against pinching activity \citep[consistent with results of e.g., ][]{Mizuno_2015,Fromm_2017_galaxies,KimLyu2018}. With higher magnetisation, the Alfv\'en speed and subsequently the magnetosonic speed increases (and hence, the fast surface moves to a larger radii: Fig.~\ref{fig:opening_angle}a), which means that the wave takes less time to travel across the jet. Therefore the oscillations have smaller wavelengths and the jet exhibits a small standard deviation in the shape over time. On the other hand, if the jet base magnetisation is low enough, current driven instabilities are not fully triggered, which is the case for model B3, where the jet is mildly magnetised and pinches are absent. 

\begin{figure} 
	\includegraphics[width=\columnwidth]{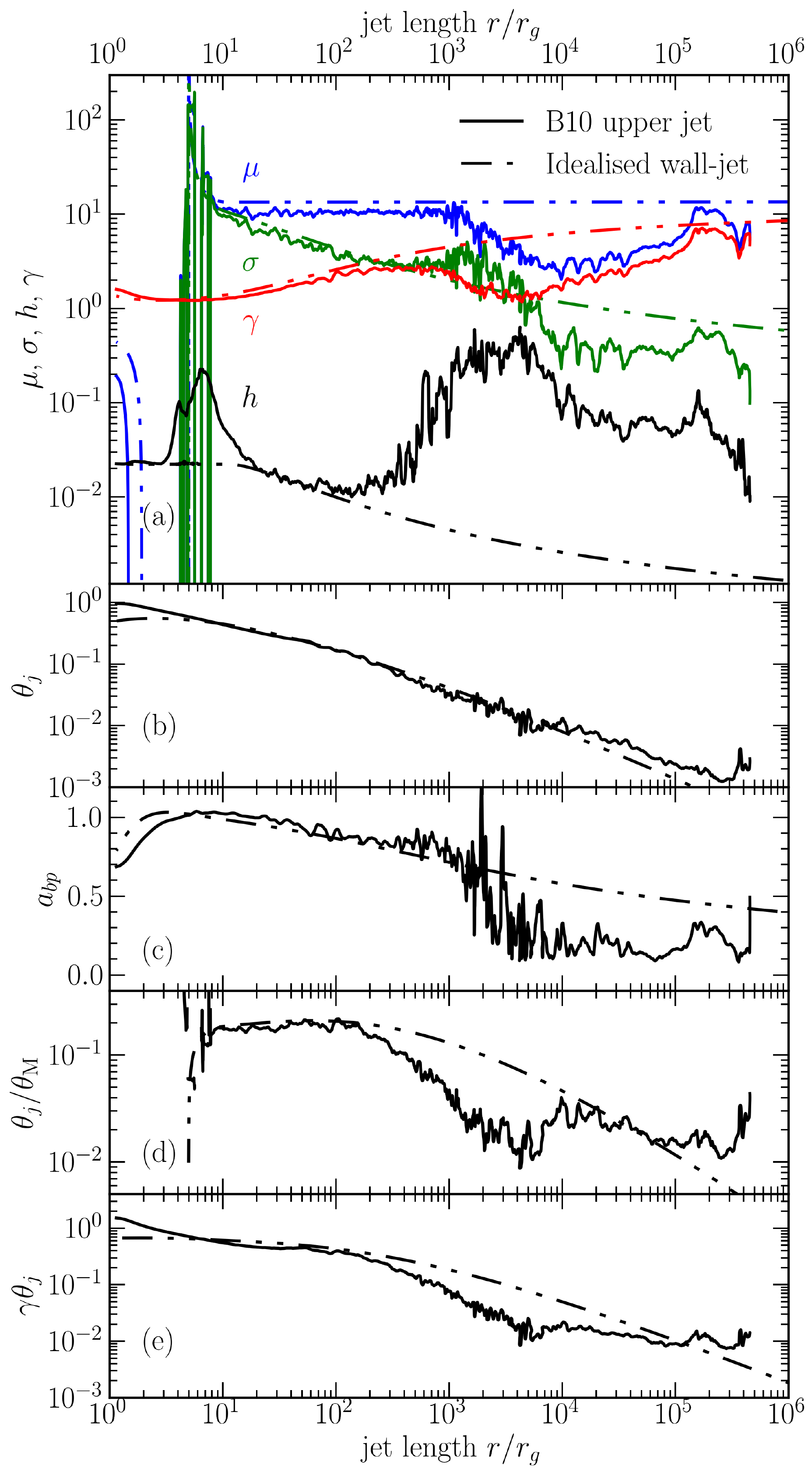}
    \caption{Over time, the disc-wind mixes with the jet and mass-loads it, significantly slowing the jet. We show the same as Fig.~\ref{fig:compare_ideal_1}, except at a later time, $t=5\times10^5 t_g$. Panel(a): There is a large drop in the specific energy profiles for the B10 upper jet as compared to earlier (Fig.~\ref{fig:compare_ideal_1}), directly affecting the acceleration profile. $\mu$ drops by almost an order of magnitude due to the increase in mass flux. Panel(b) shows that the jet collimation profile does not change by much. Panel(c): The poloidal field, on the other hand, drops due to reconnection of pinched field lines, reducing the bunching parameter. Since the jet slows down, causality is still maintained (panel e and f).}
    \label{fig:compare_ideal_2}
\end{figure}
\subsection{Gas entrainment and jet mass-loading}
\label{sec:entrainment}

In most of our simulations, the specific total energy flux $\mu$ oscillates and drops by a small amount in the pinched jet region (see e.g., Fig.~\ref{fig:compare_ideal_1}). Pinching forces the gas to move across field lines in a non-uniform fashion, which disrupts the jet's outward movement as well as causes mass-loading. Figure~\ref{fig:entrainment} shows the effect of jet mass-loading over time, as the B10 upper jet changes substantially over $t=(2-5)\times10^5t_g$. Namely, eddies trap matter in the disc-wind and travel inwards through the jet boundary during pinching, forming finger-like structures. These fingers bend as they interact with the fast moving jet interior, dissipating poloidal field lines through reconnection (Figure~\ref{fig:entrainment}, middle panels), and finally depositing surrounding gas into the jet body\footnote{Movie of mass-loading via gas entrainment in the B10 jet model: \href{https://youtu.be/1aBoNormcS0}{https://youtu.be/1aBoNormcS0}}. Figure~\ref{fig:flux_massload} shows that the mass-flux through the jet indeed increases over time. It will be interesting to test whether explicit resistivity (e.g., \citealt{ripper2019}; we rely on numerical dissipation in \hammer{}) brings any changes to the mass-loading in jets. 

The entrainment mechanism we see here might be a manifestation of the Kruskal Schwarzschild instability (KSI, \citealt{kruskal_schwarzschild_1954}), a magnetised analogue of the Rayleigh-Taylor instability. Possibly, as the jet gets pinched, acceleration towards the axis (which can be seen as an effective gravity term) causes the heavy wind to push against the jet funnel leading to the onset of the KSI and producing the finger-like structures protruding into the jet. The growth rate of KSI is proportional to the square-root of the effective gravity term \citep{Lyubarsky_2010_KSI, Gill2017_KS}, which increases only when the pinches are fully developed. The small gravity term is why KSI may appear at such a late stage ($t>10^5r_g$) in the simulation. The fingers, when extended long enough, may be susceptible to secondary Kelvin-Helmholtz instabilities and bend, much like the distortion of a gas blob immersed in a magnetised fluid under gravity, shown in Fig.~12 of \citet{Gill2017_KS}. These bent fingers eventually collapse on themselves and lead to magnetic field reconnection. In fact, laboratory experiments of a plasma jet by \citet{Moser2012} also exhibit similar hybrid kink-KSI behaviour, which sets off magnetic reconnection events.

Figure~\ref{fig:compare_ideal_2} shows how gas entrainment and jet mass-loading affects model B10 quantities shown in Fig.~\ref{fig:compare_ideal_1}. Compared to the wall-jet, specific energies $\mu$, $\sigma$ and $h$ drop and the jet slows down (panel a) as more and more gas enters the jet over time. From the marginal change in the field line shape (panel b) and the drop in the bunching parameter (panel c) as compared to Fig.~\ref{fig:compare_ideal_1}(b,c), we conclude that the entrainment leads to significant poloidal field reconnection. Ultimately, due to the increasing jet density, the transverse profile of the Lorentz factor undergoes a dramatic change, as the slow sheath region decelerates from $\gamma\sim3$ at $t=2\times10^5t_g$ (Fig.~\ref{fig:transverse_1}, see also Sec.~\ref{sec:upper:-transverse}) to $\gamma\sim 1-2$ at $t>5\times10^5t_g$, suggesting a dynamically changing sheath layer.

In order to understand how the jet behaves in various regimes, we have looked at jet quantities along field lines in different parts of the jet in model B10. In Sec.~\ref{sec:results}, we presented jet properties close to the axis (the foot-point jet opening angle $\theta_{j, \rm H}=0.44$~rad), where the mass-loading is low. We find that even though the inner jet, experiences a slowdown due to pinching, it achieves the peak Lorentz factor $\sim5$. If we look at the mid-jet (Sec.~\ref{sec:ideal:-comparison}, $\theta_{j, \rm H}=0.9$~rad), mass-loading is more prominent between $10^3-10^4r_g$ and eventually brings down $\gamma$ from 3 to near non-relativistic values at $t\sim(2-5)\times10^5t_g$. When we look at the jet as a whole, we find that the distribution of gas from mass-loading and jet velocities is similar to the two-component (spine and sheath) structure of jets \citep[e.g., ][]{ghisellini2005,Ghisellini2008} deduced from limb brightening in AGN jet observations \citep[e.g., ][]{Giroletti_2004, Nagai_2014, Hada2016, Kim_M87_2018}.

\section{Conclusions}
\label{sec:conclusions}

In this work, we use our new GPU-accelerated GRMHD \hammer{} code to investigate the largest extent disc-jet simulation performed till now, reaching over over 5 orders of magnitude in both distance and time in ultra high resolution. We start with a magnetised accretion disc around a spinning black hole that launches and accelerates a jet. This jet is self-consistently collimated by the disc-wind with the support of a large disc and qualitatively resembles the shape and acceleration profile of the M87 jet (Sec.~\ref{sec:M87}). We find that the highly collimated jet maintains lateral causal connectivity with $\gamma \theta \lesssim 0.1$, consistent with VLBI observations of AGN jets (~\ref{sec:M87}). 

Instead of the smooth outflow produced by jets in idealised models \citep[e.g., ][]{kom07,tch10a}, the interacting jet-wind interface exhibits oscillations from very near the jet origin. These oscillations appear to drive pinch instabilities at the jet's outer boundary when the jet becomes super-fast (Sec.~\ref{sec:pinch:-origins}). Pinch instabilities significantly affect jet dynamics as they not only heat up the jet via magnetic reconnection, creating a thermal pressure gradient, but also lead to mass loading of the jet, both of which decelerate the jet. The dissipation due to pinching may lead to particle acceleration, potentially explaining non-thermal synchrotron and inverse Compton hot-spots in jets \citep[e.g., ][]{nkt11,sironi2015,christie2019}. The mass-loading, over time, helps to form a distinct slow moving layer with Lorentz factor $\gamma\sim1$ that surrounds an inner fast moving jet core $\gamma\gtrsim4$ (Sec.~\ref{sec:entrainment}), resembling the spine-sheath structure seen in AGN jets \citep[e.g., ][]{Kim_M87_2018}. 

The instability causing the pinch modes is important to understand, since it may excite kink instabilities in 3D. As the jet kinks, it will interact with the ambient medium, leading to enhanced dissipation \citep[e.g., ][]{begelman_pinch_instability_1998,giannios2006}, which disrupts the jet and potentially explains the FR-I/FR-II dichotomy \citep[e.g., ][]{brom2016, tch16, Duran17, liska2019a}. Our future work will thus focus on extending these results to full 3D, utilising the AMR capability of \hammer{} to focus the resolution on the dissipative regions.

\section*{Acknowledgements}
\label{sec:acks}

We thank O. Bromberg, A. Philippov, O. Porth and E. Sobacchi for helpful discussions. KC thanks M. Lucchini, T. Beuchert, F. Krau{\ss} and T.D. Russell for providing useful insights for the background study. This research was made possible by NSF PRAC award no. 1615281 and OAC-1811605 at the Blue Waters sustained-petascale computing project and supported in part under grant no. NSF PHY-1125915. KC and SM are supported by the Netherlands Organization for Scientific Research (NWO) VICI grant (no. 639.043.513), ML is supported by the NWO Spinoza Prize and AT by Northwestern University, the TAC and NASA Einstein (grant no. PF3-140131) postdoctoral fellowships and the National Science Foundation grant 1815304 and NASA grant 80NSSC18K0565. \\

\noindent The simulation data presented in this work is available on request to AT at \href{mailto:atchekho@northwestern.edu}{atchekho@northwestern.edu}.
\bibliographystyle{mnras}
\bibliography{mybib}

\begin{thebibliography}{}
\makeatletter
\relax
\def\mn@urlcharsother{\let\do\@makeother \do\$\do\&\do\#\do\^\do\_\do\%\do\~}
\def\mn@doi{\begingroup\mn@urlcharsother \@ifnextchar [ {\mn@doi@}
  {\mn@doi@[]}}
\def\mn@doi@[#1]#2{\def\@tempa{#1}\ifx\@tempa\@empty \href
  {http://dx.doi.org/#2} {doi:#2}\else \href {http://dx.doi.org/#2} {#1}\fi
  \endgroup}
\def\mn@eprint#1#2{\mn@eprint@#1:#2::\@nil}
\def\mn@eprint@arXiv#1{\href {http://arxiv.org/abs/#1} {{\tt arXiv:#1}}}
\def\mn@eprint@dblp#1{\href {http://dblp.uni-trier.de/rec/bibtex/#1.xml}
  {dblp:#1}}
\def\mn@eprint@#1:#2:#3:#4\@nil{\def\@tempa {#1}\def\@tempb {#2}\def\@tempc
  {#3}\ifx \@tempc \@empty \let \@tempc \@tempb \let \@tempb \@tempa \fi \ifx
  \@tempb \@empty \def\@tempb {arXiv}\fi \@ifundefined
  {mn@eprint@\@tempb}{\@tempb:\@tempc}{\expandafter \expandafter \csname
  mn@eprint@\@tempb\endcsname \expandafter{\@tempc}}}

\bibitem[\protect\citeauthoryear{Akiyama et~al.,}{Akiyama
  et~al.}{2015}]{Akiyama_2015}
Akiyama K.,  et~al., 2015, \mn@doi [The Astrophysical Journal]
  {10.1088/0004-637x/807/2/150}, 807, 150

\bibitem[\protect\citeauthoryear{Angl\'es-Alc\'azar, Faucher-Gigu$\Grave{\rm
  e}$re, Quataert, Hopkins, Feldmann, Torrey, Wetzel  \& Kere$\check{\rm
  s}$}{Angl\'es-Alc\'azar et~al.}{2017}]{anglesalcazar_2017_FIRE}
Angl\'es-Alc\'azar D.,  Faucher-Gigu$\Grave{\rm e}$re C.-A.,  Quataert E.,
  Hopkins P.~F.,  Feldmann R.,  Torrey P.,  Wetzel A.,   Kere$\check{\rm s}$
  D.,  2017, \mn@doi [\mnras] {10.1093/mnrasl/slx161}, 472, L109

\bibitem[\protect\citeauthoryear{{Asada} \& {Nakamura}}{{Asada} \&
  {Nakamura}}{2012}]{asadanak2012}
{Asada} K.,  {Nakamura} M.,  2012, ApJL, 745, 5 pp

\bibitem[\protect\citeauthoryear{{Balbus} \& {Hawley}}{{Balbus} \&
  {Hawley}}{1991}]{bal91}
{Balbus} S.~A.,  {Hawley} J.~F.,  1991, \mn@doi [\apj] {10.1086/170270}, \href
  {http://adsabs.harvard.edu/cgi-bin/nph-bib_query?bibcode=1991ApJ...376..214B&db_key=AST}
  {376, 214}

\bibitem[\protect\citeauthoryear{{Barkov} \& {Baushev}}{{Barkov} \&
  {Baushev}}{2011}]{bb11}
{Barkov} M.~V.,  {Baushev} A.~N.,  2011, \mn@doi [\na]
  {10.1016/j.newast.2010.07.001}, \href
  {http://adsabs.harvard.edu/abs/2011NewA...16...46B} {16, 46}

\bibitem[\protect\citeauthoryear{{Barniol Duran}, {Tchekhovskoy}  \&
  {Giannios}}{{Barniol Duran} et~al.}{2017}]{Duran17}
{Barniol Duran} R.,  {Tchekhovskoy} A.,   {Giannios} D.,  2017, \mn@doi
  [\mnras] {10.1093/mnras/stx1165}, 469, 4957

\bibitem[\protect\citeauthoryear{{Begelman}}{{Begelman}}{1998}]{begelman_pinch_instability_1998}
{Begelman} M.~C.,  1998, \mn@doi [\apj] {10.1086/305119}, \href
  {http://adsabs.harvard.edu/abs/1998ApJ...493..291B} {493, 291}

\bibitem[\protect\citeauthoryear{Begelman \& Li}{Begelman \&
  Li}{1994}]{begelman_asymptotic_1994}
Begelman M.~C.,  Li Z.-Y.,  1994, \apj, \href
  {http://adsabs.harvard.edu.ezp1.harvard.edu/abs/1994ApJ...426..269B} {426,
  269}

\bibitem[\protect\citeauthoryear{{Beskin} \& {Kuznetsova}}{{Beskin} \&
  {Kuznetsova}}{2000}]{bk00}
{Beskin} V.~S.,  {Kuznetsova} I.~V.,  2000, Nuovo Cimento B Serie, \href
  {http://adsabs.harvard.edu/cgi-bin/nph-bib_query?bibcode=2000NCimB.115..795B&db_key=PHY}
  {115, 795}

\bibitem[\protect\citeauthoryear{{Beskin} \& {Nokhrina}}{{Beskin} \&
  {Nokhrina}}{2006}]{bes06}
{Beskin} V.~S.,  {Nokhrina} E.~E.,  2006, \mn@doi [\mnras]
  {10.1111/j.1365-2966.2006.09957.x}, \href
  {http://adsabs.harvard.edu/cgi-bin/nph-bib_query?bibcode=2006MNRAS.367..375B&db_key=AST}
  {367, 375}

\bibitem[\protect\citeauthoryear{{Beskin}, {Kuznetsova}  \& {Rafikov}}{{Beskin}
  et~al.}{1998}]{bes98}
{Beskin} V.~S.,  {Kuznetsova} I.~V.,   {Rafikov} R.~R.,  1998, \mnras, \href
  {http://adsabs.harvard.edu/abs/1998MNRAS.299..341B} {299, 341}

\bibitem[\protect\citeauthoryear{{Biretta}, {Sparks}  \& {Macchetto}}{{Biretta}
  et~al.}{1999}]{Biret1999}
{Biretta} J.~A.,  {Sparks} W.~B.,   {Macchetto} F.,  1999, \mn@doi [\apj]
  {10.1086/307499}, \href {http://adsabs.harvard.edu/abs/1999ApJ...520..621B}
  {520, 621}

\bibitem[\protect\citeauthoryear{{Blandford} \& {K{\"o}nigl}}{{Blandford} \&
  {K{\"o}nigl}}{1979}]{blandfordkonigl79}
{Blandford} R.~D.,  {K{\"o}nigl} A.,  1979, \apj, \href
  {http://adsabs.harvard.edu/abs/1979ApJ...232...34B} {232, 34}

\bibitem[\protect\citeauthoryear{{Blandford} \& {Payne}}{{Blandford} \&
  {Payne}}{1982}]{bp82}
{Blandford} R.~D.,  {Payne} D.~G.,  1982, \mnras, \href
  {http://adsabs.harvard.edu/cgi-bin/nph-bib_query?bibcode=1982MNRAS.199..883B&db_key=AST}
  {199, 883}

\bibitem[\protect\citeauthoryear{{Blandford} \& {Znajek}}{{Blandford} \&
  {Znajek}}{1977}]{bz77}
{Blandford} R.~D.,  {Znajek} R.~L.,  1977, \mnras, \href
  {http://adsabs.harvard.edu/cgi-bin/nph-bib_query?bibcode=1977MNRAS.179..433B&db_key=AST}
  {179, 433}

\bibitem[\protect\citeauthoryear{Bourne \& Sijacki}{Bourne \&
  Sijacki}{2017}]{bourne_arepo2017}
Bourne M.~A.,  Sijacki D.,  2017, \mn@doi [\mnras] {10.1093/mnras/stx2269},
  472, 4707

\bibitem[\protect\citeauthoryear{Bower, Benson, Malbon, Helly, Frenk, Baugh,
  Cole  \& Lacey}{Bower et~al.}{2006}]{bower2006}
Bower R.~G.,  Benson A.~J.,  Malbon R.,  Helly J.~C.,  Frenk C.~S.,  Baugh
  C.~M.,  Cole S.,   Lacey C.~G.,  2006, \mn@doi [\mnras]
  {10.1111/j.1365-2966.2006.10519.x}, 370, 645

\bibitem[\protect\citeauthoryear{Broderick \& Loeb}{Broderick \&
  Loeb}{2006}]{broderick_2006}
Broderick A.~E.,  Loeb A.,  2006, \mn@doi [\mnras]
  {10.1111/j.1365-2966.2006.10152.x}, 367, 905

\bibitem[\protect\citeauthoryear{{Broderick} \& {Tchekhovskoy}}{{Broderick} \&
  {Tchekhovskoy}}{2015}]{2015ApJ...809...97B}
{Broderick} A.~E.,  {Tchekhovskoy} A.,  2015, \mn@doi [\apj]
  {10.1088/0004-637X/809/1/97}, \href
  {http://adsabs.harvard.edu/abs/2015ApJ...809...97B} {809, 97}

\bibitem[\protect\citeauthoryear{Bromberg \& Levinson}{Bromberg \&
  Levinson}{2007}]{Brom2007}
Bromberg O.,  Levinson A.,  2007, \apj, 671, 678

\bibitem[\protect\citeauthoryear{{Bromberg} \& {Tchekhovskoy}}{{Bromberg} \&
  {Tchekhovskoy}}{2016}]{brom2016}
{Bromberg} O.,  {Tchekhovskoy} A.,  2016, \mn@doi [\mnras]
  {10.1093/mnras/stv2591}, 456, 1739

\bibitem[\protect\citeauthoryear{{Ceccobello}, {Cavecchi}, {Heemskerk},
  {Markoff}, {Polko}  \& {Meier}}{{Ceccobello} et~al.}{2018}]{ceccobello18}
{Ceccobello} C.,  {Cavecchi} Y.,  {Heemskerk} M.~H.~M.,  {Markoff} S.,  {Polko}
  P.,   {Meier} D.~L.,  2018, \mn@doi [MNRAS] {10.1093/mnras/stx2567}, 473,
  4417

\bibitem[\protect\citeauthoryear{{Chen}, {Yuan}  \& {Yang}}{{Chen}
  et~al.}{2018}]{2018ApJ...863L..31C}
{Chen} A.~Y.,  {Yuan} Y.,   {Yang} H.,  2018, \mn@doi [\apjl]
  {10.3847/2041-8213/aad8ab}, \href
  {http://adsabs.harvard.edu/abs/2018ApJ...863L..31C} {863, L31}

\bibitem[\protect\citeauthoryear{{Chiueh}, {Li}  \& {Begelman}}{{Chiueh}
  et~al.}{1998}]{chiueh_crab_1998}
{Chiueh} T.,  {Li} Z.-Y.,   {Begelman} M.~C.,  1998, \mn@doi [\apj]
  {10.1086/306209}, \href {http://adsabs.harvard.edu/abs/1998ApJ...505..835C}
  {505, 835}

\bibitem[\protect\citeauthoryear{Christie, Petropoulou, Sironi  \&
  Giannios}{Christie et~al.}{2018}]{christie2019}
Christie I.~M.,  Petropoulou M.,  Sironi L.,   Giannios D.,  2018, \mn@doi
  [\mnras] {10.1093/mnras/sty2636}, 482, 65

\bibitem[\protect\citeauthoryear{{Clausen-Brown}, {Savolainen}, {Pushkarev},
  {Kovalev}  \& {Zensus}}{{Clausen-Brown} et~al.}{2013}]{clausenB13}
{Clausen-Brown} E.,  {Savolainen} T.,  {Pushkarev} A.~B.,  {Kovalev} Y.~Y.,
  {Zensus} J.~A.,  2013, \mn@doi [\aap] {10.1038/nature13399}, \href
  {https://doi.org/10.1051/0004-6361/201322203} {558, A144}

\bibitem[\protect\citeauthoryear{{Colella} \& {Woodward}}{{Colella} \&
  {Woodward}}{1984}]{col84}
{Colella} P.,  {Woodward} P.~R.,  1984, Journal of Computational Physics, \href
  {http://adsabs.harvard.edu/cgi-bin/nph-bib_query?bibcode=1984JCoPh..54..174C&db_key=AST}
  {54, 174}

\bibitem[\protect\citeauthoryear{{Cowling}}{{Cowling}}{1934}]{cowling34}
{Cowling} T.~G.,  1934, \mnras, \href
  {http://adsabs.harvard.edu/abs/1934MNRAS..94..768C} {94, 768}

\bibitem[\protect\citeauthoryear{{De Villiers}, {Hawley}  \& {Krolik}}{{De
  Villiers} et~al.}{2003}]{dev03}
{De Villiers} J.-P.,  {Hawley} J.~F.,   {Krolik} J.~H.,  2003, \mn@doi [\apj]
  {10.1086/379509}, \href {http://adsabs.harvard.edu/abs/2003ApJ...599.1238D}
  {599, 1238}

\bibitem[\protect\citeauthoryear{{Doeleman} et~al.}{{Doeleman}
  et~al.}{2012}]{doel12}
{Doeleman} S.~S.,  et~al., 2012, Science, \href
  {http://adsabs.harvard.edu/abs/2012Sci...338..355D} {338, 355}

\bibitem[\protect\citeauthoryear{Drenkhahn \& Spruit}{Drenkhahn \&
  Spruit}{2002}]{Drenkhahn_spruit_2002}
Drenkhahn G.,  Spruit H.~C.,  2002, \mn@doi [A\&A]
  {10.1051/0004-6361:20020839}, 391, 1141

\bibitem[\protect\citeauthoryear{{Eichler}}{{Eichler}}{1993}]{eichler_93}
{Eichler} D.,  1993, \mn@doi [\apj] {10.1086/173464}, \href
  {http://adsabs.harvard.edu/abs/1993ApJ...419..111E} {419, 111}

\bibitem[\protect\citeauthoryear{{Fabian}}{{Fabian}}{2012}]{Fabian2012}
{Fabian} A.~C.,  2012, \mn@doi [Annual Review of Astronomy and Astrophysics]
  {10.1146/annurev-astro-081811-125521}, 50, 455

\bibitem[\protect\citeauthoryear{{Fishbone} \& {Moncrief}}{{Fishbone} \&
  {Moncrief}}{1976}]{fis76}
{Fishbone} L.~G.,  {Moncrief} V.,  1976, \apj, \href
  {http://adsabs.harvard.edu/cgi-bin/nph-bib_query?bibcode=1976ApJ...207..962F&db_key=AST}
  {207, 962}

\bibitem[\protect\citeauthoryear{Fromm, Porth, Younsi, Mizuno, De~Laurentis,
  Olivares  \& Rezzolla}{Fromm et~al.}{2017}]{Fromm_2017_galaxies}
Fromm C.~M.,  Porth O.,  Younsi Z.,  Mizuno Y.,  De~Laurentis M.,  Olivares H.,
    Rezzolla L.,  2017, \mn@doi [Galaxies] {10.3390/galaxies5040073}, 5

\bibitem[\protect\citeauthoryear{{Gammie}, {McKinney}  \& {T{\'o}th}}{{Gammie}
  et~al.}{2003}]{gam03}
{Gammie} C.~F.,  {McKinney} J.~C.,   {T{\'o}th} G.,  2003, \mn@doi [\apj]
  {10.1086/374594}, \href
  {http://adsabs.harvard.edu/cgi-bin/nph-bib_query?bibcode=2003ApJ...589..444G&db_key=AST}
  {589, 444}

\bibitem[\protect\citeauthoryear{{Gardiner} \& {Stone}}{{Gardiner} \&
  {Stone}}{2005}]{gar05}
{Gardiner} T.~A.,  {Stone} J.~M.,  2005, \mn@doi [Journal of Computational
  Physics] {10.1016/j.jcp.2004.11.016}, \href
  {http://adsabs.harvard.edu/cgi-bin/nph-bib_query?bibcode=2005JCoPh.205..509G&db_key=PHY}
  {205, 509}

\bibitem[\protect\citeauthoryear{Ghisellini \& Tavecchio}{Ghisellini \&
  Tavecchio}{2008}]{Ghisellini2008}
Ghisellini G.,  Tavecchio F.,  2008, \mn@doi [\mnras]
  {10.1111/j.1745-3933.2008.00441.x}, 385, L98

\bibitem[\protect\citeauthoryear{Ghisellini, Tavecchio  \&
  Chiaberge}{Ghisellini et~al.}{2005}]{ghisellini2005}
Ghisellini G.,  Tavecchio F.,   Chiaberge M.,  2005, \mn@doi [A\&A]
  {10.1051/0004-6361:20041404}, 432, 401

\bibitem[\protect\citeauthoryear{Giannios \& Spruit}{Giannios \&
  Spruit}{2006}]{giannios2006}
Giannios D.,  Spruit H.~C.,  2006, A\&A, 450, 887

\bibitem[\protect\citeauthoryear{Gill, Granot  \& Lyubarsky}{Gill
  et~al.}{2017}]{Gill2017_KS}
Gill R.,  Granot J.,   Lyubarsky Y.,  2017, \mn@doi [\mnras]
  {10.1093/mnras/stx3000}, 474, 3535

\bibitem[\protect\citeauthoryear{Giroletti et~al.,}{Giroletti
  et~al.}{2004}]{Giroletti_2004}
Giroletti M.,  et~al., 2004, \mn@doi [\apj] {10.1086/379663}, 600, 127

\bibitem[\protect\citeauthoryear{{Granot}, {Komissarov}  \&
  {Spitkovsky}}{{Granot} et~al.}{2011}]{Granot2011}
{Granot} J.,  {Komissarov} S.~S.,   {Spitkovsky} A.,  2011, \mn@doi [\mnras]
  {10.1111/j.1365-2966.2010.17770.x}, \href
  {http://cdsads.u-strasbg.fr/abs/2011MNRAS.411.1323G} {411, 1323}

\bibitem[\protect\citeauthoryear{Hada, Doi, Kino, Nagai, Hagiwara  \&
  Kawaguchi}{Hada et~al.}{2011}]{hada2011}
Hada K.,  Doi A.,  Kino M.,  Nagai H.,  Hagiwara Y.,   Kawaguchi N.,  2011,
  Nature, 477, 185

\bibitem[\protect\citeauthoryear{{Hada} et~al.,}{{Hada}
  et~al.}{2013}]{Hada2013}
{Hada} K.,  et~al., 2013, \apj, 775, 70

\bibitem[\protect\citeauthoryear{Hada et~al.,}{Hada et~al.}{2016}]{Hada2016}
Hada K.,  et~al., 2016, \apj, 817, 131

\bibitem[\protect\citeauthoryear{{Harrison}, {Costa}, {Tadhunter},
  {Fl{\"u}tsch}, {Kakkad}, {Perna}  \& {Vietri}}{{Harrison}
  et~al.}{2018}]{har18feed}
{Harrison} C.~M.,  {Costa} T.,  {Tadhunter} C.~N.,  {Fl{\"u}tsch} A.,  {Kakkad}
  D.,  {Perna} M.,   {Vietri} G.,  2018, \mn@doi [Nature Astronomy]
  {10.1038/s41550-018-0403-6}, 2, 198

\bibitem[\protect\citeauthoryear{{Harten}}{{Harten}}{1983}]{har83}
{Harten} A.,  1983, Journal of Computational Physics, 49, 357

\bibitem[\protect\citeauthoryear{{Hawley} \& {Krolik}}{{Hawley} \&
  {Krolik}}{2006}]{hk06}
{Hawley} J.~F.,  {Krolik} J.~H.,  2006, \mn@doi [\apj] {10.1086/500385}, \href
  {http://adsabs.harvard.edu/abs/2006ApJ...641..103H} {641, 103}

\bibitem[\protect\citeauthoryear{{Hawley}, {Guan}  \& {Krolik}}{{Hawley}
  et~al.}{2011}]{hgk11}
{Hawley} J.~F.,  {Guan} X.,   {Krolik} J.~H.,  2011, ArXiv:1103.5987, \href
  {http://adsabs.harvard.edu/abs/2011arXiv1103.5987H} {}

\bibitem[\protect\citeauthoryear{{Hirotani} \& {Okamoto}}{{Hirotani} \&
  {Okamoto}}{1998}]{1998ApJ...497..563H}
{Hirotani} K.,  {Okamoto} I.,  1998, \mn@doi [\apj] {10.1086/305479}, \href
  {http://adsabs.harvard.edu/abs/1998ApJ...497..563H} {497, 563}

\bibitem[\protect\citeauthoryear{{Hirotani} \& {Pu}}{{Hirotani} \&
  {Pu}}{2016}]{2016ApJ...818...50H}
{Hirotani} K.,  {Pu} H.-Y.,  2016, \mn@doi [\apj] {10.3847/0004-637X/818/1/50},
  \href {http://adsabs.harvard.edu/abs/2016ApJ...818...50H} {818, 50}

\bibitem[\protect\citeauthoryear{{Igumenshchev}, {Narayan}  \&
  {Abramowicz}}{{Igumenshchev} et~al.}{2003}]{igu03}
{Igumenshchev} I.~V.,  {Narayan} R.,   {Abramowicz} M.~A.,  2003, \mn@doi
  [\apj] {10.1086/375769}, \href
  {http://adsabs.harvard.edu/cgi-bin/nph-bib_query?bibcode=2003ApJ...592.1042I&db_key=AST}
  {592, 1042}

\bibitem[\protect\citeauthoryear{{Jorstad} et~al.,}{{Jorstad}
  et~al.}{2005}]{jorstad_agn_jet_2005}
{Jorstad} S.~G.,  et~al., 2005, \mn@doi [\aj] {10.1086/444593}, \href
  {http://adsabs.harvard.edu/abs/2005AJ....130.1418J} {130, 1418}

\bibitem[\protect\citeauthoryear{Jorstad et~al.,}{Jorstad
  et~al.}{2017}]{jorstad2017}
Jorstad S.~G.,  et~al., 2017, \apj, 846, 98

\bibitem[\protect\citeauthoryear{Kennel \& Coroniti}{Kennel \&
  Coroniti}{1984}]{kennel_confinement_crab_1984}
Kennel C.~F.,  Coroniti F.~V.,  1984, \apj, \href
  {http://adsabs.harvard.edu.ezp-prod1.hul.harvard.edu/abs/1984ApJ...283..694K}
  {283, 694}

\bibitem[\protect\citeauthoryear{Kim, Balsara, Lyutikov  \& Komissarov}{Kim
  et~al.}{2018a}]{KimLyu2018}
Kim J.,  Balsara D.~S.,  Lyutikov M.,   Komissarov S.~S.,  2018a, \mn@doi
  [\mnras] {10.1093/mnras/stx3065}, 474, 3954

\bibitem[\protect\citeauthoryear{{Kim} et~al.,}{{Kim}
  et~al.}{2018b}]{Kim_M87_2018}
{Kim} J.-Y.,  et~al., 2018b, A\&A, \href
  {https://doi.org/10.1051/0004-6361/201832921} {616, A188}

\bibitem[\protect\citeauthoryear{Komissarov}{Komissarov}{1994}]{kom94}
Komissarov S.~S.,  1994, \mn@doi [\mnras] {10.1093/mnras/266.3.649}, 266, 649

\bibitem[\protect\citeauthoryear{{Komissarov}}{{Komissarov}}{2001}]{kom01}
{Komissarov} S.~S.,  2001, \mn@doi [\mnras] {10.1046/j.1365-8711.2001.04863.x},
  \href {http://adsabs.harvard.edu/abs/2001MNRAS.326L..41K} {326, L41}

\bibitem[\protect\citeauthoryear{{Komissarov}}{{Komissarov}}{2005}]{kom05}
{Komissarov} S.~S.,  2005, \mn@doi [\mnras] {10.1111/j.1365-2966.2005.08974.x},
  \href
  {http://adsabs.harvard.edu/cgi-bin/nph-bib_query?bibcode=2005MNRAS.359..801K&db_key=AST}
  {359, 801}

\bibitem[\protect\citeauthoryear{{Komissarov}}{{Komissarov}}{2012}]{kom2012}
{Komissarov} S.~S.,  2012, \mn@doi [\mnras] {10.1111/j.1365-2966.2012.20609.x},
  \href {http://cdsads.u-strasbg.fr/abs/2012MNRAS.422..326K} {422, 326}

\bibitem[\protect\citeauthoryear{{Komissarov}, {Barkov}, {Vlahakis}  \&
  {K{\"o}nigl}}{{Komissarov} et~al.}{2007}]{kom07}
{Komissarov} S.~S.,  {Barkov} M.~V.,  {Vlahakis} N.,   {K{\"o}nigl} A.,  2007,
  \mn@doi [\mnras] {10.1111/j.1365-2966.2007.12050.x}, \href
  {http://adsabs.harvard.edu/abs/2007MNRAS.380...51K} {380, 51}

\bibitem[\protect\citeauthoryear{{Komissarov}, {Vlahakis}, {K{\"o}nigl}  \&
  {Barkov}}{{Komissarov} et~al.}{2009}]{kom09}
{Komissarov} S.~S.,  {Vlahakis} N.,  {K{\"o}nigl} A.,   {Barkov} M.~V.,  2009,
  \mn@doi [\mnras] {10.1111/j.1365-2966.2009.14410.x}, \href
  {http://adsabs.harvard.edu/abs/2009MNRAS.394.1182K} {394, 1182}

\bibitem[\protect\citeauthoryear{Komissarov, Vlahakis  \& Königl}{Komissarov
  et~al.}{2010}]{kom2010}
Komissarov S.~S.,  Vlahakis N.,   Königl A.,  2010, \mn@doi [\mnras]
  {10.1111/j.1365-2966.2010.16779.x}, 407, 17

\bibitem[\protect\citeauthoryear{Komissarov, Porth  \& Lyutikov}{Komissarov
  et~al.}{2015}]{kom2015}
Komissarov S.~S.,  Porth O.,   Lyutikov M.,  2015, \mn@doi [Computational
  Astrophysics and Cosmology] {10.1186/s40668-015-0013-y}, 2, 9

\bibitem[\protect\citeauthoryear{Kruskal \& Schwarzschild}{Kruskal \&
  Schwarzschild}{1954}]{kruskal_schwarzschild_1954}
Kruskal M.~D.,  Schwarzschild M.,  1954, \mn@doi [Proc. Roy. Soc. (London)]
  {http://doi.org/10.1098/rspa.1954.0120}, A223, 348

\bibitem[\protect\citeauthoryear{{Levinson} \& {Segev}}{{Levinson} \&
  {Segev}}{2017}]{2017PhRvD..96l3006L}
{Levinson} A.,  {Segev} N.,  2017, \mn@doi [\prd] {10.1103/PhysRevD.96.123006},
  \href {http://adsabs.harvard.edu/abs/2017PhRvD..96l3006L} {96, 123006}

\bibitem[\protect\citeauthoryear{{Li}, {Chiueh}  \& {Begelman}}{{Li}
  et~al.}{1992}]{li92}
{Li} Z.-Y.,  {Chiueh} T.,   {Begelman} M.~C.,  1992, \mn@doi [\apj]
  {10.1086/171597}, \href {http://adsabs.harvard.edu/abs/1992ApJ...394..459L}
  {394, 459}

\bibitem[\protect\citeauthoryear{{Liska}, {Tchekhovskoy}, {Ingram}  \& {van der
  Klis}}{{Liska} et~al.}{2018b}]{Liska2018C}
{Liska} M.,  {Tchekhovskoy} A.,  {Ingram} A.,   {van der Klis} M.,  2018b,
  preprint, \href {http://adsabs.harvard.edu/abs/2018arXiv181000883L} {}
  (\mn@eprint {arXiv} {1810.00883})

\bibitem[\protect\citeauthoryear{{Liska}, {Tchekhovskoy}  \&
  {Quataert}}{{Liska} et~al.}{2018a}]{liska2018b}
{Liska} M.,  {Tchekhovskoy} A.,   {Quataert} E.,  2018a, preprint, \href
  {http://adsabs.harvard.edu/abs/2018arXiv180904608L} {} (\mn@eprint {arXiv}
  {1809.04608})

\bibitem[\protect\citeauthoryear{{Liska}, {Hesp}, {Tchekhovskoy}, {Ingram},
  {van der Klis}  \& {Markoff}}{{Liska} et~al.}{2018c}]{liska2018a}
{Liska} M.,  {Hesp} C.,  {Tchekhovskoy} A.,  {Ingram} A.,  {van der Klis} M.,
  {Markoff} S.,  2018c, \mn@doi [\mnras] {10.1093/mnrasl/slx174}, 474, L81

\bibitem[\protect\citeauthoryear{{Liska}, {Hesp}, {Tchekhovskoy}, {Ingram},
  {van der Klis}  \& {Markoff}}{{Liska} et~al.}{2019}]{liska2019a}
{Liska} M.,  {Hesp} C.,  {Tchekhovskoy} A.,  {Ingram} A.,  {van der Klis} M.,
  {Markoff} S.,  2019, preprint, \href
  {http://adsabs.harvard.edu/abs/2019arXiv190105970L} {} (\mn@eprint {arXiv}
  {1901.05970})

\bibitem[\protect\citeauthoryear{Lister et~al.,}{Lister
  et~al.}{2016}]{lister2016}
Lister M.~L.,  et~al., 2016, ApJ, 152, 12

\bibitem[\protect\citeauthoryear{Lucchini, Krau{\ss}, Markoff, Crumley  \&
  Connors}{Lucchini et~al.}{2018}]{lucchini19_BLLac}
Lucchini M.,  Krau{\ss} F.,  Markoff S.,  Crumley P.,   Connors R. M.~T.,
  2018, \mn@doi [\mnras] {10.1093/mnras/sty2929}, 482, 4798

\bibitem[\protect\citeauthoryear{{Lyubarsky}}{{Lyubarsky}}{2009}]{lyub09}
{Lyubarsky} Y.,  2009, \mn@doi [\apj] {10.1088/0004-637X/698/2/1570}, \href
  {http://adsabs.harvard.edu/abs/2009ApJ...698.1570L} {698, 1570}

\bibitem[\protect\citeauthoryear{Lyubarsky}{Lyubarsky}{2010}]{Lyubarsky_2010_KSI}
Lyubarsky Y.,  2010, \mn@doi [\apj] {10.1088/2041-8205/725/2/l234}, 725, L234

\bibitem[\protect\citeauthoryear{Magorrian et~al.}{Magorrian
  et~al.}{1998}]{Magorrian98}
Magorrian J.,  et~al., 1998, \apj, \href
  {http://adsabs.harvard.edu/abs/1998A%26A...331L...1S} {115, 2285}

\bibitem[\protect\citeauthoryear{Markoff}{Markoff}{2010}]{Markoff2010}
Markoff S.,  2010, From Multiwavelength to Mass Scaling: Accretion and Ejection
  in Microquasars and AGN.
Springer Berlin Heidelberg, Berlin, Heidelberg, pp 143--172,
  \mn@doi{10.1007/978-3-540-76937-8_6}, \url
  {https://doi.org/10.1007/978-3-540-76937-8_6}

\bibitem[\protect\citeauthoryear{Markoff, Falcke  \& Fender}{Markoff
  et~al.}{2001}]{mar01}
Markoff S.,  Falcke H.,   Fender R.~P.,  2001, Astron. Astrophys., 372, L25

\bibitem[\protect\citeauthoryear{Markoff, Nowak  \& Wilms}{Markoff
  et~al.}{2005}]{mar05}
Markoff S.,  Nowak M.~A.,   Wilms J.,  2005, \apj, 635, 1203

\bibitem[\protect\citeauthoryear{{McKinney}}{{McKinney}}{2005}]{mck05}
{McKinney} J.~C.,  2005, \mn@doi [\apjl] {10.1086/468184}, \href
  {http://adsabs.harvard.edu/cgi-bin/nph-bib_query?bibcode=2005ApJ...630L...5M&db_key=AST}
  {630, L5}

\bibitem[\protect\citeauthoryear{{McKinney}}{{McKinney}}{2006}]{mck06jf}
{McKinney} J.~C.,  2006, \mn@doi [\mnras] {10.1111/j.1365-2966.2006.10256.x},
  \href
  {http://adsabs.harvard.edu/cgi-bin/nph-bib_query?bibcode=2006MNRAS.368.1561M&db_key=AST}
  {368, 1561}

\bibitem[\protect\citeauthoryear{{Meier}}{{Meier}}{2012}]{BOOK2012bhae.book}
{Meier} D.~L.,  2012, Black Hole Astrophysics: The Engine Paradigm.
Springer, Verlag Berlin Heidelberg

\bibitem[\protect\citeauthoryear{{Meier}, {Edgington}, {Godon}, {Payne}  \&
  {Lind}}{{Meier} et~al.}{1997}]{meier97}
{Meier} D.~L.,  {Edgington} S.,  {Godon} P.,  {Payne} D.~G.,   {Lind} K.~R.,
  1997, \mn@doi [\nat] {10.1038/41034}, \href
  {http://adsabs.harvard.edu/abs/1997Natur.388..350M} {388, 350}

\bibitem[\protect\citeauthoryear{{Mertens}, {Lobanov}, {Walker}  \&
  {Hardee}}{{Mertens} et~al.}{2016}]{mertens2016}
{Mertens} F.,  {Lobanov} A.~P.,  {Walker} R.~C.,   {Hardee} P.~E.,  2016,
  \mn@doi [\aap] {10.1051/0004-6361/201628829}, 595, A54

\bibitem[\protect\citeauthoryear{{Mizuno}, {Yamada}, {Koide}  \&
  {Shibata}}{{Mizuno} et~al.}{2004}]{mizuno04}
{Mizuno} Y.,  {Yamada} S.,  {Koide} S.,   {Shibata} K.,  2004, \mn@doi [\apj]
  {10.1086/423949}, \href {http://adsabs.harvard.edu/abs/2004ApJ...615..389M}
  {615, 389}

\bibitem[\protect\citeauthoryear{Mizuno, G{\'{o}}mez, Nishikawa, Meli, Hardee
  \& Rezzolla}{Mizuno et~al.}{2015}]{Mizuno_2015}
Mizuno Y.,  G{\'{o}}mez J.~L.,  Nishikawa K.-I.,  Meli A.,  Hardee P.~E.,
  Rezzolla L.,  2015, \mn@doi [The Astrophysical Journal]
  {10.1088/0004-637x/809/1/38}, 809, 38

\bibitem[\protect\citeauthoryear{{Moll}, {Spruit}  \& {Obergaulinger}}{{Moll}
  et~al.}{2008}]{Moll2008}
{Moll} R.,  {Spruit} H.~C.,   {Obergaulinger} M.,  2008, \mn@doi [\aap]
  {10.1051/0004-6361:200810523}, \href
  {http://adsabs.harvard.edu/abs/2008A%26A...492..621M} {492, 621}

\bibitem[\protect\citeauthoryear{Moser \& Bellan}{Moser \&
  Bellan}{2012}]{Moser2012}
Moser A.~L.,  Bellan P.~M.,  2012, \mn@doi [\nat] {10.1038/nature10827}, 482,
  379

\bibitem[\protect\citeauthoryear{Nagai et~al.,}{Nagai
  et~al.}{2014}]{Nagai_2014}
Nagai H.,  et~al., 2014, \mn@doi [\apj] {10.1088/0004-637x/785/1/53}, 785, 53

\bibitem[\protect\citeauthoryear{{Nakamura} \& {Asada}}{{Nakamura} \&
  {Asada}}{2013}]{nak2013}
{Nakamura} M.,  {Asada} K.,  2013, \apj, 775, 118

\bibitem[\protect\citeauthoryear{Nakamura et~al.,}{Nakamura
  et~al.}{2018}]{Nakamura_2018}
Nakamura M.,  et~al., 2018, \mn@doi [\apj] {10.3847/1538-4357/aaeb2d}, 868, 146

\bibitem[\protect\citeauthoryear{{Narayan}, {Igumenshchev}  \&
  {Abramowicz}}{{Narayan} et~al.}{2003}]{nia03}
{Narayan} R.,  {Igumenshchev} I.~V.,   {Abramowicz} M.~A.,  2003, \pasj, \href
  {http://adsabs.harvard.edu/abs/2003PASJ...55L..69N} {55, L69}

\bibitem[\protect\citeauthoryear{{Narayan}, {Kumar}  \&
  {Tchekhovskoy}}{{Narayan} et~al.}{2011}]{nkt11}
{Narayan} R.,  {Kumar} P.,   {Tchekhovskoy} A.,  2011, \mn@doi [\mnras]
  {10.1111/j.1365-2966.2011.19197.x}, \href
  {http://adsabs.harvard.edu/abs/2011MNRAS.416.2193N} {416, 2193}

\bibitem[\protect\citeauthoryear{{Noble}, {Gammie}, {McKinney}  \& {Del
  Zanna}}{{Noble} et~al.}{2006}]{nob06}
{Noble} S.~C.,  {Gammie} C.~F.,  {McKinney} J.~C.,   {Del Zanna} L.,  2006,
  \mn@doi [\apj] {10.1086/500349}, \href
  {http://adsabs.harvard.edu/cgi-bin/nph-bib_query?bibcode=2006ApJ...641..626N&db_key=AST}
  {641, 626}

\bibitem[\protect\citeauthoryear{Parfrey, Philippov  \& Cerutti}{Parfrey
  et~al.}{2019}]{Parfrey2019}
Parfrey K.,  Philippov A.,   Cerutti B.,  2019, \mn@doi [Phys. Rev.~Lett.]
  {10.1103/PhysRevLett.122.035101}, 122, 035101

\bibitem[\protect\citeauthoryear{Plavin, Kovalev, Pushkarev  \& Lobanov}{Plavin
  et~al.}{2019}]{plavin2019}
Plavin A.~V.,  Kovalev Y.~Y.,  Pushkarev A.~B.,   Lobanov A.~P.,  2019,
  preprint, \href {http://adsabs.harvard.edu/abs/2018arXiv181102544P} {}
  (\mn@eprint {arXiv} {1811.02544})

\bibitem[\protect\citeauthoryear{{Polko}, {Meier}  \& {Markoff}}{{Polko}
  et~al.}{2010}]{pol10}
{Polko} P.,  {Meier} D.~L.,   {Markoff} S.,  2010, \mn@doi [ApJ]
  {10.1093/mnras/stt2155}, 723, 1343

\bibitem[\protect\citeauthoryear{Porth \& Komissarov}{Porth \&
  Komissarov}{2015}]{porth2015}
Porth O.,  Komissarov S.~S.,  2015, \mn@doi [\mnras] {10.1093/mnras/stv1295},
  452, 1089

\bibitem[\protect\citeauthoryear{Proga \& Zhang}{Proga \&
  Zhang}{2006}]{proga06}
Proga D.,  Zhang B.,  2006, \mn@doi [\mnras]
  {10.1111/j.1745-3933.2006.00189.x}, 370, L61

\bibitem[\protect\citeauthoryear{{Ptitsyna} \& {Neronov}}{{Ptitsyna} \&
  {Neronov}}{2016}]{2016A&A...593A...8P}
{Ptitsyna} K.,  {Neronov} A.,  2016, \mn@doi [\aap]
  {10.1051/0004-6361/201527549}, \href
  {http://adsabs.harvard.edu/abs/2016A%26A...593A...8P} {593, A8}

\bibitem[\protect\citeauthoryear{Pu, Nakamura, Hirotani, Mizuno, Wu  \&
  Asada}{Pu et~al.}{2015}]{Pu2015_SS}
Pu H.-Y.,  Nakamura M.,  Hirotani K.,  Mizuno Y.,  Wu K.,   Asada K.,  2015,
  \apj, 801, 56

\bibitem[\protect\citeauthoryear{{Pushkarev}, {Kovalev}, {Lister}  \&
  {Savolainen}}{{Pushkarev} et~al.}{2009}]{pushkarev2009}
{Pushkarev} A.~B.,  {Kovalev} Y.~Y.,  {Lister} M.~L.,   {Savolainen} T.,  2009,
  \mn@doi [\aap] {10.1051/0004-6361/200913422}, \href
  {http://adsabs.harvard.edu/abs/2009A%26A...507L..33P} {507, L33}

\bibitem[\protect\citeauthoryear{Pushkarev, Kovalev, Lister  \&
  Savolainen}{Pushkarev et~al.}{2017}]{Pushkarev2017}
Pushkarev A.~B.,  Kovalev Y.~Y.,  Lister M.~L.,   Savolainen T.,  2017, \mn@doi
  [\mnras] {10.1093/mnras/stx854}, 468, 4992

\bibitem[\protect\citeauthoryear{Ressler, Tchekhovskoy, Quataert  \&
  Gammie}{Ressler et~al.}{2017}]{res17}
Ressler S.~M.,  Tchekhovskoy A.,  Quataert E.,   Gammie C.~F.,  2017, \mn@doi
  [\mnras] {10.1093/mnras/stx364}, 467, 3604

\bibitem[\protect\citeauthoryear{{Ripperda}, {Porth}, {Sironi}  \&
  {Keppens}}{{Ripperda} et~al.}{2019}]{ripper2019}
{Ripperda} B.,  {Porth} O.,  {Sironi} L.,   {Keppens} R.,  2019, \mn@doi
  [\mnras] {10.1093/mnras/stz387}, \href
  {https://ui.adsabs.harvard.edu/\#abs/2019MNRAS.485..299R} {485, 299}

\bibitem[\protect\citeauthoryear{Romero, Boettcher, Markoff  \&
  Tavecchio}{Romero et~al.}{2017}]{Romero2017}
Romero G.~E.,  Boettcher M.,  Markoff S.,   Tavecchio F.,  2017, \mn@doi [Space
  Science Reviews] {10.1007/s11214-016-0328-2}, 207, 5

\bibitem[\protect\citeauthoryear{Russell et~al.,}{Russell
  et~al.}{2013}]{russellD2013}
Russell D.~M.,  et~al., 2013, \mn@doi [\mnras] {10.1093/mnras/sts377}, 429, 815

\bibitem[\protect\citeauthoryear{{Sauty}, {Trussoni}  \& {Tsinganos}}{{Sauty}
  et~al.}{2004}]{Sauty2004}
{Sauty} C.,  {Trussoni} E.,   {Tsinganos} K.,  2004, \mn@doi [A\&A]
  {10.1051/0004-6361:20035790}, 421, 797

\bibitem[\protect\citeauthoryear{Sijacki, Springel, Di~Matteo  \&
  Hernquist}{Sijacki et~al.}{2007}]{Sijacki2007}
Sijacki D.,  Springel V.,  Di~Matteo T.,   Hernquist L.,  2007, \mn@doi
  [\mnras] {10.1111/j.1365-2966.2007.12153.x}, 380, 877

\bibitem[\protect\citeauthoryear{{Silk} \& {Rees}}{{Silk} \&
  {Rees}}{1998}]{SilkRees98}
{Silk} J.,  {Rees} M.~J.,  1998, \aap, \href
  {http://adsabs.harvard.edu/abs/1998A%26A...331L...1S} {331, L1}

\bibitem[\protect\citeauthoryear{Sironi \& Spitkovsky}{Sironi \&
  Spitkovsky}{2009}]{Sironi2009}
Sironi L.,  Spitkovsky A.,  2009, \mn@doi [\apj]
  {https://doi.org/10.1088/0004-637X/698/2/1523}, 698, 1523

\bibitem[\protect\citeauthoryear{Sironi \& Spitkovsky}{Sironi \&
  Spitkovsky}{2014}]{Sironi_2014}
Sironi L.,  Spitkovsky A.,  2014, \mn@doi [\apj] {10.1088/2041-8205/783/1/l21},
  783, L21

\bibitem[\protect\citeauthoryear{{Sironi}, {Petropoulou}  \&
  {Giannios}}{{Sironi} et~al.}{2015}]{sironi2015}
{Sironi} L.,  {Petropoulou} M.,   {Giannios} D.,  2015, \mn@doi [\mnras]
  {10.1093/mnras/stv641}, 450, 183

\bibitem[\protect\citeauthoryear{Sobacchi \& Lyubarsky}{Sobacchi \&
  Lyubarsky}{2018a}]{sobacchi_2018a}
Sobacchi E.,  Lyubarsky Y.~E.,  2018a, \mn@doi [\mnras]
  {10.1093/mnras/stx2592}, 473, 2813

\bibitem[\protect\citeauthoryear{Sobacchi \& Lyubarsky}{Sobacchi \&
  Lyubarsky}{2018b}]{sobacchi_2018b}
Sobacchi E.,  Lyubarsky Y.~E.,  2018b, \mn@doi [\mnras]
  {10.1093/mnras/sty2231}, 480, 4948

\bibitem[\protect\citeauthoryear{Springel, di Matteo  \& Hernquist}{Springel
  et~al.}{2005}]{springel2005}
Springel V.,  di Matteo T.,   Hernquist L.,  2005, \mn@doi [\mnras]
  {10.1111/j.1365-2966.2005.09238.x}, 361, 776

\bibitem[\protect\citeauthoryear{Spruit, Foglizzo  \& Stehle}{Spruit
  et~al.}{1997}]{spruit97}
Spruit H.~C.,  Foglizzo T.,   Stehle R.,  1997, \mn@doi [\mnras]
  {10.1093/mnras/288.2.333}, 288, 333

\bibitem[\protect\citeauthoryear{Spruit, Daigne  \& Drenkhahn}{Spruit
  et~al.}{2001}]{spruitdaignedrenkhan_2001}
Spruit H.~C.,  Daigne F.,   Drenkhahn G.,  2001, \mn@doi [A\&A]
  {10.1051/0004-6361:20010131}, 369, 694

\bibitem[\protect\citeauthoryear{Tayler}{Tayler}{1957}]{Tayler_1957}
Tayler R.~J.,  1957, \mn@doi [Proc. Phys. Soc. B]
  {10.1088/0370-1301/70/11/305}, 70, 1049

\bibitem[\protect\citeauthoryear{{Tchekhovskoy} \& {Bromberg}}{{Tchekhovskoy}
  \& {Bromberg}}{2016}]{tch16}
{Tchekhovskoy} A.,  {Bromberg} O.,  2016, \mn@doi [\mnras]
  {10.1093/mnrasl/slw064}, 461, L46

\bibitem[\protect\citeauthoryear{{Tchekhovskoy}, {McKinney}  \&
  {Narayan}}{{Tchekhovskoy} et~al.}{2008}]{tch08}
{Tchekhovskoy} A.,  {McKinney} J.~C.,   {Narayan} R.,  2008, \mn@doi [\mnras]
  {10.1111/j.1365-2966.2008.13425.x}, \href
  {http://adsabs.harvard.edu/abs/2008MNRAS.388..551T} {388, 551}

\bibitem[\protect\citeauthoryear{{Tchekhovskoy}, {McKinney}  \&
  {Narayan}}{{Tchekhovskoy} et~al.}{2009}]{tch09}
{Tchekhovskoy} A.,  {McKinney} J.~C.,   {Narayan} R.,  2009, \mn@doi [\apj]
  {10.1088/0004-637X/699/2/1789}, 699, 1789

\bibitem[\protect\citeauthoryear{{Tchekhovskoy}, {Narayan}  \&
  {McKinney}}{{Tchekhovskoy} et~al.}{2010a}]{tch10b}
{Tchekhovskoy} A.,  {Narayan} R.,   {McKinney} J.~C.,  2010a, New Astronomy,
  \href {http://adsabs.harvard.edu/abs/2009arXiv0909.0011T} {15, 749}

\bibitem[\protect\citeauthoryear{{Tchekhovskoy}, {Narayan}  \&
  {McKinney}}{{Tchekhovskoy} et~al.}{2010b}]{tch10a}
{Tchekhovskoy} A.,  {Narayan} R.,   {McKinney} J.~C.,  2010b, \mn@doi [\apj]
  {10.1088/0004-637X/711/1/50}, \href
  {http://adsabs.harvard.edu/abs/2010ApJ...711...50T} {711, 50}

\bibitem[\protect\citeauthoryear{{Tchekhovskoy}, {Narayan}  \&
  {McKinney}}{{Tchekhovskoy} et~al.}{2011}]{tch11}
{Tchekhovskoy} A.,  {Narayan} R.,   {McKinney} J.~C.,  2011, \mn@doi [\mnras]
  {10.1111/j.1745-3933.2011.01147.x}, \href
  {http://adsabs.harvard.edu/abs/2011MNRAS.418L..79T} {418, L79}

\bibitem[\protect\citeauthoryear{{Tomimatsu}}{{Tomimatsu}}{1994}]{tom94}
{Tomimatsu} A.,  1994, \pasj, \href
  {http://cdsads.u-strasbg.fr/abs/1994PASJ...46..123T} {46, 123}

\bibitem[\protect\citeauthoryear{{Vlahakis}}{{Vlahakis}}{2004}]{vla04}
{Vlahakis} N.,  2004, \apj, \href
  {http://adsabs.harvard.edu/cgi-bin/nph-bib_query?bibcode=2004ApJ...600..324V&db_key=AST}
  {600, 324}

\bibitem[\protect\citeauthoryear{{Vlahakis} \& {K{\"o}nigl}}{{Vlahakis} \&
  {K{\"o}nigl}}{2003a}]{vk03a}
{Vlahakis} N.,  {K{\"o}nigl} A.,  2003a, \apj, \href
  {http://adsabs.harvard.edu/cgi-bin/nph-bib_query?bibcode=2003ApJ...596.1080V&db_key=AST}
  {596, 1080}

\bibitem[\protect\citeauthoryear{{Vlahakis} \& {K{\"o}nigl}}{{Vlahakis} \&
  {K{\"o}nigl}}{2003b}]{vla03a}
{Vlahakis} N.,  {K{\"o}nigl} A.,  2003b, \mn@doi [\apj] {10.1086/378226}, \href
  {http://adsabs.harvard.edu/abs/2003ApJ...596.1080V} {596, 1080}

\bibitem[\protect\citeauthoryear{Weinberger et~al.,}{Weinberger
  et~al.}{2018}]{Weinberger2019_illustris}
Weinberger R.,  et~al., 2018, \mn@doi [\mnras] {10.1093/mnras/sty1733}, 479,
  4056

\makeatother
\end{thebibliography}

%\input{ms.bbl}

%%%%%%%%%%%%%%%%% APPENDICES %%%%%%%%%%%%%%%%%%%%%

\appendix
\section{Local adaptive time-stepping}
\label{sec:LAT}
\begin{figure} 
	\includegraphics[width=\columnwidth]{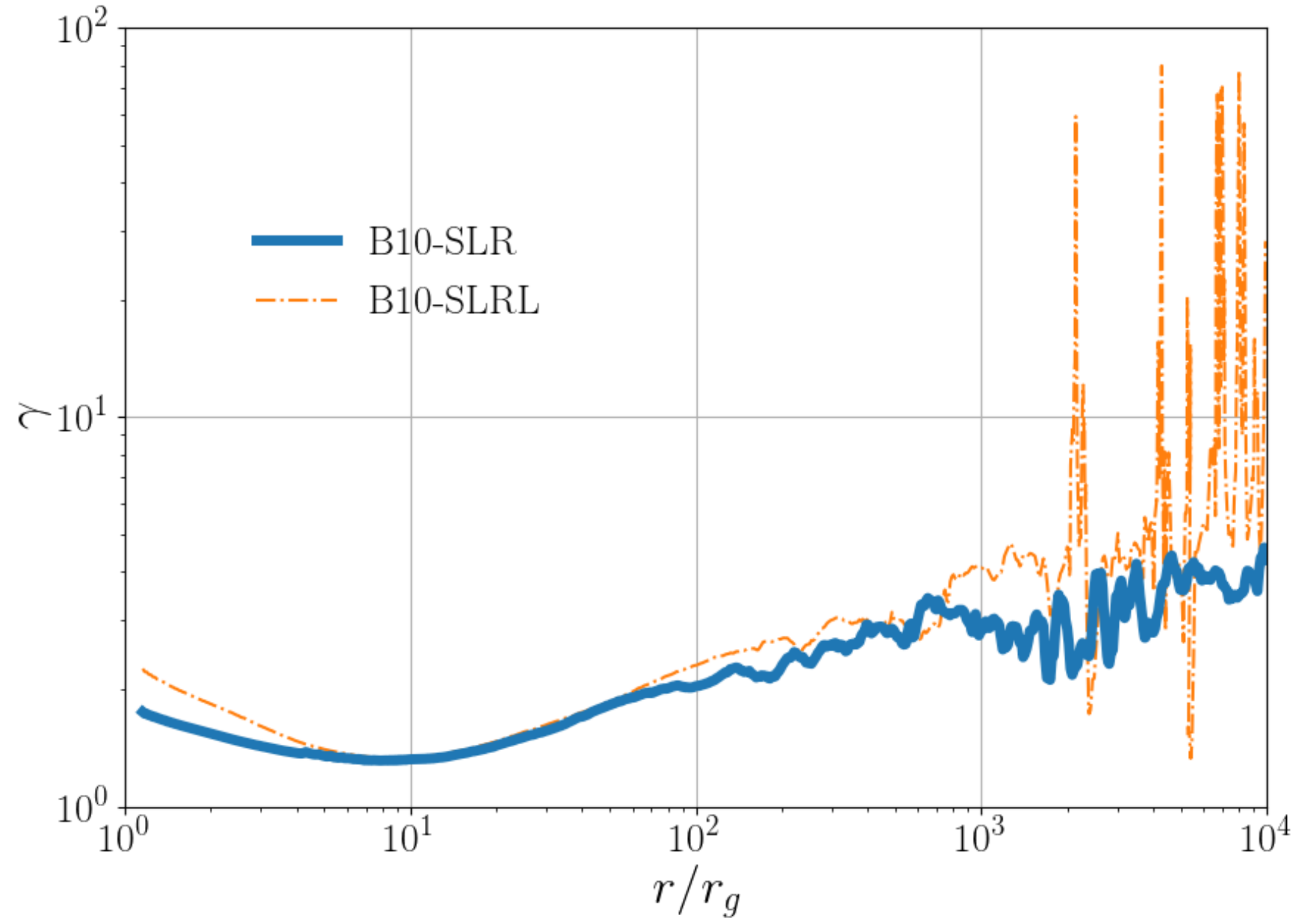}
    \caption{The Lorentz factor along a field line for our lowest resolution models, B10-SLR (with LAT implemented; in red solid) and B10-SLRL (without LAT, see text; in black dashed dotted) at $t=4\times10^4 t_g$. The spikes in $\gamma$ correspond to inversion failures and are, therefore, unphysical. Using LAT, we reduce these failures and achieve a simulation speedup by a factor $3-5$.}
    \label{fig:LAT}
\end{figure}

We have implemented in our block based AMR code \hammer{} \citep{liska2018a} a so-called local adaptive time-stepping (LAT) routine. In addition to evolving higher spatial refinement levels with a smaller time-step, similar to AMR codes with a hierarchical time-stepping routine, LAT can also use different timesteps for blocks with the same spatial refinement level. Since most GRMHD simulations utilise a logarithmic spaced spherical grid, where cell sizes are small close to the black hole and large further away from the black hole, this can speed up the simulation by an additional factor $3-5$. In a future publication we will describe the detailed implementation of the LAT algorithm and show excellent scaling on pre-exascale GPU clusters. 

LAT can also increase numerical accuracy by reducing the number of conserved to primitive variable inversions \citep{nob06}. Namely, as one moves away from the black hole on a logarithmic spaced spherical grid, the timescale of the problem increases. Evolving the outer grid with the same timestep as the inner grid leads to many unnecessary variable inversions. To illustrate that this produces noise in the outer grid, we produced two simulations with the same initial conditions at a resolution of $640\times 256 \times 1$ (as lower resolutions naturally produce more noise): one with LAT enabled (Model B10-SLR) and one with LAT switched off (named B10-SLRL). Figure~\ref{fig:LAT} shows that there are unphysical spikes in the Lorentz factor for model B10-SLRL caused by variable inversion failures in the outer jet region, which are absent in model B10-SLR. This confirms that LAT has the potential to increase speed and numerical accuracy.

\section{Grid shape}
\label{sec:grid}
\begin{figure}
\begin{subfigure}{\columnwidth}
	\includegraphics[width=\columnwidth]{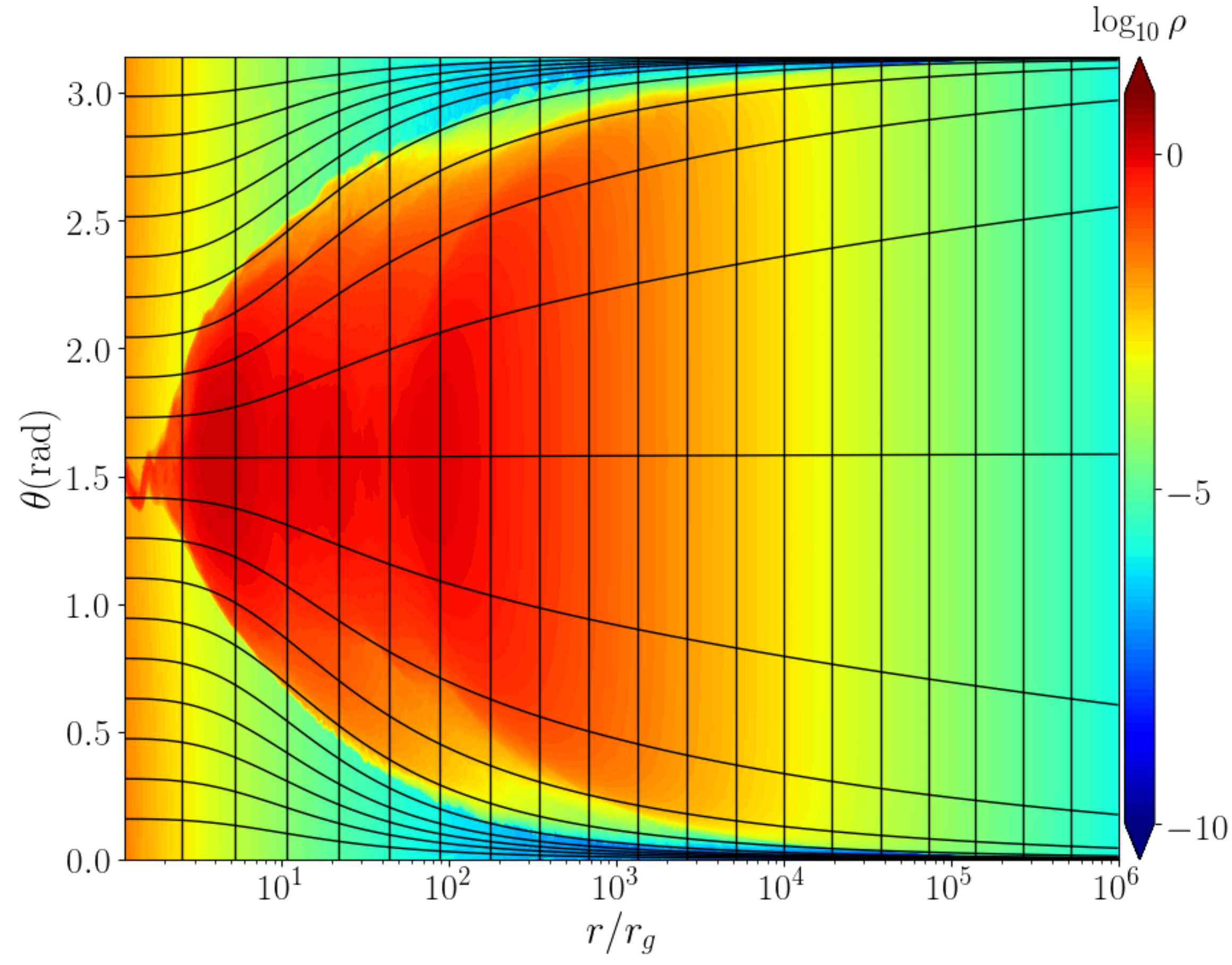}
\end{subfigure}
\begin{subfigure}{\columnwidth}
	\includegraphics[width=\columnwidth]{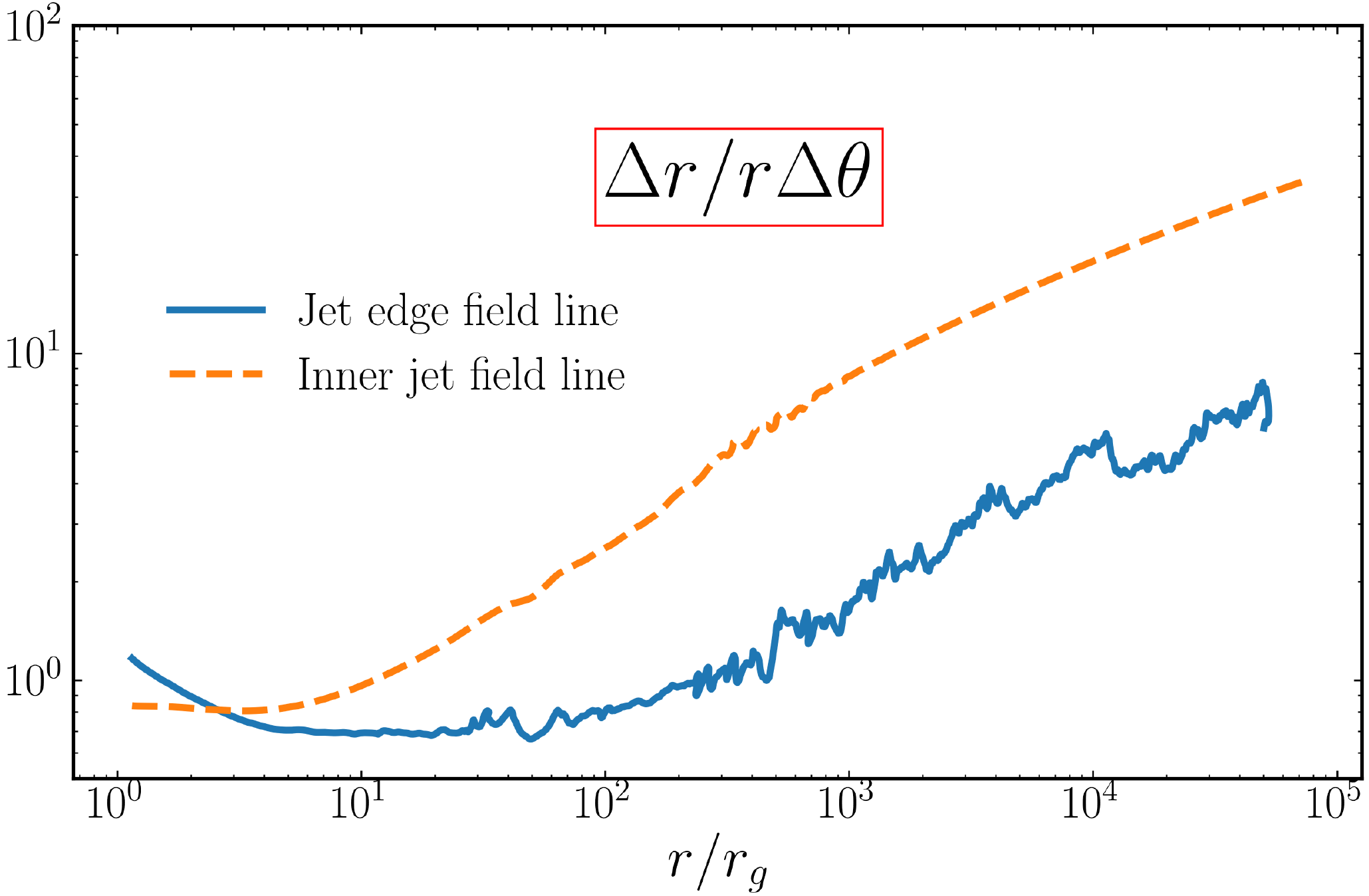}
\end{subfigure}
    \caption{\textit{Top}: A colour map of density at $t=0$ for fiducial model B10 overplotted by grid lines (black). As designed, the grid follows the shape of the jet. \textit{Bottom}: the cell aspect ratio $\Delta r/r \Delta \theta$ for two field lines, one in the inner jet and the other near the jet-edge. To resolve the jet's microstructure in both dimensions this ratio is ideally kept below $10$.} 
    \label{fig:grid}
\end{figure} 
We have designed a grid that can track the shape of the jet over $5$ orders of magnitude in distance (Fig.~\ref{fig:grid}, top). Furthermore, the grid keeps the cell aspect ratio in the outer jet below $10$ (Fig.~\ref{fig:grid}, bottom), such that turbulent eddies at the disc-jet boundary remain resolved. The internal grid coordinates ($x_0$, $x_1$, $x_2$, $x_3$) are related to the real physical coordinates ($t$, $r$, $\theta$, $\phi$) as follows,
\begin{align}
t&=x_0,\\
r&=\exp\left(x_1^{n_r}\right),\\
\theta&=A_1\pi x_1+\pi(1-A_1)\left[A_3x_2^{A_2}+\frac{1}{2\pi}\sin\left(\pi+2\pi A_3x_2^{A_2}\right)\right],\\
\phi&=x_3,
\end{align}

\noindent where $A_1=[1+g_{1j}(\log_{10}r)^{g_{2j}}]^{-1}$, $A_2=g_{3j}\log_{10}r+g_{4j}$ and $A_3=0.5^{1-A_2}$. The parameters $g_{1j}=0.8$, $g_{2j}=3.0$, $g_{3j}=0.5$ and $g_{4j}=1.0$ are used to focus resolution on the jet, while $n_r=0.95$ is used to focus extra resolution on outer parts of the grid. 

\section{Convergence of jet properties}
\label{sec:convergence}

\begin{figure} 
	\includegraphics[width=\columnwidth]{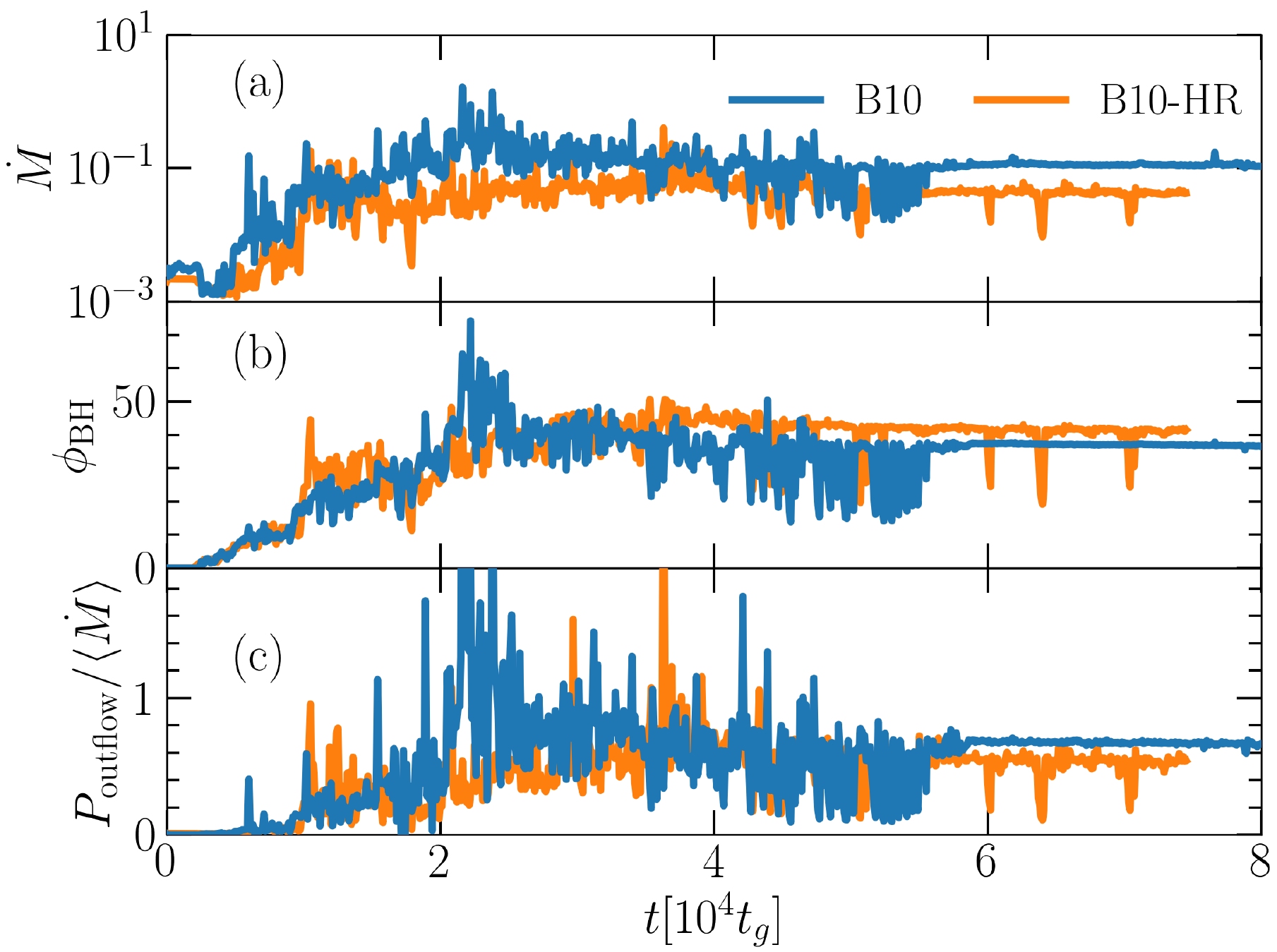}
    \caption{Accretion rate $\dot{M}$, normalised magnetic flux $\phi_{\rm BH}$ and normalised jet power from model B10 is compared with the high resolution B10-HR model. The parameters evolve similarly which gives us confidence that the B10 model is sufficiently converged. $\dot{M}$ averaged over $5-7.5\times10^4t_g$ is taken as the normalisation.}
    \label{fig:convergence:-time}
\end{figure}

\begin{figure} 
	\includegraphics[width=\columnwidth]{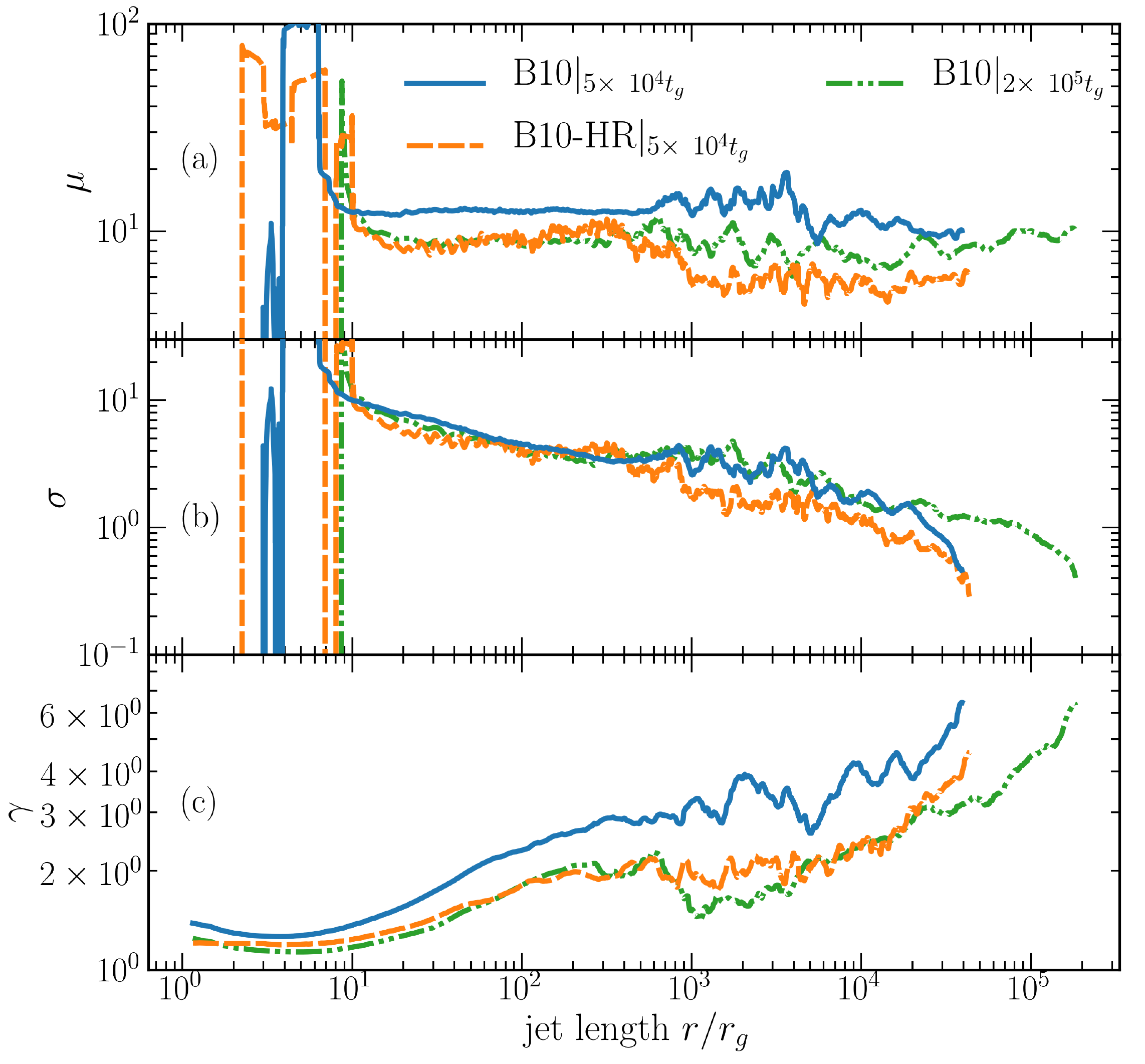}
    \caption{The specific total energy $\mu$, magnetisation $\sigma$ and the Lorentz factor $\gamma$ along a field line close to the upper jet axis from model B10 is compared with the high resolution B10-HR model at $5\times10^4t_g$. Additionally, the parameters for B10 at $2\times10^5t_g$ is also shown. Resolving the pinch instabilities is thus important as the jet is mass-loaded over time, vastly changing dynamics.}
    \label{fig:convergence:-energetics}
\end{figure}

Here we compare the time and spatial evolution of jets produced by the fiducial model B10 and its exceedingly high resolution version B10-HR. B10-HR has a resolution of $18000\times1200\times1$ and extends till $10^5r_g$, with all other parameters the same as B10. Figure~\ref{fig:convergence:-time} shows that the accretion properties of the B10 disc-jet system converges very well with respect to B10-HR. However, Fig.~\ref{fig:convergence:-energetics} shows that the jet in B10-HR gets mass-loaded ($\mu$ drops) earlier than the jet in B10, presumably because mass-loading is more efficient at higher resolutions, which capture the small scale eddies better. 

%%%%%%%%%%%%%%%%%%%%%%%%%%%%%%%%%%%%%%%%%%%%%%%%%%

\bsp	% typesetting comment
\label{lastpage}
\end{document}